\documentclass{LMCS}

\def\doi{9(2:01)2013}
\lmcsheading%
{\doi}
{1--39}
{}
{}
{Sep.~28, 2012}
{Apr.~\phantom02, 2013}
{}

\usepackage{enumerate}
\usepackage{hyperref}
\usepackage[english]{babel}
\usepackage{latexsym,amssymb,stmaryrd}
\usepackage{amsmath}
\usepackage{url}
\usepackage{xspace}
\usepackage{pst-node, pst-tree}

\theoremstyle{plain}\newtheorem{satz}[thm]{Satz}
\def\eg{{\em e.g.}}
\def\cf{{\em cf.}}

\newcommand{\coment}[1]{}
\newcommand{\remarque}[1]{\marginpar{\scriptsize #1}}

\def\frew#1#2#3#4#5#6#7#8{
\setbox0=\hbox{$#6 #7 #1 #8$}%
\setbox1=\hbox{$#6 #7 #2 #8$}%
\ifdim \wd0>\wd1 \rlap{\rlap{\hbox to \wd0{#5}}%
                            {\hbox to\wd0{\hfil\lower #3\box1\relax\hfil}}}{\raise #4\box0}%
\else \rlap{\rlap{\hbox to \wd1{#5}}{\hbox to\wd1{\hfil\raise #4\box0\relax\hfil}}}{\lower #3\box1}%
\fi
}

\def\fstep#1#2#3#4#5{\mathchoice{\frew{#1}{#2}{1.1ex}{1.1ex}{#5}{\scriptstyle}{#3}{#4}}%
                                {\frew{#1}{#2}{0.82ex}{1.1ex}{#5}{\scriptstyle}{#3}{#4}}%
                                {\frew{#1}{#2}{0.51ex}{0.82ex}{#5}{\scriptscriptstyle}{#3}{#4}}%
                                {\frew{#1}{#2}{0.51ex}{0.69ex}{#5}{\scriptscriptstyle}{#3}{#4}}}

\newcommand{\lrstep}[2]{\mathrel{\fstep{#1}{#2}{\;\>}{\>\>\;}{\rightarrowfill}}}

\def\upder#1{\!\mathrel{\mathop{\null\rightarrow}\limits^{#1}}}

\def\final{\ensuremath{\mathsf{f}}}
\newcommand{\height}{\ensuremath{h}}
\newcommand{\Pos}{\ensuremath{\mathit{Pos}}}

\newcommand{\term}{\ensuremath{\mathtt{term}}}
\def\A{\ensuremath{\mathcal{A}}}

\def\F{\ensuremath{\Sigma}}
\def\X{\mathcal{X}}
\def\T{\ensuremath{\mathcal{T}}}
\def\U{\ensuremath{\mathcal{U}}}
\def\L{\ensuremath{\mathcal{L}}}
\def\rootp{\ensuremath{\lambda}}

\def\TA{\texttt{TA}\xspace}
\def\TAGED{\texttt{TAGED}\xspace}
\def\TACBB{\texttt{TACBB}\xspace}
\def\TACBBF{\texttt{TAB}\xspace}
\def\TAGBCF{\texttt{TABG}\xspace}
\def\TAGC{\texttt{TAG}\xspace}
\def\PCTAGC{\TAGC^{\wedge}}

\def\PCTAGBCF{\ensuremath{\mathtt{TABG}^{\wedge}}}
\def\HAGC{\texttt{HAG}\xspace}
\def\UTASC{\texttt{UTASC}\xspace}
\def\ta{\ensuremath{\mathit{ta}}}

\def\tacbbf{\ensuremath{\mathit{tab}}}

\def\MSO{\ensuremath\mathtt{MSO}}
\def\EMSO{\ensuremath\exists\mathtt{MSO}}
\def\FO2{\ensuremath{\mathtt{FO}^2}}
\newcommand{\ofarity}{\mathbin{:}}
\newcommand{\ext}{\mathtt{curry}}
\def\true{\mathit{true}}
\def\false{\mathit{false}}

\begin{document}

\title[TA with Local and Global Constraints]{Decidable Classes of Tree Automata Mixing Local and Global Constraints Modulo Flat Theories\rsuper*}

\author[L.~Bargu\~n\'o]{Luis Bargu\~n\'o\rsuper a}	
\address{{\lsuper{a,b,c}}Universitat Polit\`ecnica de Catalunya, Jordi Girona 1, Barcelona, Spain}	
\email{\{luisbargu,ccreuslopez\}@gmail.com, ggodoy@lsi.upc.edu}  

\author[C.~Creus]{Carles Creus\rsuper b}	
\address{\vskip-6 pt}	
\thanks{{\lsuper{a,b,c}}The first three authors were supported by
the FORMALISM project (TIN2007-66523), 
by the SweetLogics project (TIN2010-21062-C02-01), and
by an FI-DGR grant.}	

\author[G.~Godoy]{Guillem Godoy\rsuper c}	
\address{\vskip-6 pt}	

\author[F.~Jacquemard]{Florent Jacquemard\rsuper d}	
\address{{\lsuper d}INRIA Saclay, LSV-CNRS/ENS Cachan}	
\email{florent.jacquemard@lsv.ens-cachan.fr}  
\thanks{{\lsuper{d,e}}The last two authors were supported by the
Future and Emerging Technologies (FET) program
under the 
FET-Open grant agreement FOX number FP7-ICT-23359,
and by the INRIA ARC 2010 project ACCESS}	

\author[C.~Vacher]{Camille Vacher\rsuper e}	
\address{{\lsuper e}LIFL, Univ. Lille I, INRIA Lille, 40 avenue Halley, 59650 Villeneuve d'Ascq, France}	
\email{vacher@lsv.ens-cachan.fr}  



\ACMCCS{[{\bf  Theory of computation}]:  Formal languages and automata
  theory---Tree languages;  Logic---Higher order logic} 

\titlecomment{{\lsuper*}A preliminary version~\cite{DBLP:conf/lics/BargunoCGJV10}
appeared in the Proceedings of the 25th Annual IEEE Symposium on
Logic In Computer Science (LICS 2010).
Here, we generalize the results by allowing the constraints to be interpreted
modulo a flat equational theory, and make the results
stronger and easier to follow by presenting
completely new proofs for the part on arithmetic
constraints.
}


\begin{abstract}
  \noindent 
We define a class of ranked tree automata TABG
generalizing both the tree automata with local tests between brothers
of Bogaert and Tison (1992)
and with global equality and disequality constraints
(TAGED) of Filiot et al. (2007).
TABG can test for equality and disequality modulo a given flat equational theory
between brother subterms and between subterms whose positions are defined by the states reached during a computation. In particular, TABG can check that all the subterms reaching a given state are distinct. 
This constraint is related to monadic key constraints for XML documents, meaning that every two distinct positions of a given type have different values.

We prove decidability of the emptiness problem for TABG. This solves, in particular, 
the open question of the decidability of emptiness for TAGED.
We further extend our result by allowing global arithmetic constraints for counting the number of occurrences of some state or the number of different 
equivalence classes of subterms (modulo a given flat equational theory)
reaching some state during a computation. We also adapt the model to unranked ordered terms.
As a consequence of our results for TABG, we prove the decidability of a fragment of the monadic second order logic on trees extended with predicates for equality and disequality between subtrees, 
and cardinality.
\end{abstract}

\maketitle

\section{Introduction}
Tree automata techniques are widely used in several domains
like automated deduction (see e.g.~\cite{tata}),
static analysis of programs~\cite{BouajjaniT05} or protocols~\cite{VermaG07,FeuilladeGT04},
and XML processing~\cite{Schwentick07}.
However, a severe limitation of standard tree automata (\TA) is
that they are not able to test for equality (isomorphism) or disequality between 
subterms in an input term.
For instance, the language of terms matching a non-linear pattern
such as $f(x,x)$ is not regular 
(i.e.\ there exists no \TA recognizing this language).
Let us illustrate how this limitation can be problematic 
in the  context of XML documents processing.
XML documents are commonly represented as labeled trees,
and they can be 
constrained by {XML schemas}, 
which define both typing restrictions and integrity constraints.
All the typing formalisms currently used for XML are based on finite tree automata.
The key constraints for databases are common integrity constraints 
expressing that every two distinct positions of a given type have different values. 
This is typically the kind of constraints that can not be characterized by \TA.

One first approach to overcome this limitation of \TA consists in 
adding the possibility to make equality or disequality tests at each step 
of the computation of the automaton.
The tests are performed \emph{locally}, between subterms at a bounded
distance from the current
computation position in the input term.
The emptiness problem, i.e.\ whether the language recognized by a given automaton
is empty, is undecidable with such tests~\cite{Mongy81}.
A decidable subclass is obtained by restricting 
the tests to sibling subterms~\cite{BogaertTison92}
(see~\cite{tata} for a survey).

Another approach was proposed more recently 
in~\cite{FiliotTalbotTison07,FiliotTalbotTison08}
with the definition of tree automata with \emph{global} 
equality and disequality tests  (\TAGED).
The \TAGED do not perform the tests during the computation steps
but globally on the term, at the end of the computation,
at positions which are defined by the states reached during the computation. 
For instance, they can express that all the subterms that reached a
given state $q$ are equal, 
or that every two subterms that reached respectively the states $q$
and $q'$ are different. Nevertheless, arbitrary disequalities are
not allowed in \TAGED, since such $q$ and $q'$ must be different.
The emptiness has been shown decidable for several subclasses of 
\TAGED~\cite{FiliotTalbotTison07,FiliotTalbotTison08},
but the decidability of emptiness for the whole class remained a challenging open question.

\medskip

In this paper, we answer this question positively, 
for a class of tree recognizers more general than $\TAGED$.
We propose (in Section~\ref{section-definition}) 
a class of tree automata with 
local constraints between siblings and global constraints ($\TAGBCF$)
which significantly extends $\TAGED$ in several directions:
$(i)$ $\TAGBCF$ combine global constraints a la $\TAGED$
with \emph{local} equality and disequality constraints between brother 
subterms a la~\cite{BogaertTison92},  
$(ii)$ the equality and disequality constraints are treated
modulo a given flat equational theory
(here flat means that both sides of the equation have the same variables
and height, and that this height is bounded by $1$),
allowing to consider relations more general than syntactic equalities and
disequalities, like e.g.\ structural equalities and disequalities,
$(iii)$ testing global disequality constraints 
between subterms that reached the same state is allowed
(such test specify key constraints, which are not expressible with $\TAGED$),
$(iv)$ the global constraints are arbitrary Boolean combinations
(including negation) of atomic equality and disequality
(in $\TAGED$, only conjunction of atoms are allowed, without negation).

In Section~\ref{section-arithmetic}, 
we consider the addition to $\TAGBCF$ of global counting constraints on
the number $|q|$ of occurrences of a given state $q$ in a computation,
or the number $\|q\|$ of distinct
equivalence classes (modulo the flat theory) of subterms 
reaching a given state $q$ in a computation. 
These counting constraints are only allowed to compare states to constants,
like in $|q| \leq 5$ or $\|q\| + 2 \|q'\| \geq 9$
(with counting constraints being able to compare state cardinalities, 
like in $|q| = |q'|$, the emptiness problem becomes undecidable).
Using this formalism as an intermediate step, 
we show that negative literals and disjunctions
can be eliminated without loss of generality 
in the global constraints of $\TAGBCF$,
i.e.\ that $\TAGBCF$ whose global constraints are restricted to be
conjunctions of positive literals
(namely positive conjunctive $\TAGBCF$)
have already the same expressiveness of the full $\TAGBCF$ class.
In particular, the counting constraints do not improve the expressiveness of $\TAGBCF$.

Our main result, presented in Section~\ref{section-emptiness},
is that emptiness is decidable for positive conjunctive $\TAGBCF$
(and hence for $\TAGBCF$).
The decision algorithm uses an involved pumping argument:
every sufficiently large term recognized by the given $\TAGBCF$ 
can be reduced by an operation of parallel pumping into a
smaller term which is still recognized.
The existence of the bound for the minimum accepted term
is based on a particular well quasi-ordering.

We show
that the emptiness decision algorithm of Section~\ref{section-emptiness} 
can also be applied to a generalization of 
the subclass $\TAGC$ of $\TAGBCF$ without the local constraints
computing on unranked ordered labeled trees (Section~\ref{section-unranked}).
This demonstrates the robustness of the method. 

As an application of our results,
in Section~\ref{section-logic} we present
a (strict) extension of the monadic second order logic on trees
whose existential fragment corresponds exactly to $\TAGC$.
In particular, we conclude its decidability.

\subsection*{Related Work}

\TAGBCF is a strict (decidable) extension of \TAGC and \TA
with local equality and disequality constraints, since the
expressiveness of both subclasses is incomparable (see e.g.~\cite{JKV-lata09}).

The tree automata model of~\cite{BogaertTison92} 
has been generalized from ranked trees to
unranked ordered trees
into a decidable class called $\UTASC$~\cite{KL07,LW09}.
In unranked trees, the number of brothers (under a position) is unbounded,
and $\UTASC$ transitions use $\MSO$ formulae (on words) with 2 free variables
in order to select the sibling positions
to be tested for equality and disequality.
The decidable generalization of $\TAGC$ to unranked ordered trees proposed in
Section~\ref{section-unranked} and the automata of~\cite{KL07,LW09}
are incomparable. The combination of both formalisms could be the object of a further study.

Another way to handle subterm equalities is to use 
automata computing on DAG representation 
of terms~\cite{Charatonik99,SivaNarendranRusinowitch05dag}.
This model is incomparable to $\TAGC$ whose constraints 
are conjunctions of equalities~\cite{JKV-lata09}.
The decidable extension of \TA with one tree shaped
memory~\cite{ComonCortier05}   
can simulate $\TAGC$ with equality constraints only, 
providing that at most one state per run can be used to test 
equalities~\cite{FiliotTalbotTison07}.
%

We show in Section~\ref{section-tagc} that
the $\TAGBCF$ strictly generalize the $\TAGED$ 
of~\cite{FiliotTalbotTison07,FiliotTalbotTison08}.
The latter have been introduced as a tool to 
decide a fragment of the spatial logic $\mathtt{TQL}$ \cite{FiliotTalbotTison07}.
Decidable subclasses of $\TAGED$ were also shown decidable
in correspondence with fragments of monadic second order logic on the tree
extended with predicates for subtree (dis)equality tests.
In Section~\ref{section-logic}, we generalize this correspondence
to $\TAGC$ and a more natural extension of $\MSO$.

There have been several approaches to extend \TA
with arithmetic constraints on cardinalities $|q|$ described above:
the constraints can be added to transitions in order to count between 
siblings~\cite{MuschollSS-PODS03,DalZilioLugiez06}
(in this case we could call them \emph{local} by analogy with equality tests)
or they can be \emph{global}~\cite{Klaedtke02parikhautomata}.
We compare in Section~\ref{section-arithmetic}
the latter approach (closer to our settings)
with our extension of $\TAGBCF$,
with respect to emptiness decision.
To our knowledge, this is the first time that 
arithmetic constraints on cardinalities of the form $\| q \|$ are studied.



\section{Preliminaries}\label{section-preliminaries}

\subsection{Terms, Positions, Replacements} 

We use the standard notations for terms and positions, see~\cite{Allthat}.
A \emph{signature} $\F$ is a finite set of function symbols with arity.
We sometimes denote $\F$ explicitly as 
$\{f_1\ofarity a_1,\ldots, f_n\ofarity a_n\}$
where $f_1,\ldots,f_n$ are the function symbols, and
$a_1,\ldots,a_n$ are the corresponding arities, or as
$\{f_1,\ldots,f_n\}$ when the arities are omitted.
We denote the subset of function symbols of $\F$ of arity $m$ as $\F_m$.
The set of (ranked) \emph{terms} over the signature $\F$
is defined recursively as
$\T(\F) := \{ f(t_1,\ldots, t_m) \mid f\ofarity m \in \F, t_1,\ldots,
t_m \in \T(\F) \}$. Note that the base case of this definition is
$\{f \mid f\ofarity 0 \in \F\}$, which coincides with $\F_0$ by
omitting the arity. Elements of this subset are called
constants.

Positions in terms are denoted by sequences of natural numbers.
With $\rootp$ we denote the empty sequence (root position), and $p.p'$ denotes
the concatenation of positions $p$ and $p'$.
The set of positions of a term is defined recursively as 
$\Pos\bigl(f(t_1,\ldots,t_m)\bigr)=\{\rootp\}\cup\{i.p\;|\;i\in\{1,\ldots,m\}\wedge p\in\Pos(t_i)\}$.
A term $t \in \T(\F)$ 
can be seen as a function from its set of positions $\Pos(t)$ into $\F$.
For this reason, the symbol labeling the position $p$ in $t$ shall be denoted by $t(p)$.
By $p<p'$ and $p\leq p'$ we denote that $p$ is a proper prefix of $p'$,
and that $p$ is a prefix of $p'$, respectively. In these cases, $p'$
is necessarily of the form $p.p''$, and we define $p'-p$ as $p''$.
Two positions $p_1,p_2$
incomparable with respect to the prefix ordering are called \emph{parallel},
and it is denoted by $p_1\parallel p_2$.
The \emph{subterm} of $t$ at position $p$, denoted $t|_p$,
is defined recursively as $t|_\rootp=t$ and
$f(t_1,\ldots,t_m)|_{i.p}=t_i|_p$.
The replacement in $t$ of the subterm at position $p$ by $s$, denoted $t[s]_p$,
is defined recursively as $t[s]_\rootp=s$ and
$f(t_1,\ldots,t_{i-1},t_i,t_{i+1},\ldots,t_m)[s]_{i.p}=
f(t_1,\ldots,t_{i-1},t_i[s]_p,t_{i+1},\ldots,t_m)$.
The \emph{height} of a term $t$, denoted $\height(t)$, 
is the maximal length of a position of $\Pos(t)$.
In particular, the length of $\rootp$ is $0$.

\subsection{Tree automata}

A \emph{tree automaton} ($\TA$, see e.g.~\cite{tata})
is a tuple $\A = \langle Q, \F,F, \Delta\rangle$
where $Q$ is a finite set of \emph{states}, $\F$ is a signature,
$F \subseteq Q$ is a subset of final (or accepting) states and $\Delta$ is 
a set of \emph{transition} rules of the form $f(q_1,\ldots,q_m) \to q$
where $f\ofarity m \in \F$, $q_1,\ldots,q_m, q \in Q$.
Sometimes, we shall refer to $\A$ as a subscript of its components,
like in $Q_\A$ to indicate that this is the set of states of $\A$.

A \emph{run} of $\A$ is a pair $r=\langle t,M\rangle$ 
where $t$ is a term in $\T(\F)$ and
$M:\Pos(t)\to \Delta_\A$ is a mapping satisfying the following
statement for each $p \in \Pos(t)$:
if $t|_p$ is written of the form $f(t_1,\ldots,t_m)$, and
$M(p.1),\ldots,M(p.m)$ are rules with right-hand side states
$q_1,\ldots,q_m\in Q_\A$, respectively, then $M(p)$ is a rule of
the form $f(q_1,\ldots,q_m)\to q$ for some $q\in Q_\A$.
We write $r(p)$ for the right-hand side state
of $M(p)$, and say that
$r$ is a run of $\A$ on $t$. 
Moreover, by $\term(r)$ we refer to $t$,
and by ${\tt symbol}(r)$ we refer to $t(\rootp)$.
The run $r$ is called \emph{successful} (or \emph{accepting})  
if $r(\rootp)$ is in $F_\A$.
The language $\L(\A)$ of $\A$ is the set of terms $t$ for which there 
exists a successful run of $\A$.
A language $L$ is called \emph{regular} if there exists a \TA $\A$
satisfying $L=\L(\A)$.
For facility of explanations, we shall use term-like notations for runs
defined as follows in the natural way.
For a run $r=\langle t,M\rangle$,
by $\Pos(r)$ we denote $\Pos(t)$, and
by $\height(r)$ we denote $\height(t)$.
Similarly, by $r|_p$ we denote
the run $\langle t|_p,M|_p\rangle$, where $M|_p$ is defined
as $M|_p(p')=M(p.p')$ for each $p'$ in $\Pos(t|_p)$,
and say that $r|_p$ is a subrun of $r$.
Moreover, for a run $r'=\langle t',M'\rangle$
such that the states $r'(\lambda)$ and $r(p)$ coincide,
by $r[r']_p$ we denote the run
$\langle t[t']_p,M[M']_p\rangle$,
where $M[M']_p$ is defined
as $M[M']_p(p.p')=M'(p')$ for each $p'$ in $\Pos(t')$,
and as $M[M']_p(p')=M(p')$ for each $p'$ with $p\not\leq p'$.

\subsection{Tree automata with local constraints between brothers}

A \emph{tree automaton with constraints between brothers} 
(defined in~\cite{BogaertTison92} and called $\TACBB$ 
in~\cite{tata})
is a tuple $\A = \langle Q, \F, F, \Delta\rangle$ 
where $Q$, $\F$ and $F$ are defined as for \TA, 
but with the difference that $\Delta$ is
a set of constrained rules of the form
$f(q_1,\ldots,q_m) \upder{C} q$, where $C$ is a set of equalities
and disequalities of the form $i\approx j$ or $i\not\approx j$
for $i,j\in\{1,\ldots,m\}$.
We call $C$ a local {\em constraint between brothers}.
By $\ta(\A)$ we define the \TA obtained from $\A$ by removing all constraints
from $\Delta$.

A \emph{run} of a $\TACBB$ $\A$ is a pair $r=\langle t,M\rangle$ 
defined similarly to the case of \TA; 
$t$ is a term in $\T(\F)$ and 
the mapping $M:\Pos(t)\to \Delta_\A$ satisfies the following
statement for each $p \in \Pos(t)$:
if $t|_p$ is written of the form $f(t_1,\ldots,t_m)$, and
$M(p.1),\ldots,M(p.m)$ are rules with right-hand side states
$q_1,\ldots,q_m\in Q_\A$, respectively, then $M(p)$ is a rule of
the form $f(q_1,\ldots,q_m)\upder{C} q$ for some $q\in Q_\A$ and
constraint between brothers $C$. Moreover, for each equality
$i\approx j$ in $C$, $t_i=t_j$ holds, and for each disequality
$i\not\approx j$ in $C$, $t_i\not=t_j$ holds.
The notions of successful run and recognized language
are defined for $\TACBB$ analogously
to the case of \TA.

\subsection{Term equations}

Given a set of variables $\X$,
the set of (ranked) \emph{terms} over $\F$ and $\X$
is defined as $\T(\F\cup\X)$ by considering arity $0$
for the elements of $\X$. A substitution $\sigma$ is a mapping
from variables to terms $\sigma:\X\to\T(\F\cup\X)$. It
is also considered as a function from
arbitrary terms to terms $\sigma:\T(\F\cup\X)\to\T(\F\cup\X)$
by the recursive definition
$\sigma(f(t_1,\ldots,t_m))=f(\sigma(t_1),\ldots,\sigma(t_m))$
for every function symbol $f$ and subterms $t_1,\ldots,t_m$.

An equation between terms is an unordered pair of terms
denoted $l\approx r$. Given a set of equations $E$
and two terms $s,t$, we say that $s$ and $t$ are equivalent
modulo $E$, denoted $s=_E t$, if there exist
terms $s_1,s_2,\ldots,s_n, n\geq 1$ satisfying the
following statement: $s=s_1$, $s_n=t$,
and for each $i\in\{1,\ldots,n-1\}$, there exists an
equation $l\approx r$ in $E$, a substitution $\sigma$,
and a position $p$, such that $s_i|_p=\sigma(l)$ and
$s_{i+1}=s_i[\sigma(r)]_p$.
A {\em flat equation} is an equation $l\approx r$ where $l$
and $r$ are terms satisfying $\height(l)=\height(r)\leq 1$,
and any variable $x$ occurs
in $l$ if and only if $x$ occurs in $r$. A {\em flat theory} is
a set of flat equations.

The following technical lemma shows that equivalence
modulo a flat theory is preserved by certain replacements
of subterms. It will be useful in Section~\ref{section-emptiness}.

\begin{lem}\label{lemma-nightmare}
Let $E$ be a flat theory.
Let $s=f(s_1,\ldots,s_n)$, $t=g(t_1,\ldots,t_m)$,
$s'=f(s_1',\ldots,s_n')$ and $t'=g(t_1',\ldots,t_m')$
be terms satisfying the following conditions:
\begin{iteMize}{$\bullet$}
\item For each $i\in\{1,\ldots,n\}$, 
$(s_i\in\F_0\Leftrightarrow s_i'\in\F_0)$ and
$(s_i,s_i'\in\F_0\Rightarrow s_i=_E s_i')$ hold.
\item For each $j\in\{1,\ldots,m\}$,
$(t_j\in\F_0\Leftrightarrow t_j'\in\F_0)$ and
$(t_j,t_j'\in\F_0\Rightarrow t_j=_E t_j')$ hold.
\item For each $i\in\{1,\ldots,n\}$ and
$j\in\{1,\ldots,m\}$, 
$(s_i'=_E t_j'\Leftrightarrow s_i=_E t_j)$ holds.
\end{iteMize}
Then, $s=_E t\Leftrightarrow s'=_E t'$ holds.
\end{lem}

\proof
We prove the left-to-right direction only.
The other one is analogous by swapping the roles of
$s$ and $t$ by the roles of $s'$ and $t'$, respectively.

Since $s=_E t$ holds, there exist 
terms $u_1,u_2,\ldots,u_k, k\geq 1$ satisfying the
following statement: $s=u_1$, $u_k=t$,
and for each $i\in\{1,\ldots,k-1\}$, there exists an
equation $l\approx r$ in $E$, a substitution $\sigma$,
and a position $p$, such that $u_i|_p=\sigma(l)$ and
$u_{i+1}=u_i[\sigma(r)]_p$.

We prove the statement by induction on $k$. For $k=1$,
$s=t$ holds. Thus, $g$ is $f$, $m$ is $n$, and for each
$i\in\{1,\ldots,n\}$,
$s_i=t_i$ holds. In particular,
each $s_i=_E t_i$ holds. Therefore, each $s_i'=_E t_i'$
also holds, and hence $s'=f(s_1',\ldots,s_n')=_E
f(t_1',\ldots,t_n')=t'$ holds.

Now, assume $k>1$. Let $l\approx r$, $p$ and $\sigma$ be the rule,
position and substitution satisfying 
$u_1|_p=\sigma(l)$ and
$u_2=u_1[\sigma(r)]_p$. Recall that $u_1$ is $s$.
First, suppose that $p$ is not $\rootp$. Then, $p$ is
of the form $j.p'$ for some $j\in\{1,\ldots,n\}$ and position $p'$.
Note that $u_2|_j=_E u_1|_j$ holds, and for each $i\in\{1,\ldots,n\}\setminus\{j\}$,
$u_2|_i=u_1|_i$ holds.
Thus, $u_2$ is of the form
$f(v_1,\ldots,v_n)$ and for each $i\in\{1,\ldots,n\}$,
$v_i=_E s_i$ holds. Moreover,
since $E$ is a flat theory, the step at $p$ preserves the height,
and hence,
for each $i\in\{1,\ldots,n\}$, $v_i\in\F_0\Leftrightarrow s_i\in\F_0$
and $v_i,s_i\in\F_0\Rightarrow v_i=_E s_i$ hold.
From the statement of the lemma, the following conditions
follow:
\begin{iteMize}{$\bullet$}
\item For each $i\in\{1,\ldots,n\}$, 
$(v_i\in\F_0\Leftrightarrow s_i'\in\F_0)$ and
$(v_i,s_i'\in\F_0\Rightarrow v_i=_E s_i')$ hold.
\item For each $j\in\{1,\ldots,m\}$,
$(t_j\in\F_0\Leftrightarrow t_j'\in\F_0)$ and
$(t_j,t_j'\in\F_0\Rightarrow t_j=_E t_j')$ hold.
\item For each $i\in\{1,\ldots,n\}$ and
$j\in\{1,\ldots,m\}$, 
$(s_i'=_E t_j'\Leftrightarrow v_i=_E t_j)$ holds.
\end{iteMize}
By induction hypothesis, $f(s_1',\ldots,s_n')=_E g(t_1',\ldots,t_m')$
holds, and we are done.

Now, consider the case where $p$ is $\rootp$. In this case
$s=u_1=\sigma(l)$, and $u_2=\sigma(r)$. Since $E$ is
a flat theory, $l$ and $r$ are of the form
$f(\alpha_1,\ldots,\alpha_n)$ and $h(\beta_1,\ldots,\beta_\mu)$,
where either $n,\mu>0$ or $n=\mu=0$, and $\alpha_1,\ldots,\alpha_n,
\beta_1,\ldots,\beta_\mu$ are either constants or variables.
Moreover, a variable occurs in $l$ if and only if it occurs in $r$.
Note that $\sigma(\alpha_1)=s_1,\ldots,\sigma(\alpha_n)=s_n$ holds.
We call $v_1=\sigma(\beta_1),\ldots,v_\mu=\sigma(\beta_\mu)$.
Note that $u_2=h(v_1,\ldots,v_\mu)$.
We define terms $v_1',\ldots,v_\mu'$ as follows for each
$i$ in $\{1,\ldots,\mu\}$. If $v_i$ is a constant, then we
define $v_i'$ as $v_i$. Otherwise, if $v_i$ is not a constant,
then $\beta_i$ is a variable $x$. Since $E$ is a flat theory,
some $\alpha_j$ (we choose any) must be $x$. In this case
we define $v_i'$ as $s_j'$. With these definitions, the
following conditions follow:
\begin{iteMize}{$\bullet$}
\item For each $i\in\{1,\ldots,\mu\}$, 
$(v_i\in\F_0\Leftrightarrow v_i'\in\F_0)$ and
$(v_i,v_i'\in\F_0\Rightarrow v_i=_E v_i')$ hold.
\item For each $j\in\{1,\ldots,m\}$,
$(t_j\in\F_0\Leftrightarrow t_j'\in\F_0)$ and
$(t_j,t_j'\in\F_0\Rightarrow t_j=_E t_j')$ hold.
\item For each $i\in\{1,\ldots,\mu\}$ and
$j\in\{1,\ldots,m\}$, 
$(v_i'=_E t_j'\Leftrightarrow v_i=_E t_j)$ holds.
\end{iteMize}
By induction hypothesis, $h(v_1',\ldots,v_\mu')=_E g(t_1',\ldots,t_m')$
holds.

Now, let $s_1'',\ldots,s_n''$ be defined as follows
for each $i$ in $\{1,\ldots,n\}$. If  $s_i'$ is not a constant
then define $s_i''$ as $s_i'$. Otherwise, if $s_i'$ is a constant,
then define $s_i''$ as $s_i$. 
By the condition $(s_i,s_i'\in\F_0\Rightarrow s_i=_E s_i')$
we have that
$f(s_1',\ldots,s_n')=_E f(s_1'',\ldots,s_n'')$ holds.
Moreover, the same rule $l\approx r$ can be used to prove
$f(s_1'',\ldots,s_n'')=_E h(v_1',\ldots,v_\mu')$. Hence,
$f(s_1',\ldots,s_n')=_E f(s_1'',\ldots,s_n'')=_E
h(v_1',\ldots,v_\mu')=_E g(t_1',\ldots,t_m')$ holds, and we
are done.
\qed

\subsection{Well quasi-orderings}

A {\em well quasi-ordering}~\cite{Gallier91}
$\leq$ on a set $S$ is
a reflexive and transitive relation such that
any infinite sequence of elements $e_1,e_2,\ldots$
of $S$ contains an increasing pair $e_i\leq e_j$ with $i<j$.


\section{Tree Automata with Global Constraints} 
\label{section-tagc} \label{section-definition}
\noindent
In this subsection, we define a class of tree automata with global constraints
strictly generalizing 
both the $\TACBB$ of~\cite{BogaertTison92} 
and the $\TAGED$ of~\cite{FiliotTalbotTison08}.
The generalization consists in considering more general global
constraints, and interpreting all the constraints modulo a flat
equational theory.


As an intermediate step, we define an extension of
the $\TACBB$ of~\cite{BogaertTison92} 
where the local constraints between brothers
are considered modulo a flat equational theory.
\begin{defi}
A \emph{tree automaton with constraints between brothers modulo a flat theory} 
($\TACBBF$) is a tuple $\A=\langle Q, \F, F, \Delta, E\rangle$ 
where $\langle Q, \F, F, \Delta \rangle$ is a $\TACBB$ 
and $E$ is a flat equational theory.
\end{defi}
By $\ta(\A)$ we denote $\ta(\langle Q, \F, F, \Delta\rangle)$.

A \emph{run} of a $\TACBBF$ $\A=\langle Q, \F, F, \Delta, E\rangle$
is a pair $r = \langle t,M\rangle$ 
defined analogously to a run of a $\TACBB$, except that
the constraints between brothers are interpreted modulo $E$.
More specifically, 
for each position $p$ in $\Pos(t)$,
if $t|_p$ is written of the form $f(t_1,\ldots,t_m)$, and
$M(p.1),\ldots,M(p.m)$ are rules with right-hand side states
$q_1,\ldots,q_m\in Q$, respectively, then $M(p)$ is a transition rule of
$\Delta_\A$ of the form $f(q_1,\ldots,q_m)\upder{C} q$ 
for some $q\in Q$ and
constraint between brothers $C$. Moreover, for each equality
$i\approx j$ in $C$, $t_i =_E t_j$ holds, and for each disequality
$i \not\approx j$ in $C$, $t_i \neq_E t_j$ holds.
The notions of successful run and recognized language
are defined for $\TACBBF$ analogously
to the case of \TA.

\medskip
We further extend this class $\TACBBF$ with
global equality and disequality constraints
generalizing those of $\TAGED$~\cite{FiliotTalbotTison08}.

\begin{defi} \label{definition-TABG}
A \emph{tree automaton with global and brother constraints
modulo a flat theory} ($\TAGBCF$)
is a tuple $\A = \langle Q, \F, F, \Delta, E, C\rangle$
where $\langle Q, \F, F, \Delta, E\rangle$ is a $\TACBBF$, 
denoted $\tacbbf(\A)$, and
$C$ is a Boolean combination of atomic constraints
of the form $q \approx q'$ or $q \not\approx q'$, where $q, q' \in Q$.
\end{defi}
By $\ta(\A)$ we denote $\ta(\tacbbf(\A))$.

A \emph{run} of a $\TAGBCF$ $\A=\langle Q, \F, F, \Delta, E, C\rangle$
is a run $r=\langle t,M\rangle$
of $\tacbbf(\A)$ such that
$r$ satisfies $C$, denoted $r \models C$,
where the satisfiability of constraints is defined as follows.
For atomic constraints,
$r \models q \approx q'$ (respectively $r \models q \not\approx q'$) holds
if and only if
for all different positions $p, p' \in \Pos(t)$ such that
$M(p) = q$ and $M(p') = q'$,
$t|_{p} =_E t|_{p'}$ (respectively $t|_{p} \not=_E t|_{p'}$) holds.
This notion of satisfiability is extended to Boolean combinations as usual.
As for \TA, we say that $r$ is a run of $\A$ on $t$.
A run $r$ of $\A$ on $t \in \T(\F)$ is \emph{successful} (or \emph{accepting})  
if $r(\rootp)\in F$.
The language $\L(\A)$ of $\A$ is the set of terms $t$ for which there 
exists a successful run of $\A$.

It is important to note that the semantics of 
$\neg(q \approx q')$ and $q \not\approx q'$ differ,
as well as the semantics of $\neg(q \not\approx q')$ and $q \approx q'$.
This is because we have a ``for all'' quantifier in both definitions
of semantics of $q \approx q'$ and $q \not\approx q'$.

\medskip

\begin{figure}
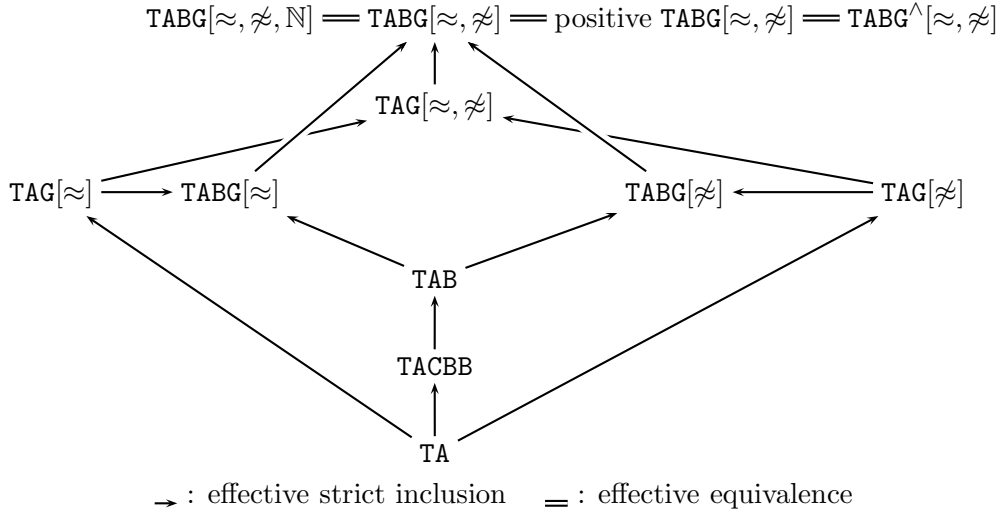

\[
\psmatrix[colsep=7mm,rowsep=7mm]
 & \TAGBCF[\approx,\not\approx,\mathbb{N}] & \TAGBCF[\approx,\not\approx] & 
     \mathrm{positive}\;\TAGBCF[\approx,\not\approx] & \PCTAGBCF[\approx,\not\approx]\\
 & & \TAGC[\approx,\not\approx]\\
\TAGC[\approx] & \TAGBCF[\approx] &  & \TAGBCF[\not\approx] & \TAGC[\not\approx]\\
 & & \TACBBF\\
 & & \TACBB\\
 & & \TA\\
\endpsmatrix
\psset{nodesep=3pt,arrows=->}
\ncline{6,3}{5,3}
\ncline{5,3}{4,3}
\ncline{6,3}{3,1}
\ncline{6,3}{3,5}
\ncline{4,3}{3,2}
\ncline{4,3}{3,4}
\ncline{3,1}{2,3}
\ncline{3,1}{3,2}
\ncline{3,5}{2,3}
\ncline{3,5}{3,4}
\ncline[border=2pt]{3,2}{1,3}
\ncline[border=2pt]{3,4}{1,3}
\ncline{2,3}{1,3}
\ncline[doubleline=true,arrows=-]{1,2}{1,3}
\ncline[doubleline=true,arrows=-]{1,3}{1,4}
\ncline[doubleline=true,arrows=-]{1,4}{1,5}
\]
\begin{center}
$\psmatrix[colsep=3mm,rowsep=1mm] \;  & \; \endpsmatrix\ncline[arrows=->]{1,1}{1,2}$ :
 effective strict inclusion
\quad
$\psmatrix[colsep=3mm,rowsep=1mm] \;  & \; \endpsmatrix
  \ncline[doubleline=true]{1,1}{1,2}$ :
 effective equivalence
\end{center}
\caption{Decidable classes of \TA with local and global constraints}
\label{fig-classes}
\end{figure}

Let us introduce some notations, 
summarized in Figure~\ref{fig-classes} 
that we use below to characterize some classes of tree automata related to $\TAGBCF$
(Figure~\ref{fig-classes} also refers to 
 a class defined in Section~\ref{section-arithmetic}).
A $\TAGBCF$ $\A$ is called
\emph{positive} if $C_\A$ is a disjunction of conjunctions of atomic constraints
and it is called \emph{positive conjunctive} if $C_\A$ is a conjunction
of atomic constraints.
The subclass of 
positive conjunctive $\TAGBCF$ is
denoted by 
$\PCTAGBCF$.

We recall that a $\TACBBF$ where all the constraints are empty
is just a \TA.
%
For a $\TAGBCF$ $\A$, when the theory $E_\A$ is empty and $\tacbbf(\A)$ is just a \TA,
we say that $\A$ is just a tree automaton with global constraints
(\TAGC). 
Its subclass with 
positive conjunctive
constraints is denoted 
$\PCTAGC$.

With the notation $\TAGBCF[\tau_1,\ldots,\tau_m]$, we characterize
the class of tree automata with global and brother constraints
modulo a flat theory
whose global constraints are Boolean combination of atomic constraints
of types $\tau_1,\ldots,\tau_m$.
The types $\approx$ and $\not\approx$ denote respectively the
atomic constraints of the form $q \approx q'$ and $q \not\approx q'$, 
where $q, q'$ are states.
For instance, the abbreviation $\TAGBCF$ used in Definition~\ref{definition-TABG}
stands for $\TAGBCF[\approx, \not\approx]$.
This notation is extended to the 
positive conjunctive fragment by 
$\PCTAGBCF[\tau_1,\ldots,\tau_k]$
and to the fragment without local constraints between brother, 
by $\TAGC[\tau_1,\ldots,\tau_k]$.

\subsection{Expressiveness}
The class of regular languages 
is strictly included in the class of $\TAGBCF$ languages
due to the constraints.

\begin{exa} \label{ex:ftt}
Let $\F = \{ a \ofarity 0, f \ofarity 2 \}$.
The set $\{ f(t, t) \mid t \in \T(\Sigma) \}$ 
is not a regular tree language
(this can be shown using a classical \emph{pumping} argument).

However, 
it is recognized by the following $\TACBBF$:
\[
\bigl\langle \{ q_0, q_\final \}, \F, \{ q_\final \}, 
\{ a \to q_0, f(q_0, q_0) \to q_0,
   f(q_0, q_0) \lrstep{1 \approx 2}{} q_\final \}, \emptyset \bigr\rangle,
\]
and it is also recognized by the following $\TAGC[\approx]$:
\[
\A = \bigl\langle \{ q_0, q_1, q_\final \}, \F, \{ q_\final \}, 
\{ a \to q_0 \mathbin{|} q_1, f(q_0, q_0) \to q_0 \mathbin{|} q_1, f(q_1, q_1) \to q_\final 
\}, \emptyset, q_1 \approx q_1 
\bigr\rangle,
\]
where $t \to q \mathbin{|} q_r$ is an abbreviation for $t \to q$ and $t \to q_r$.
\noindent
An example of successful run of $\A$ on $t = f(f(a,a), f(a,a))$ 
is $q_\final\bigl(q_1(q_0,q_0),q_1(q_0,q_0)\bigr)$, where we use
term-like notation for marking the reached state at each position.
\end{exa}

\noindent
Moreover, the $\TAGED$ of~\cite{FiliotTalbotTison08} are also a
particular case of $\TAGC[\approx,\not\approx]$, 
since they can be redefined in our setting as
restricted $\PCTAGC[\approx,\not\approx]$,
where the equational theory is empty, and
where $q$ and $q'$ are required to be distinct in
any atomic constraint of the form $q \not\approx q'$.

Reflexive disequality constraints such as $q \not\approx q$ 
correspond to monadic \emph{key constraints} for XML documents, 
meaning that every two distinct positions of type $q$ have different values. 
A state $q$ of a $\TAGC[\approx,\not\approx]$ 
can be used for instance to characterize unique identifiers
as in the following example, 
which presents a $\TAGC[\approx,\not\approx]$ 
whose language cannot be recognized by a $\TAGED$.
This example will be referred several times
in Section~\ref{section-emptiness}, in order to illustrate the definitions
used in the decision procedure of the emptiness problem for $\TAGC[\approx,\not\approx]$.

\begin{figure}
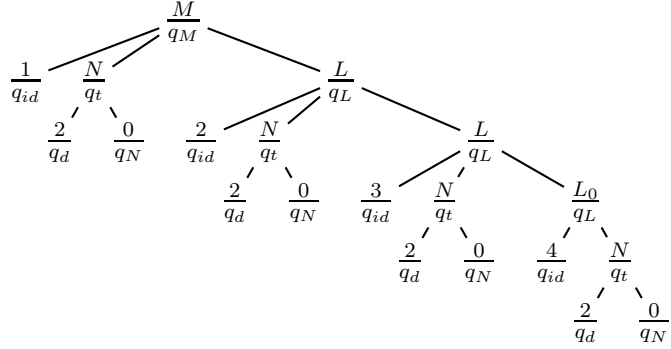

\[
\def\dedge{\ncline[linestyle=dotted]}
\psset{nodesep=2pt,levelsep=8mm,treesep=5mm}
\pstree{\TR{\frac{M}{q_M}}}{%
 \TR{\frac{1}{q_{id}}}
 \pstree{\TR{\frac{N}{q_t}}}{\TR{\frac{2}{q_{d}}} \TR{\frac{0}{q_{N}}}}
  \pstree{\TR{\frac{L}{q_L}}}{%
  \TR{\frac{2}{q_{id}}}
  \pstree{\TR{\frac{N}{q_t}}}{\TR{\frac{2}{q_{d}}} \TR{\frac{0}{q_{N}}}}
  \pstree{\TR{\frac{L}{q_L}}}{%
   \TR{\frac{3}{q_{id}}}
   \pstree{\TR{\frac{N}{q_t}}}{\TR{\frac{2}{q_{d}}} \TR{\frac{0}{q_{N}}}}
   \pstree{\TR{\frac{L_0}{q_L}}}{%
    \TR{\frac{4}{q_{id}}}
    \pstree{\TR{\frac{N}{q_t}}}{\TR{\frac{2}{q_{d}}} \TR{\frac{0}{q_{N}}}}}}}}
\]
\caption{Term and successful run (Example~\ref{example-menu}).}
\label{figure-menu}
\end{figure}

\begin{exa} \label{example-menu}
The $\TAGC[\approx,\not\approx]$ of our running example
accepts (in state $q_M$) lists of dishes called menus,
where every dish is associated with
one identifier (state $q_{id}$) 
and the time needed to cook it (state $q_t$).
%
We have other states accepting digits ($q_d$),
numbers ($q_N$) and lists of dishes ($q_{L}$). 

The $\TAGC[\approx,\not\approx]$ 
$\A = \langle Q,\F, F,\Delta, \emptyset, C \rangle$
is defined as follows:
\noindent
$\F = \{ 0,\ldots, 9 \ofarity 0, N, L_0\ofarity 2, L, M \ofarity 3\}$,
$Q = \{ q_d, q_N, q_{id}, q_{t},q_L, q_M \}$,
$F=\{ q_M \}$,
and
$
\Delta = \{ i\to q_d \mathbin{|} q_N \mathbin{|} q_{id} \mathbin{|} q_{t} : 0 \leq i \leq 9 \}
$
$\cup$ 
$
\{ N(q_d, q_N) \to q_N \mathbin{|} q_{id} \mathbin{|} q_{t},
   L_0(q_{id}, q_t) \to q_L, L(q_{id}, q_t, q_L) \to q_L, 
   M(q_{id}, q_t, q_L) \to q_M \} 
$.

The constraint $C$ 
ensures that all the identifiers of the dishes in a menu are pairwise distinct
(i.e.\ that $q_{id}$ is a key) and 
that the time to cook is the same for all dishes:
$C = q_{id}\not\approx q_{id}\wedge q_t \approx q_t$.
%
%
\noindent A term in $\L(\A)$ together
with an associated successful run
are depicted in Figure~\ref{figure-menu}.
\end{exa}

Althought this is a simple exercise,
let us establish formally that 
$\TAGC[\approx,\not\approx]$ are strictly more expressive than $\TAGED$.

\begin{lem} \label{example-not-taged}
The class of languages recognized by $\PCTAGC[\approx,\not\approx]$
strictly includes the class of languages recognized by
$\TAGED$.
\end{lem}

\proof
Since a \TAGED is just a $\PCTAGC[\approx,\not\approx]$ where
no constraint of the form $q\not\approx q$ occurs, the inclusion
holds. In order to see that it is strict, it suffices
to show a language $L$ which can be recognized by a
$\PCTAGC[\approx,\not\approx]$ but not by a \TAGED.

Let $\F = \{ a \ofarity 0, s \ofarity 1, f \ofarity 2 \}$.
The set $L$ of terms of $\T(\F)$
of the form $f(s^{n_1}(a), f(s^{n_2}(a),$
$\ldots, f(s^{n_k}(a), a)\ldots))$, 
such that $k \geq 0$ and the natural numbers $n_i$, for $i \leq k$,
are pairwise distinct,
is recognized by the following $\PCTAGC[\approx,\not\approx]$:

\[
\left\langle 
\{ q_a, q, q_\final \},
\F,
\{ q_\final \},
\left\{ 
\begin{array}{l}
a \to q_a \mathbin{|} q \mathbin{|} q_\final,\\
s(q_a) \to q_a \mathbin{|} q,\\
f(q, q_\final) \to q_\final 
\end{array}
\right\},
\emptyset,
q \not\approx q
\right\rangle.
\]

Assume that there exists a $\PCTAGC[\approx,\not\approx]$ $\A$ 
without reflexive disequality constraints of the form $q \not\approx q$
(i.e.\ a \TAGED), recognizing this language $L$. 
Then, there exists an accepting run $r$ of $\A$ on the term 
$t = f(s(a), f(s^2(a), \dots f(s^{|Q_\A| +1}(a),a)\ldots))\in L$.
Therefore, $r \models C_\A$ 
(the global constraint of $\A$, which is positive by hypothesis).

There are two different positions $p_i =
\overbrace{2.2.\ldots.2}^{i}.1$ and
$p_j = \overbrace{2.2.\ldots.2}^{j}.1$,
$0\leq i< j \leq |Q_\A|$ such that $r(p_i) = r(p_j)$. 
Let us show that $r' = r[r|_{p_i}]_{p_j}$ is an accepting run of $\A$ on 
$t' = t[t|_{p_i}]_{p_j}$. 
Since $r(p_i) = r(p_j)$ and $r$ is a run of $\A$ on $t$,
$r'$ is a run of $\ta(\A)$ on $t'$.
Hence, it suffices to prove that the constraint $C_\A$
is satisfied by $r'$. Consider a position $p$ of the form $2.2.\ldots.2$
with $|p|<j$. We start by proving that any atomic constraint
involving $r'(p)$ is satisfied. Note that $r'(p)=r(p)$ holds, and that
the subterm $t|_p$ has only this occurrence in $t$.
Thus, any atomic constraint involving $r(p)$ and a state $q$
occurring in $r$ is necessarily of the form $r(p)\not\approx q$.
Since any state occurring in $r'$ occurs also in $r$,
any atomic constraint involving $r'(p)$ and a state $q$
occurring in $r'$ is of the form $r'(p)\not\approx q$.
Moreover, the subterm $t'|_p$ has only this occurrence
in $t'$. Thus, such a constraint is satisfied.
Now consider two different positions $p_1,p_2$ which are not
of the form described above. It remains to see that any atomic
constraint involving $r'(p_1)$ and $r'(p_2)$ is satisfied.
In the case where $r'|_{p_1}$ and $r'|_{p_2}$ are different,
this is a direct consequence of the fact that both
subruns $r'|_{p_1}$ and $r'|_{p_2}$ are also subruns of $r$ at different
positions. Otherwise, in the case where $r'|_{p_1}$ and $r'|_{p_2}$
are the same subrun, then, $r'(p_1)=r'(p_2)$ holds,
and any atomic constraint involving
$r'(p_1)$ and $r'(p_2)$ must be of the form $r'(p_1)\approx r'(p_2)$
because $\A$ has no reflexive disequalities. Thus, the
atomic constraint is also satisfied in this case.
\qed

The following example shows a \TAGBCF recognizing a language
that cannot be recognized by a $\TAGC[\approx,\not\approx]$.
The proof is a simple exercise and it is left to the reader.

\begin{exa} \label{example-btta}
Assume that the terms of Example~\ref{example-menu}
are now used to record the activity of a restaurant.
To this end, we transform the \TAGC of example~\ref{example-menu}
into a \TAGBCF as follows.
First, in order to simplify the example
we omit the restriction that all cooking times
coincide, i.e.\ $C = q_{id}\not\approx q_{id}$.
Second,
we add a new argument of type $q_t$ to $L_0$, $L$ and $M$, 
so that the old argument $q_t$ characterizes the theoretical time to cook,
and the new $q_t$ characterizes
the real time that was needed to cook the dish.
Let us replace the transitions with $L_0$, $L$ and $M$ in input by
$L_0(q_{id}, q_t, q_t) \lrstep{2 \approx 3}{} q_L$, 
$L_0(q_{id}, q_t, q_t) \lrstep{2 \not\approx 3}{} q'_L$, 
$L(q_{id}, q_t, q_t, q_L) \lrstep{2 \approx 3}{} q_L$, 
$L(q_{id}, q_t, q_t, q_L) \lrstep{2 \not\approx 3}{} q'_L$, 
$M(q_{id}, q_t, q_t, q_L) \lrstep{2 \approx 3}{} q_M$, 
$M(q_{id}, q_t, q_t, q_L) \lrstep{2 \not\approx 3}{} q'_M$, 
where $q'_L$ is a new state meaning that there was an anomaly.
We also add a transition
$L(q_{id}, q_t, q_t, q'_L) \to q'_L$ to propagate $q'_L$
and $M(q_{id}, q_t, q_t, q'_L) \to q'_M$.

By keeping the set of final states as $\{q_M\}$,
the recognized language of the $\TAGBCF$ obtained is 
the set of records well cooked, 
i.e.\ such that for all dishes, the real time to cook
is equal to the theoretical time.
By redefining the set of final states as $\{q_M'\}$,
the recognized language is the set of records with an anomaly.
\end{exa}

%
%

\subsection{Decision Problems}
\noindent
The \emph{membership} is the problem to decide,
given a term $t \in \T(\F)$ and a $\TAGBCF$ $\A$ over $\F$
whether $t \in \L(\A)$.
\begin{prop} \label{theorem-membership}
Membership is NP-complete for $\TAGBCF$, by assuming that
the maximum arity of the signature $\F$ is a constant for the problem.
\end{prop}
\proof
In order to prove that this problem is in NP, given
a $\TAGBCF$ $\A =  \langle Q, \F, F, \Delta, E, C \rangle$
and a term $t \in \T(\F)$, we can
non-deterministically
guess a function $M$ from $\Pos(t)$ into $\Delta$,
and check that $\langle t,M\rangle$ is a successful run of $\A$ on $t$.
The checking can be performed in polynomial time. In particular, testing
equivalence modulo $E$ can be performed in polynomial time using
a dynamic programming scheme, by assuming that the maximum arity
of $\F$ is a constant of the problem, which is a usual assumption.
More general results are given
in~\cite{DBLP:conf/lics/Nieuwenhuis96,DBLP:journals/iandc/ComonHJ94}.
For NP-hardness, 
\cite{FiliotTalbotTison08,JKV-lata09}
present PTIME reductions of the satisfiability of Boolean expressions
into membership for $\PCTAGC[\approx]$ whose constraints are conjunctions of 
equalities of the form $q \approx q$.
\qed

Recall that for plain \TA, membership is in PTIME.

\medskip

The \emph{universality} is the
problem to decide, given a $\TAGBCF$ $\A$ over $\F$, whether $\L(\A) = \T(\F)$.
It is known to be undecidable already for a small subclass of $\TAGC$.
%
\begin{prop}\cite{FiliotTalbotTison08,JKV-lata09} \label{prop-universality}
Universality is undecidable for $\PCTAGC[\approx]$.
\end{prop}
%
The following consequence is a new result for $\TAGED$.

\begin{prop} \label{prop-regularity}
It is undecidable 
whether the language of a given 
$\PCTAGC[\approx]$ 
is regular.
\end{prop}
\proof
We show that universality is reducible to regularity
using a new function symbol $f$
with arity $2$, and any non-regular
language $L$ which is recognizable by a $\PCTAGC[\approx]$
(such a language exists).

Let $\A$ be an input of universality for $\PCTAGC[\approx]$ and let 
\[ 
L' = \bigl\{ f(t_1,t_2) \mid t_1 \in \T(\F)\wedge t_2 \in L \bigr\}
\cup \bigl\{ f(t_1,t_2) \mid t_1 \in \L(\A)\wedge t_2\in\T(\F) \bigr\}.
\]
It is possible to compute a new $\PCTAGC[\approx]$ $\A'$
recognizing the language $L'$
(see Lemma~\ref{lem:union}).
Thus, in order to conclude, it suffices to
show that $\L(\A) = \T(\F)$ if and only if $\L(\A')$ is regular.
For this purpose let us first define the 
quotient of a term language $R$ by a term $s$ with respect to
a function symbol $f$:
$R/s := \{ t \mid f(s,t) \in R \}$.
This operation preserves regular languages:
for all $s$ and $f$, if $R$ is regular then $R/s$ is regular.

If $\L(\A) = \T(\F)$, then 
$\L(\A')$ is $\bigl\{ f(t_1,t_2) \mid t_1, t_2 \in \T(\F) \bigr\}$,
which is regular.
Assume that $\L(\A) \neq \T(\F)$ and let $s \in \T(\F) \setminus \L(\A)$.
By construction, $\L(\A')/s = L$ which is not regular.
Hence $\L(\A')$ is not regular.
\qed

\medskip\noindent
The \emph{emptiness} is the problem to decide,
given a $\TAGBCF$ $\A$, whether $\L(\A) = \emptyset$.
The proof that it is decidable for $\TAGBCF$ is rather involved and 
is presented in Section~\ref{section-emptiness}.



\section{Arithmetic Constraints and Reduction to $\PCTAGBCF$}
\label{section-arithmetic}

This section has two goals. The first goal is to present an
extension of $\TAGBCF$ by allowing certain global arithmetic
constraints. They are interesting by themselves since they allow the
representation of several natural properties in a simple way.
The second goal is to show that the class of $\TAGBCF$ languages
coincides (in expressiveness) with the class of $\PCTAGBCF$ languages. 
In other words, for each $\TAGBCF$ there exists a $\PCTAGBCF$ recognizing the
same language. 
This reduction will be very useful in
Section~\ref{section-emptiness} in order to prove decidability
of emptiness of $\TAGBCF$.

The reason for presenting both results in the same
section is that arithmetic constraints simplify the
task of transforming a $\TAGBCF$ into a $\PCTAGBCF$
representing the same language. This is because negations
can be replaced by arithmetic constraints with an equivalent meaning
in a first intermediate step, and such constraints are easier
to deal with.

All this work is developed in Subsection~\ref{subsection-naturalconstraints}.
Before that, in Subsection~\ref{subsection-integerconstraints}
we present a more general form of arithmetic constraints for which
emptiness is undecidable. The motivation of this first
subsection is to show the limits
of positive results in this setting, and to justify the limited
form of the constraints in Subsection~\ref{subsection-naturalconstraints}.

\subsection{Global Integer Linear Constraints}
\label{subsection-integerconstraints}

Let $Q$ be a set of states.
A \emph{linear inequality} over $Q$ is an expression of the form
$\displaystyle\mathop\sum_{q\in Q} a_q \cdot |q| \geq a$
or 
$\displaystyle\sum_{q\in Q} a_q \cdot \|q\| \geq a$
where every $a_q$ and $a$ belong to $\mathbb{Z}$.
We consider the above linear inequalities
as atomic constraints of tree automata with global constraints,
and denote by $|.|_\mathbb{Z}$ and $\|.\|_\mathbb{Z}$ their respective types.
The type $\mathbb{Z}$ denotes $|.|_\mathbb{Z}$ and $\|.\|_\mathbb{Z}$ together.

Using the notation introduced in Section~\ref{section-tagc},
$\TAGBCF[\approx, \not\approx, |.|_\mathbb{Z}, \|.\|_\mathbb{Z}]$ 
(or $\TAGBCF[\approx, \not\approx, \mathbb{Z}]$)
denotes the class of tree automata with global and brother constraints modulo a flat theory
of the form $\A = \langle Q, \F, F, \Delta, E, C\rangle$
such that $\langle Q, \F, F, \Delta, E\rangle$ is a $\TACBBF$
(denoted $\tacbbf(\A)$) 
and $C$ is a Boolean combination of atomic constraints
which can be linear inequalities as above
or equality or disequality constraints
of the form $q \approx q'$ or $q \not\approx q'$, with $q, q' \in Q$.

Let $\A$ be a $\TAGBCF[\approx, \not\approx, |.|_\mathbb{Z}, \|.\|_\mathbb{Z}]$ over $\F$
and with state set $Q$
and flat equational theory $E$, 
let $r$ be a run of $\tacbbf(\A)$ on a term $t \in \T(\F)$ and
let $q \in Q$. 
Intuitively, the interpretation of $|q|$ with respect to $r$ is the number
of occurrences of $q$ in $r$, i.e.\ the number of positions
$p$ holding $r(p)=q$. The interpretation of
$\|q\|$ with respect to $r$ is the number of different subterms (modulo $E$)
in $t$ reaching state $q$ with $r$, i.e.\ the maximum number
of positions $p_1,p_2,\ldots,p_n$ holding $r(p_1)=r(p_2)=\ldots=r(p_m)=q$
and such that the terms $t|_{p_1},t|_{p_2},\ldots,t|_{p_n}$ are
pairwise different (modulo $E$).
More formally, the interpretations of $|q|$ and $\| q \|$ with respect to $r$
(and $t$) are defined, respectively, by the following cardinalities:
\[ \llbracket\, |q|\, \rrbracket_{r} = \bigl| \{ p\mid p\in \Pos(t)\;\wedge\;r(p) = q \} \bigr|\]
\[ \llbracket\, \|q\|\, \rrbracket_{r} = 
\bigl| \{ [ t|_p ]_E \mid p\in \Pos(t)\;\wedge\;r(p) = q \} \bigr|.\]
This permits to define the satisfiability of linear inequalities
with respect to  $r$ and $t$: $r \models \displaystyle\mathop\sum_{q\in Q} a_q \cdot |q| \geq a$
holds if and only if
$\displaystyle\mathop\sum_{q\in Q} a_q \cdot \llbracket\, |q|\, \rrbracket_{r} \geq a$ holds,
and
$r \models \displaystyle\sum_{q\in Q} a_q \cdot \|q\| \geq a$
holds if and only if
$\displaystyle\mathop\sum_{q\in Q} a_q \cdot \llbracket\, \|q\|\, \rrbracket_{r} \geq a$ holds.
The satisfiability of the global constraint $C_\A$ of $\A$ by $r$, 
denoted $r \models C_\A$ is defined accordingly, and if $r \models C_\A$
then $r$ is called a run of $\A$.
A run of $\A$ on $t \in \T(\F)$ is \emph{successful} (or \emph{accepting})  
if $r(\rootp)\in F_\A$.
The language $\L(\A)$ of $\A$ is the set of terms $t$ for which there 
exists a successful run of $\A$.

\begin{exa}
Let us add a new argument to the dishes of 
the menu of Example~\ref{example-menu}
which represents the price coded on two digits by a term $N(d_1,d_0)$.
We add a new state $q_p$ for the type of prices,
and other states
$q_\mathit{cheap}$,
$q_\mathit{moderate}$,
$q_\mathit{expensive}$,
$q_\mathit{chic}$ describing price level ranges, and transitions
$0 | 1 \to q_\mathit{cheap}$,
$2 | 3 \to q_\mathit{moderate}$,
$4 | 5 | 6 \to q_\mathit{expensive}$,
$7 | 8 | 9 \to q_\mathit{chic}$
and $N(q_\mathit{cheap}, q_d) \to q_p$, \dots.
The price is a new argument of $L_0$, $L$ and $M$,
hence we replace the transitions with these symbols in input by
$L_0(q_{id}, q_t, q_p) \to q_L$, 
$L(q_{id}, q_t, q_p, q_L) \to q_L$, 
$M(q_{id}, q_t, q_p, q_L) \to q_M$. 
We can use a linear inequality 
$|q_\mathit{cheap}| + |q_\mathit{moderate}| -
 |q_\mathit{expensive}| - |q_\mathit{chic}| \geq 0$
to characterize the moderate menus, 
and 
$|q_\mathit{expensive}| + |q_\mathit{chic}| \geq 6$
to characterize the menus with too many expensive dishes.
A linear inequality $\| q_p \| \leq 1$ expresses that all the dishes
have the same price.
\end{exa}

The class $\TAGC[\,|.|_\mathbb{Z}]$ has been studied under different names
(e.g.\ Parikh automata in~\cite{Klaedtke02parikhautomata},
linear constraint tree automata in~\cite{BojanczykMSS09})
and it has a decidable emptiness test.
Indeed, the set of successful runs of a given \TA with state set $Q$
is a context-free language (seeing runs as words of $Q^*$),
and the Parikh projection 
(the set of tuples over $\mathbb{N}^{|Q|}$ whose components are the 
$\llbracket\, |q|\, \rrbracket_{r}$ for every run $r$)
of such a language is a semi-linear set.
The idea for deciding emptiness for a $\TAGC[\,|.|_\mathbb{Z}]$ $\A$ 
is to compute this semi-linear set and 
to test the emptiness of its intersection with 
the set of solutions in $\mathbb{N}^{|Q|}$ of $C_\A$,
the arithmetic constraint of $\A$
(a Boolean combination of linear inequalities of type $|.|_\mathbb{Z}$)
which is also semi-linear.
This can be done in NPTIME, see~\cite{BojanczykMSS09}.

To our knowledge, the class $\TAGC[\,\|.\|_\mathbb{Z}]$ 
with global constraints counting the number of distinct subterms in each state,
has not been studied,
even modulo an empty theory.

\medskip
Combining constraints of type $\approx$ and counting constraints
of type $|.|_\mathbb{Z}$ however leads to undecidability.
\begin{thm}
Emptiness is undecidable for $\PCTAGC[\approx,|.|_\mathbb{Z}]$.
\end{thm}
\proof
%
We consider the  
Hilbert's tenth problem, that is, solvability of an input
equation $P=0$ where $P$ is a polynomial with integer
coefficients and variables ranging over the natural numbers.
This problem is known undecidable, and with the addition of
new variables it is easily reducible
to a question of the form 
$\exists x_1 \ldots \exists x_n: e_1\wedge\ldots\wedge e_m$, where $x_1,\ldots,x_n$
are variables ranging over the natural numbers, and
$e_1,\ldots,e_m$ are equations that are either of the form
$x_j+x_k=x_t$ or $x_j*x_k=x_t$ or $x_j=1$ or $x_j=0$.
We reduce this last problem to emptiness of
$\PCTAGC[\approx,|.|_\mathbb{Z}]$.

We consider an instance
$\varphi\equiv\exists x_1\ldots\exists x_n: e_1\wedge\ldots\wedge e_m$.
Without loss of generality, we assume that
$e_1,\ldots,e_{m'}$ for $m'\leq m$ are all the equations
of the form $x_j*x_k=x_t$, and that for each of such
equations, the indexes $j,k,t$ are different.
We will construct a
$\PCTAGC[\approx,|.|_\mathbb{Z}]$
$\A$ such that $\varphi$ is true if and only if
$\L(\A)$ is not empty.

\begin{figure}
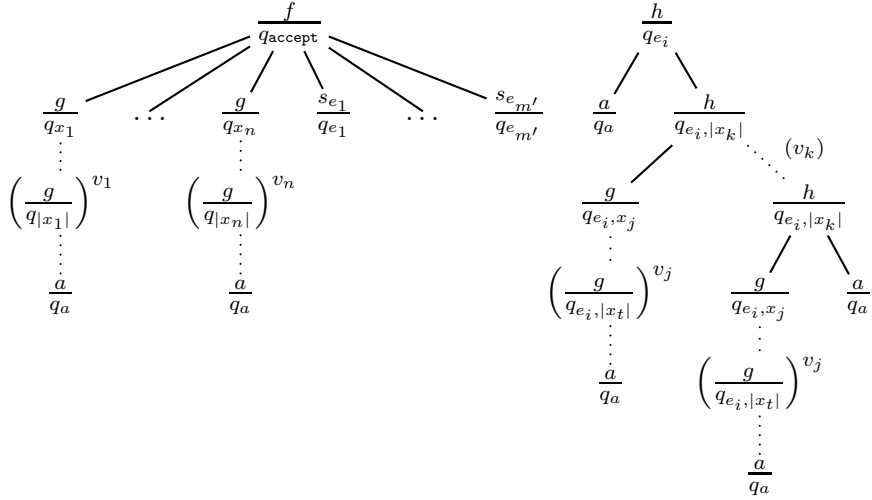

\[
\def\dedge{\ncline[linestyle=dotted]}
\psset{nodesep=2pt,levelsep=12mm,treesep=7mm}
\pstree{\TR{\frac{f}{q_{\tt accept}}}}{%
 \pstree{\TR{\frac{g}{q_{x_1}}}}{%
  \pstree{\TR[edge=\dedge]{\left(\frac{g}{q_{|x_1|}}\right)^{v_1}}}{%
   \TR[edge=\dedge]{\frac{a}{q_a}}}}
 \TR{\ldots}
 \pstree{\TR{\frac{g}{q_{x_n}}}}{%
  \pstree{\TR[edge=\dedge]{\left(\frac{g}{q_{|x_n|}}\right)^{v_n}}}{%
   \TR[edge=\dedge]{\frac{a}{q_a}}}}
 \TR{\frac{s_{e_1}}{q_{e_1}}}
 \TR{\ldots}
 \TR{\frac{s_{e_{m'}}}{q_{e_{m'}}}}}
\pstree{\TR{\frac{h}{q_{e_i}}}}{%
 \TR{\frac{a}{q_a}}
 \pstree{\TR{\frac{h}{q_{e_i,|x_k|}}}}{%
  \pstree{\TR{\frac{g}{q_{e_i,x_j}}}}{%
   \pstree{\TR[edge=\dedge]{\left(\frac{g}{q_{e_i,|x_t|}}\right)^{v_j}}}{%
    \TR[edge=\dedge]{\frac{a}{q_a}}}}
  \pstree{\TR[edge=\dedge]{\frac{h}{q_{e_i,|x_k|}}}\trput{^{(v_k)}}}{%
   \pstree{\TR{\frac{g}{q_{e_i,x_j}}}}{%
    \pstree{\TR[edge=\dedge]{\left(\frac{g}{q_{e_i,|x_t|}}\right)^{v_j}}}{%
     \TR[edge=\dedge]{\frac{a}{q_a}}}}
   \TR{\frac{a}{q_a}}}}}
\]
\caption{Accepting run of
$s=f(g^{v_1+1}(a),\ldots,g^{v_n+1}(a),s_{e_1},\ldots,s_{e_{m'}})$ and
the subrun of $s_{e_i}$, where $e_i$ is of the form $x_j*x_k=x_t$.}
\label{figure-acceptingrun}
\end{figure}

Since the construction of $\A$ is technical,
let us give first some intuitions
(see Figure~\ref{figure-acceptingrun}).
Consider a possible assignment $x_1:=v_1,\ldots,x_n:=v_n$. 
A concrete run of $\A$ will be able
to check whether this assignment proves that $\varphi$ is true,
and only accept the corresponding term if the answer is positive.
In this run, there will be $v_1$ occurrences of state $q_{|x_1|}$,
$v_2$ occurrences of state $q_{|x_2|}$, and so on.
Equations of the form $x_j+x_k = x_t$, $x_j=1$ and $x_j=0$
can directly be checked by constraints of the form
$|q_{|x_j|}|+|q_{|x_k|}| = |q_{|x_t|}|$, 
$|q_{|x_j|}| = 1$ and $|q_{|x_j|}| = 0$.

For each equation $e_i$ of the form $x_j*x_k=x_t$ there will be
$v_k$ occurrences of a state called $q_{e_i,|x_k|}$. 
This is ensured by the constraint $|q_{e_i,|x_k|}| = |q_{|x_k|}|$.
Under each of these occurrences,  
there will be the same term, 
reaching a state $q_{e_i,x_j}$,
and containing $v_j$ occurrences of a state $q_{e_i,|x_t|}$.
The uniqueness of this term,
as well as the number of occurrences of $q_{e_i,|x_t|}$,
are both ensured by an equality constraint $q_{x_j}\approx q_{e_i,x_j}$.
In summary, there will
be $v_j*v_k$ occurrences of state $q_{e_i,|x_t|}$. The satisfiability
of the equation $x_j*x_k=x_t$ will be checked by the constraint
$|q_{|x_t|}|=|q_{e_i,|x_t|}|$.

The components of the $\PCTAGC[\approx,|.|_\mathbb{Z}]$
$\A = \langle Q, \F, F, \Delta, C\rangle$ are defined as follows:
\[\begin{array}{rcl}
Q&=&\{q_{\tt accept},\;q_a\}\cup \{q_{|x_j|},\;q_{x_j}\;\big|\;j\in\{1,\ldots,n\}\}
\cup \{q_{e_i}\;\big|\;i\in\{1,\ldots,m'\}\}\cup\\
&&\{q_{e_i,x_j},\;q_{e_i,|x_t|},\;q_{e_i,|x_k|}\;\big|\;
i\in\{1,\ldots,m'\},e_i\equiv x_j*x_k=x_t\}\\
\F&=&\{a:0,\;g:1,\;h:2,\;f:n+m'\}\\
F&=&\{q_{\tt accept}\}\\
\Delta&=&\{a\to q_a,\;
f(q_{x_1},\ldots,q_{x_n},q_{e_1},\ldots,q_{e_{m'}})\to q_{\tt accept}\}\cup\\
&&\{g(q_a)\to q_{|x_j|},\;
g(q_a)\to q_{x_j},\;
g(q_{|x_j|})\to q_{|x_j|},\;
g(q_{|x_j|})\to q_{x_j}\;\big|\;j\in\{1,\ldots,n\}\}\cup\\
&&\{g(q_a)\to q_{e_i,|x_t|},\;
g(q_a)\to q_{e_i,x_j},\;
g(q_{e_i,|x_t|})\to q_{e_i,|x_t|},\;
g(q_{e_i,|x_t|})\to q_{e_i,x_j},\\
&&\ \;h(q_{e_i,x_j},q_a)\to q_{e_i,|x_k|},\;
h(q_{e_i,x_j},q_{e_i,|x_k|})\to q_{e_i,|x_k|},\;
h(q_a,q_{e_i,|x_k|})\to q_{e_i},\\
&&\ \;h(q_a,q_a)\to q_{e_i}\;\big|\;i\in\{1,\ldots,m'\},e_i\equiv x_j*x_k=x_t\}
\end{array}
\]
\[
\begin{array}{rcl}
C &=&
\bigwedge_{m' < i \leq m, e_i \equiv x_j+x_k = x_t}
|q_{|x_j|}|+|q_{|x_k|}|=|q_{|x_t|}|\; \wedge \\
& & \bigwedge_{m'< i \leq m, e_i\equiv x_j=1} |q_{|x_j|}|=1\; \wedge \\
& & \bigwedge_{ m' < i \leq m, e_i\equiv x_j=0 } |q_{|x_j|}|=0\; \wedge\\
& & 
\bigwedge_{1 \leq i \leq m',e_i\equiv x_j*x_k=x_t}
\bigl( |q_{e_i,|x_t|}|=|q_{|x_t|}| \wedge |q_{e_i,|x_k|}|=|q_{|x_k|}| \wedge
q_{x_j}\approx q_{e_i,x_j} \bigr)
\end{array}
\]\medskip

It remains to prove that 
$\varphi$ is true if and only if
$\L(\A)$ is not empty. 
To this end, let us first assume that
$x_1:=v_1,\ldots,x_n:=v_n$ is a solution of $\varphi$.
In order to simplify the presentation, we denote the term
$h(a,h(s,h(s,\ldots,h(s,a)\ldots)))$, with $k$ occurrences of $s$, 
by $h[a,s,\ldots(k)\ldots,s,a]$,
and given an equation $e_i\equiv x_j*x_k=x_t$, we denote
the term $h[a,g^{v_j+1}(a),\ldots(v_k)\ldots,g^{v_j+1}(a),a]$ by $s_{e_i}$. 
Let us consider the term
$s=f(g^{v_1+1}(a),\ldots,g^{v_n+1}(a),s_{e_1},\ldots,s_{e_{m'}})$.
It is not difficult to see that the run of
Figure~\ref{figure-acceptingrun} is
an accepting run of $s$.
Note that for each equation $e_i\equiv x_j*x_k=x_t$,
the constraints $|q_{e_i,|x_t|}|=|q_{|x_t|}|,\;|q_{e_i,|x_k|}|=|q_{|x_k|}|,\;
q_{x_j}\approx q_{e_i,x_j}$ are satisfied, since
$x_j:=v_j,\;x_k:=v_k,\;x_t:=v_t$ satisfies the equation.

Now, assume that there is an accepting run $r$ of $\A$ on a term $s$.
Since $r$ is accepting, the transition rule
$f(q_{x_1},\ldots,q_{x_n},q_{e_1},\ldots,q_{e_{m'}})\to q_{\tt
accept}$ is applied at the root of $s$. 
According to the form of the rules involving
$q_{x_1},\ldots,q_{x_n}$, it holds that 
$s$ is of the form $s=f(g^{v_1+1}(a),\ldots,g^{v_n+1}(a),s_{e_1},\ldots,s_{e_{m'}})$,
for some natural numbers $v_1,\ldots,v_n$ and some terms
$s_{e_1},\ldots,s_{e_{m'}}$. 
Moreover, the states $q_{|x_1|},\ldots,q_{|x_n|}$ have $v_1,\ldots,v_n$ occurrences,
respectively. 
It remains to see that the assignment
$x_1:=v_1,\ldots,x_n:=v_n$ makes $\varphi$ true.
The satisfiability of a constraint of the form
$|q_{|x_j|}|+|q_{|x_k|}|=|q_{|x_t|}|$ (or $|q_{|x_j|}|=1$ or $|q_{|x_j|}|=0$)
implies that $v_j+v_k=v_t$ (or $v_j=1$ or $v_j=0$), thus
an equation of the form $x_j+x_k=x_t$ (or $x_j=1$ or $x_j=0$)
holds with this assignment. 
It remains to see that every equation
$e_i$ of the form $x_j*x_k=x_t$ also holds with this assignment.
According to the form of the rules of $\A$ and the satisfiability of the constraints
$|q_{e_i,|x_k|}|=|q_{|x_k|}|,q_{x_j}\approx q_{e_i,x_j}$, 
the term $s_{e_i}$ is of the form
$h[a,g^{v_j+1}(a),\ldots(v_k)\ldots,g^{v_j+1}(a),a]$. 
Moreover, $|q_{e_i,|x_t|}|$ has $v_j*v_k$ occurrences.
Therefore, by the satisfiability of the constraint
$|q_{e_i,|x_t|}|=|q_{|x_t|}|$, it follows $v_j*v_k=v_t$, and hence
the equation $x_j*x_k=x_t$ holds with this assignment, and we are done.
\qed

\subsection{Global Natural Linear Constraints}
\label{subsection-naturalconstraints}

We present now a restriction on linear inequalities
which enables a decidable emptiness test when
combined with $\approx$ and $\not\approx$ as global constraints.
A \emph{natural linear inequality} over $Q$
is a linear inequality as above whose coefficients $a_q$ and $a$
all have the same sign.
We call them natural since it is equivalent to consider inequalities
in both directions whose coefficients are all non-negative,
like $\mathop\sum a_q \cdot |q| \leq a$,
with $a_q, a \in \mathbb{N}$, to refer to 
$\sum  -a_q \cdot |q| \geq  -a$.
We also consider linear equalities 
$\mathop\sum a_q \cdot |q| = a$,
with $a_q, a \in \mathbb{N}$, to refer to
a conjunction of two natural linear inequalities.

%
The types of the natural linear inequalities are denoted by 
$|.|_\mathbb{N}$ and $\|.\|_\mathbb{N}$.
Below, we shall abbreviate these two types by $\mathbb{N}$.

The main difference between the linear inequalities of type
$|.|_\mathbb{Z}$ and $|.|_{\mathbb{N}}$
(and respectively $\|.\|_\mathbb{Z}$ and $\|.\|_{\mathbb{N}}$)
is that the former permits to compare the respective number 
of occurrences of two states, like e.g.\ in $|q| \leq |q'|$,
whereas the latter only permits to compare the number 
of occurrences of one state 
(or a sum of the number occurrences of several states with coefficients)
to a constant as e.g.\ in $|q| \leq 4$ or $|q| + 2|q'| \leq 9$.

\medskip
In the rest of the subsection we show that
$\TAGBCF[\approx,\not\approx, \mathbb{N}]$
has the same expressiveness as
$\PCTAGBCF[\approx,\not\approx]$.
The proof works in several steps:
\begin{iteMize}{$\bullet$}
\item First, we define the notion of normalized
$\TAGBCF[\approx,\not\approx, \mathbb{N}]$,
that is a $\TAGBCF[\approx,\not\approx, \mathbb{N}]$
with a constraint being a disjunction of conjunctions
of literals in a simple form.
\item Second, we remove
negative literals of the form
$\neg(q\approx q')$ or $\neg(q\not\approx q')$,
obtaining a list of
$\PCTAGBCF[\approx,\not\approx, \mathbb{N}]$
such that the union of their languages coincides with
the language of the original $\TAGBCF[\approx,\not\approx,
\mathbb{N}]$. In this step we use arithmetic constraints
for simulating the removed negative literals.
\item Third, we remove arithmetic literals of type $\|.\|_{\mathbb{N}}$,
obtaining a new list of
$\PCTAGBCF[\approx,\not\approx, |.|_{\mathbb{N}}]$
such that the union of their languages coincides with
the language of the original $\TAGBCF[\approx,\not\approx,
\mathbb{N}]$. 
In this step we use positive literals
of types $\approx$, $\not\approx$, and $|.|_{\mathbb{N}}$ 
in order to simulate the removed
literals of type $\|.\|_{\mathbb{N}}$.
\item Fourth, we remove arithmetic literals of type $|.|_{\mathbb{N}}$,
obtaining a new list of
$\PCTAGBCF[\approx,\not\approx]$
such that the union of their languages coincides with
the language of the original $\TAGBCF[\approx,\not\approx,
\mathbb{N}]$. In this step, new states are used for counting
the amount of occurrences of original states.
\item Finally, we show that $\PCTAGBCF[\approx,\not\approx]$ are
closed under union. Hence, we obtain a single
$\PCTAGBCF[\approx,\not\approx]$ whose language coincides
with the one of the original $\TAGBCF[\approx,\not\approx,
\mathbb{N}]$.
\end{iteMize}

\begin{defi}
Let $\A = \langle Q, \F, F, \Delta, E, C\rangle$
be a $\TAGBCF[\approx,\not\approx, \mathbb{N}]$.
The constraint $C$ is \emph{normalized}
if it is either $\true$
or $\false$ or a disjunction of 
conjunctions of literals, where all arithmetic literals are
positive.
\end{defi}

Remember that the form of the positive arithmetic literals
can be either $a_1\|q_1\|+\ldots+a_n\|q_n\|\otimes k$
or $a_1|q_1|+\ldots+a_n|q_n|\otimes k$, 
with $\otimes$ in $\{\geq,\leq,=\}$, $n>0$, $k \geq 0$
and strictly positive $a_1,\ldots,a_n$.

\begin{lem}
Any $\TAGBCF[\approx,\not\approx, \mathbb{N}]$
can be effectively transformed into a normalized $\TAGBCF[\approx,\not\approx,
\mathbb{N}]$ with the same equational theory and preserving the language.
\end{lem}
\proof
First, by applying de Morgan laws, negations are moved inwards so that
each negation is applied to just an atom.
Second, negative arithmetic literals are made positive by
simple transformations: inequalities are inverted and
equalities become disjunctions of inequalities.
Third, strict inequalities are converted into non-strict
by adding or subtracting 1 to a side. Fourth, by applying
simple arithmetic
operations all such literals are made of the required form
$a_1\|q_1\|+\ldots+a_n\|q_n\|\otimes k$
or $a_1|q_1|+\ldots+a_n|q_n|\otimes k$, for $\otimes$
in $\{\geq,\leq,=\}$, $n>0$ and strictly positive
$a_1,\ldots,a_n$. In this step, a trivially false literal
is replaced by $\false$, and a trivially true literal
is replaced by $\true$. Finally, by applying
the standard transformation into disjunctive conjunctive normal form
we get the desired result.
\qed

In order to remove negative equality and disequality literals 
and positive arithmetic constraints,
we use the idea of inserting new
states which are {\em synonyms}
of existing states. Intuitively, a synonym is
a new state $\hat{q}$ that behaves analogous to
an existing state $\bar{q}$,
i.e.\ the rules and constraints are modified such
that the relation of $\hat{q}$ with the other states 
is the same as for $\bar{q}$. 
Nevertheless, the constraints
are further modified to ensure that, whenever $\bar{q}$ occurs
in an execution, $\hat{q}$ also occurs. Moreover,
all subterms reaching $\hat{q}$ are the same (or
equivalent modulo the relation induced by
the flat theory), but are different from (non-equivalent to)
the ones reaching $\bar{q}$.
This way, an execution of the original automaton with occurrences of
$\bar{q}$ can be transformed into an execution of the new automaton,
where the occurrences of a concrete subterm
(up to the equivalence relation) reaching $\bar{q}$ in
the original execution now reach $\hat{q}$ instead.

\begin{defi}\label{definition-synonym}
Let $\A = \langle Q, \F, F, \Delta, E, C\rangle$
be a $\TAGBCF[\approx,\not\approx, \mathbb{N}]$.
Let $\bar{q}$ be a state in $Q$. Let $\hat{q}$ be a state not in $Q$.

We define $F_{\bar{q}\leadsto \hat{q}}$ as $F$ if $\bar{q}$ is
not in $F$, and as $F\cup\{\hat{q}\}$ if $\bar{q}$ is in $F$.

We define $\Delta_{\bar{q}\leadsto \hat{q}}$ as the set of rules
obtained from the rules of $\Delta$ with all possible replacements of
occurrences of $\bar{q}$ by $\hat{q}$. 
More formally,
$\Delta_{\bar{q}\leadsto \hat{q}}$ is
$\{ f(q_1',\ldots,q_n')\to q_{n+1}' \mid
\exists f(q_1,\ldots,q_n)\to q_{n+1}\in\Delta :
\forall i\in\{1,\ldots,n+1\}: (q_i=q_i'\vee (q_i=\bar{q}\wedge q_i'=\hat{q}))\}$.

We define $C_{\bar{q}\leadsto \hat{q}}$ as the constraint
$\bigl( (\|\bar{q}\|=0 \wedge \|\hat{q}\|=0)\vee
 (\|\hat{q}\|=1 \wedge 
 \bar{q}\not\approx\hat{q}) \bigr)\wedge C'$,
where $C'$ is obtained from the normalization of $C$ by
replacing each literal by a new formula according to the following description.

\begin{iteMize}{$\bullet$}
\item
Each literal $(q_1\approx q_2)$ is replaced by the conjunction
of the literals of the set
$\bigl\{q_1'\approx q_2' \bigm| ((q_1'=q_1\vee(q_1=\bar{q}\wedge q_1'=\hat{q}))\wedge
(q_2'=q_2\vee(q_2=\bar{q}\wedge q_2'=\hat{q})))\bigr\}$.
\item
Each literal $(q_1\not\approx q_2)$ is replaced by the conjunction
of the literals of the set
$\bigl\{q_1'\not\approx q_2'\bigm| ((q_1'=q_1\vee(q_1=\bar{q}\wedge q_1'=\hat{q}))\wedge
(q_2'=q_2\vee(q_2=\bar{q}\wedge q_2'=\hat{q})))\bigr\}$.
\item
Each literal $\neg(q_1\approx q_2)$ is replaced by the disjunction
of the literals of the set
$\bigl\{\neg(q_1'\approx q_2')\bigm| ((q_1'=q_1\vee(q_1=\bar{q}\wedge q_1'=\hat{q}))\wedge
(q_2'=q_2\vee(q_2=\bar{q}\wedge q_2'=\hat{q})))\bigr\}$.
\item
Each literal $\neg(q_1\not\approx q_2)$ is replaced by the disjunction
of the literals of the set
$\bigl\{\neg(q_1'\not\approx q_2')\bigm| ((q_1'=q_1\vee(q_1=\bar{q}\wedge q_1'=\hat{q}))\wedge
(q_2'=q_2\vee(q_2=\bar{q}\wedge q_2'=\hat{q})))\bigr\}$.
\item
Each occurrence of $|\bar{q}|$ is replaced by
$|\bar{q}|+|\hat{q}|$, and
each occurrence of
$\|\bar{q}\|$ is replaced by $\|\bar{q}\|+\|\hat{q}\|$.\vspace{3 pt}
\end{iteMize}

\noindent We define $\A_{\bar{q}\leadsto\hat{q}}$ as
$\langle Q\cup\{\hat{q}\}, \F, F_{\bar{q}\leadsto \hat{q}},
\Delta_{\bar{q}\leadsto \hat{q}}, E,
C_{\bar{q}\leadsto \hat{q}}\rangle$.

We write $(F_{\bar{q}\leadsto\hat{q}})_{\bar{q}'\leadsto\hat{q}'}$
for $\hat{q}\not=\bar{q}'$ and $\hat{q}\not=\hat{q}'$ more
succinctly as $F_{\bar{q},\bar{q}'\leadsto\hat{q},\hat{q}'}$,
and similarly for
$\Delta_{\bar{q},\bar{q}'\leadsto\hat{q},\hat{q}'}$,
$C_{\bar{q},\bar{q}'\leadsto\hat{q},\hat{q}'}$ and
$\A_{\bar{q},\bar{q}'\leadsto\hat{q},\hat{q}'}$.
\end{defi}




The condition $(\|\bar{q}\|=0 \wedge \|\hat{q}\|=0)$ added to
$C_{\bar{q}\leadsto \hat{q}}$ is necessary to satisfy
$\L(\A_{\bar{q}\leadsto\hat{q}})=\L(\A)$, as it is proved
in Lemma~\ref{lemma-synonympreserveslanguage}. This lemma
is not used in the rest of the article, since the introduction
of synonyms is combined with other constraints in
further transformations. Nevertheless, we preserve 
Lemma~\ref{lemma-synonympreserveslanguage} since its proof
gives intuition about the definition of synonyms, and
the arguments are similar to other ones appearing later.


\begin{lem}\label{lemma-synonympreserveslanguage}
Let $\A = \langle Q, \F, F, \Delta, E, C\rangle$
be a $\TAGBCF[\approx,\not\approx, \mathbb{N}]$.
Let $\bar{q}$ be a state in $Q$. Let $\hat{q}$ be a state not in $Q$.

Then, $\L(\A_{\bar{q}\leadsto\hat{q}})=\L(\A)$.
\end{lem}

\proof
Accepting runs of $\A$ having no occurrence of $\bar{q}$ are also
accepting runs of $\A_{\bar{q}\leadsto\hat{q}}$. An accepting run
of $\A$ having occurrences of $\bar{q}$ can be converted into
an accepting run of $\A_{\bar{q}\leadsto\hat{q}}$ by choosing
one subterm $t$ reaching $\bar{q}$ and replacing $\bar{q}$ by $\hat{q}$
at all positions with subterms equivalent to $t$ by the relation
induced by $E$.

Accepting runs of $\A_{\bar{q}\leadsto\hat{q}}$
can be converted into accepting runs of $\A$ by
replacing each occurrence of $\hat{q}$ by $\bar{q}$.
\qed



The following lemma makes use of synonyms in order to
remove a negative literal of the form $\neg(\bar{q}\approx\bar{q}')$
preserving the language.
The next one, Lemma~\ref{lemma-removenegnotapprox}, analogously permits to remove a negative literal
of the form $\neg(\bar{q}\not\approx\bar{q}')$.

\begin{lem}\label{lemma-removenegapprox}
Let $\A = \langle Q, \F, F, \Delta, E, C\rangle$
be a $\TAGBCF[\approx,\not\approx, \mathbb{N}]$.
Let $\bar{q}$, $\bar{q}'$ be states in $Q$. 
Let $\hat{q}$, $\hat{q}'$ be distinct states not in $Q$.
Let $C$ be of the form $\neg(\bar{q}\approx\bar{q}')\wedge C'$.
Let $\A'$ be
$\langle Q\cup\{\hat{q},\hat{q}'\}, \F, F_{\bar{q},\bar{q}'\leadsto
\hat{q},\hat{q}'}, \Delta_{\bar{q},\bar{q}'\leadsto \hat{q},\hat{q}'}, E,
(\|\hat{q}\|=1\wedge\|\hat{q}'\|=1\wedge\hat{q}\not\approx\hat{q}')\wedge C'_{\bar{q},\bar{q}'\leadsto \hat{q},\hat{q}'}\rangle$.

Then, $\L(\A')=\L(\A)$ holds.
\end{lem}
\proof
Accepting runs of $\A$ can be converted into accepting runs of $\A'$
as follows. First, we choose two subterms $\bar{t}$ and $\bar{t}'$ different modulo the equivalence
relation induced by $E$ and reaching $\bar{q}$ and $\bar{q}'$,
respectively. Note that these terms must exist in order to satisfy the
literal $\neg(\bar{q}\approx\bar{q}')$ of $C$. Second, we replace
$\bar{q}$ by $\hat{q}$
at all the positions with subterms equivalent to $\bar{t}$
by the relation induced by $E$.
Similarly, we replace
$\bar{q}'$ by $\hat{q}'$
at all the positions with subterms equivalent to
$\bar{t}'$ by the relation induced by $E$. This way,
the subconstraint 
$\|\hat{q}\|=1\wedge\|\hat{q}'\|=1\wedge\hat{q}\not\approx\hat{q}'$
is satisfied, but also
$C'_{\bar{q},\bar{q}'\leadsto \hat{q},\hat{q}'}$ is satisfied.

Accepting runs of $\A'$ can be converted into accepting runs of $\A$
by replacing each occurrence of $\hat{q}$ by $\bar{q}$, and
each occurrence of $\hat{q}'$ by $\bar{q}'$. Note that
the subconstraint
$\|\hat{q}\|=1\wedge\|\hat{q}'\|=1\wedge\hat{q}\not\approx\hat{q}'$
ensures the existence of such occurrences, and with subterms which
are different modulo the equivalence relation induced by $E$.
Thus, the literal $\neg(\bar{q}\approx\bar{q}')$ of $C$ is
satisfied. The constraint $C'$ is also satisfied.
\qed

\begin{lem}\label{lemma-removenegnotapprox}
Consider the same assumptions as in Lemma~\ref{lemma-removenegapprox}, 
except that $C$ is of the form $\neg(\bar{q}\not\approx\bar{q}')\wedge C'$
and the constraint of $\A'$ is
$(\|\hat{q}\|=1\wedge\|\hat{q}'\|=1\wedge
\hat{q}\approx\hat{q}')\wedge C'_{\bar{q},\bar{q}'\leadsto \hat{q},\hat{q}'}$

Then, $\L(\A')=\L(\A)$ holds.
\end{lem}
\proof
Analogous to the proof of Lemma~\ref{lemma-removenegnotapprox}.
\qed

The following definition will be used to remove literals
of type $\|.\|_{\mathbb{N}}$.

\begin{defi}
Let $C$ be a constraint, and let $k$ be a natural number.
By $C_{\|\bar{q}\|\leadsto k}$ we define the constraint obtained
from $C$ by replacing all occurrences of $\|\bar{q}\|$ by $k$.
\end{defi}

The following two lemmas show how to remove
literals of the form $\|q\|=1$ or $\|q\|=0$ preserving the language.

\begin{lem}\label{lemma-remove1}
Let $\A = \langle Q, \F, F, \Delta, E, C\rangle$
be a $\TAGBCF[\approx,\not\approx, \mathbb{N}]$.
Let $\bar{q}$ be a state in $Q$.
Let $C$ be of the form $\|\bar{q}\|=1\wedge C'$.
Let $\A'$ be
$\langle Q, \F, F, \Delta, E,
|\bar{q}|\geq 1\wedge \bar{q}\approx\bar{q}\wedge C'_{\|\bar{q}\|\leadsto 1}\rangle$.

Then, $\L(\A')=\L(\A)$ holds.
\end{lem}
\proof
Accepting runs of $\A'$ and $\A$ coincide because the constraints
$C$ and $C_{\A'}$ have the same semantics.
\qed

\begin{lem}\label{lemma-remove0}
Let $\A = \langle Q, \F, F, \Delta, E, C\rangle$
be a $\TAGBCF[\approx,\not\approx, \mathbb{N}]$.
Let $\bar{q}$ be a state in $Q$.
Let $C$ be of the form $\|\bar{q}\|=0\wedge C'$.
Let $\A'$ be
$\langle Q, \F, F, \Delta, E,
|\bar{q}|=0\wedge C'_{\|\bar{q}\|\leadsto 0}\rangle$.

Then, $\L(\A')=\L(\A)$ holds.
\end{lem}

\proof
Accepting runs of $\A'$ and $\A$ coincide because the constraints
$C$ and $C_{\A'}$ have the same semantics.
\qed

Now, we will use the above lemmas in order to iteratively remove all
negative literals and the arithmetic literals of type 
$\|.\|_{\mathbb{N}}$. Each removal step is not defined for
arbitrary normalized $\TAGBCF[\approx,\not\approx, \mathbb{N}]$, but
just for normalized conjunctive $\TAGBCF[\approx,\not\approx,
\mathbb{N}]$. For this reason, we first describe how to transform
a given normalized $\TAGBCF[\approx,\not\approx, \mathbb{N}]$ into a list
of normalized conjunctive
$\TAGBCF[\approx,\not\approx, \mathbb{N}]$ such that, the union of
their languages coincides with the language of the original
$\TAGBCF[\approx,\not\approx, \mathbb{N}]$.

\begin{defi} \label{definition-subdivision}
Let $\A = \langle Q, \F, F, \Delta, E, C\rangle$
be a normalized $\TAGBCF[\approx,\not\approx, \mathbb{N}]$,
such that $C$ is of the form $C_1\vee C_2\vee\ldots\vee C_n$
for conjunctive constraints $C_1,C_2,\ldots,C_n$.
Let $\A_1 = \langle Q, \F, F, \Delta, E, C_1\rangle,
\A_2 = \langle Q, \F, F, \Delta, E, C_2\rangle,
\ldots, \A_n = \langle Q, \F, F, \Delta, E, C_n\rangle$.
These automata are conjunctive and normalized
and, moreover, $\L(\A)=\L(\A_1)\cup\L(\A_2)\cup\ldots\cup\L(\A_n)$
holds. We say that $\A_1,\A_2,\ldots,\A_n$ is the {\em subdivision}
of $\A$.
\end{defi}

Iteratively, we will transform a list of normalized conjunctive
$\TAGBCF[\approx,\not\approx, \mathbb{N}]$ into a new list
of automata of the same kind but with simplified constraints, preserving the language.
In order to show that this process terminates,
we define a measure on normalized conjunctive
$\TAGBCF[\approx,\not\approx, \mathbb{N}]$ which will
decrease at each step. 
Moreover, a case with minimal measure
corresponds to a positive $\TAGBCF[\approx,\not\approx, |.|_{\mathbb{N}}]$.
This measure is a pair of natural numbers which depends on
the constraint $C$ of the
normalized conjunctive $\TAGBCF[\approx,\not\approx, \mathbb{N}]$.
In the first component we have the amount of negative literals in $C$.
In the second component we have the addition of the isolated
constants in all arithmetic literal constraints
of type $\|.\|_{\mathbb{N}}$ plus the number of uses of the function
symbol $\|.\|_{\mathbb{N}}$.

\begin{defi}
We define the measure of a normalized conjunctive constraint $C$,
denoted $\langle C\rangle$ as a pair of
natural numbers. We describe it by distinguishing the
following cases.
\begin{iteMize}{$\bullet$}
\item If $C$ is of
the form $q_1\approx q_2$ or $q_1\not\approx q_2$,
then its measure is $\langle 0,0\rangle$.
\item
If $C$ is of
the form $\neg(q_1\approx q_2)$ or $\neg(q_1\not\approx q_2)$,
then its measure is $\langle 1,0\rangle$.
\item
If $C$ is of the form
$(a_1\|q_1\|+\ldots+a_n\|q_n\|\otimes k)$,
where $\otimes$ is in $\{=,\geq,\leq\}$,
then its measure is $\langle 0,n+k\rangle$.
\item
If $C$ is of the form
$(a_1|q_1|+\ldots+a_n|q_n|\otimes k)$,
where $\otimes$ is in $\{=,\geq,\leq\}$,
then its measure is $\langle 0,0\rangle$.
\item If $C$ is either $\true$ or $\false$,
then its measure is $\langle 0,0\rangle$.
\item If $C$ is a conjunction of two or more literals
$l_1\wedge l_2\wedge\ldots\wedge l_n$ with measures
$\langle a_1,b_1\rangle,$ $\langle a_2,b_2\rangle,\ldots,
\langle a_n,b_n\rangle$, then its measure
is $\langle a_1+a_2+\ldots+a_n,b_1+b_2+\ldots+b_n\rangle$.\vspace{3 pt}
\end{iteMize}

\noindent Let $\A = \langle Q, \F, F, \Delta, E, C\rangle$
be a normalized conjunctive $\TAGBCF[\approx,\not\approx, \mathbb{N}]$.
The measure of $\A$, denoted $\langle \A\rangle$
is defined as $\langle C\rangle$.

We say that $\A_1$ is bigger than $\A_2$ (or, equivalently, that $\A_2$
is smaller than $\A_1$), denoted $\A_1>\A_2$ (or $\A_2<\A_1$),
if the measure of $\A_1$ is bigger (or smaller) than the measure of $\A_2$,
according to the lexicographic extension of the relation $>$
of natural numbers.
\end{defi}

The following lemma shows that any normalized conjunctive
$\TAGBCF[\approx,\not\approx, \mathbb{N}]$ with non-minimal
measure can be transformed into a list of
$\TAGBCF[\approx,\not\approx, \mathbb{N}]$ of the same
kind with smaller measures and preserving the language.

\begin{lem}\label{lemma-stepreducemeasure}
Let $\A = \langle Q, \F, F, \Delta, E, C\rangle$ 
be a normalized conjunctive $\TAGBCF[\approx,\not\approx,\mathbb{N}]$
whose measure is not $\langle 0,0\rangle$.

Then one can construct normalized conjunctive
$\TAGBCF[\approx,\not\approx, \mathbb{N}]$
$\A_1,\ldots,\A_n$ with the same equational theory $E$,
each of them having a measure smaller than $\langle\A\rangle$
and such that $\L(\A)=\L(\A_1)\cup\ldots\cup\L(\A_n)$ holds.
\end{lem}

\proof
In the case where $C$ has some negative literal $\neg(q\approx q')$
or $\neg(q\not\approx q')$, the transformations described in
Lemmas~\ref{lemma-removenegapprox} and
\ref{lemma-removenegnotapprox} give a new 
$\TAGBCF[\approx,\not\approx, \mathbb{N}]$ $\A'$, and 
the subdivision $\A_1,\ldots,\A_n$ of the normalization of $\A'$
(as defined in Definition~\ref{definition-subdivision})
is such that the constraints $C_{\A_1},\ldots,C_{\A_n}$ have
one less negative literal than $C$. 
Thus, the measure of each of these automata is smaller
than the measure of $\A$.

In the case where $C$ has no negative literals of the
form $\neg(q\approx q')$ or $\neg(q\not\approx q')$,
its measure is of the form $\langle 0,m\rangle$ for $m>0$.
It follows that there is at least one literal of the form
$(a\|\bar{q}\|+ \mathop\sum a_i \cdot \|q_i\| \otimes k)$,
where $\otimes$ is in $\{=,\geq,\leq\}$.
We consider a new state $\hat{q}$ and the automaton
$\A_{\bar{q}\leadsto\hat{q}}$. Its constraint
$C_{\bar{q}\leadsto\hat{q}}$ is of the form
$\bigl( (\|\bar{q}\|=0 \wedge \|\hat{q}\|=0)\vee
(\|\hat{q}\|=1\wedge \bar{q}\not\approx\hat{q})\bigr)\wedge C'$. 
Note that, according to Definition~\ref{definition-synonym},
$C'$ is a conjunction because there are no negative literals of
the form $\neg(q\approx q')$ or $\neg(q\not\approx q')$ in $C$.
Thus, $C_{\bar{q}\leadsto\hat{q}}$ can be rewritten as the disjunction
of two conjunctions $C_1$ and $C_2$, where 
$C_1$ is $\|\bar{q}\|=0 \wedge \|\hat{q}\|=0\wedge C'$
and
$C_2$ is $\|\hat{q}\|=1\wedge \bar{q}\not\approx\hat{q} \wedge C'$.
Hence, the subdivision of
the normalization of $\A_{\bar{q}\leadsto\hat{q}}$ are
the automata $\A_1,\A_2$ obtained from $\A_{\bar{q}\leadsto\hat{q}}$
by replacing its constraint by $C_1$ and $C_2$, respectively.
The measures of $C_1$ and $C_2$ may be bigger than the one of $C$. 
In order to conclude, for each case we show that additional transformations can
be applied to $\A_1$ and $\A_2$, producing automata with smaller
measures than the one of $\A$ and preserving the represented language.

\begin{iteMize}{$\bullet$}
\item
The literals of $C_1$ of type $\|.\|_{\mathbb{N}}$ are
$\|\bar{q}\|=0$ and $\|\hat{q}\|=0$, and
those obtained from the literals of $C$ of type
$\|.\|_{\mathbb{N}}$ by replacing $\|\bar{q}\|$ by
$\|\bar{q}\|+\|\hat{q}\|$.
Note that original literals of the form
$(a\|\bar{q}\|+\mathop\sum a_i \cdot \|q_i\|\otimes k)$ 
have been converted into
$(a\|\bar{q}\|+a\|\hat{q}\|+ \mathop\sum a_i \cdot \|q_i\| \otimes k)$, 
and recall that there is at least one literal of this form in $C$.
Applying to $\A_1$ the transformation described in
Lemma~\ref{lemma-remove0} for $\bar{q}$ and $\hat{q}$, 
each one of the above
literals is transformed into $(a\cdot 0 + a\cdot 0 + \mathop\sum a_i \cdot \|q_i\|\otimes k)$,
which has a smaller measure than the original literal
$(a\|\bar{q}\|+\mathop\sum a_i \cdot \|q_i\|\otimes k)$. 
Moreover, the literals $\|\bar{q}\|=0$ and $\|\hat{q}\|=0$ are converted
into $|\bar{q}|=0$ and $|\hat{q}|=0$, respectively. In summary,
the measure of $((C_1)_{\bar{q}\leadsto 0})_{\hat{q}\leadsto 0}$
is smaller than the one of $C$.

\item
Similarly, the literals of $C_2$ of type $\|.\|_{\mathbb{N}}$ are
$\|\hat{q}\|=1$ and
those obtained from the literals of $C$ of type
$\|.\|_{\mathbb{N}}$ by replacing $\|\bar{q}\|$ by
$\|\bar{q}\|+\|\hat{q}\|$.
As above, note that original literals of the form
$(a\|\bar{q}\|+\mathop\sum a_i \cdot \|q_i\|\otimes k)$ have been transformed into
$(a\|\bar{q}\|+a\|\hat{q}\|+\mathop\sum a_i \cdot \|q_i\|\otimes k)$, and recall
that there is at least one literal of this form in $C$.
Applying to $C_2$ the transformation described in
Lemma~\ref{lemma-remove1} for $\hat{q}$, 
each one of the above literals is converted into 
$(a\cdot \|\bar{q}\|+a\cdot 1+\mathop\sum a_i \cdot \|q_i\|\otimes k)$.
The normalization of such a literal is the normalization of
$(a\cdot \|\bar{q}\|+\mathop\sum a_i \cdot \|q_i\| \otimes k-a)$, which might be already
normalized or must be replaced by $\true$ or $\false$ in order
to normalize it, depending on $k-a$ and $\otimes$. 
In every case,
the resulting literal
has a smaller measure than the original literal
$(a\|\bar{q}\|+\mathop\sum a_i \cdot \|q_i\| \otimes k)$. 
Moreover, the literal $\|\hat{q}\|=1$ is replaced
by $|\hat{q}|\geq 1\;\wedge\;\hat{q}\approx\hat{q}$
as a consequence of the transformation
of Lemma~\ref{lemma-remove1}. 
To summarize, the measure of $(C_2)_{\hat{q}\leadsto 1}$
is smaller than the one of $C$.
\qed\end{iteMize}

\begin{cor}\label{corollary-removenegativeandclasscounting}
Let $\A = \langle Q, \F, F, \Delta, E, C\rangle$ 
be a $\TAGBCF[\approx,\not\approx, \mathbb{N}]$.

Then, one can construct some
$\PCTAGBCF[\approx,\not\approx, |.|_{\mathbb{N}}]$
$\A_1,\ldots,\A_n$ with the same equational theory $E$
such that $\L(\A)=\L(\A_1)\cup\ldots\cup\L(\A_n)$.
\end{cor}
\proof
Without loss of generality, the constraint $C$ can be assumed
to be normalized. The subdivision of $\A$ is a collection
of normalized conjunctive $\TAGBCF[\approx,\not\approx, \mathbb{N}]$
such that the union of their languages coincides with $\L(\A)$.

By iterated application of
the Lemma~\ref{lemma-stepreducemeasure}
to each automaton of the subdivision,
combined with the fact that the ordering on
measures is well founded, we conclude to the 
effective existence of normalized conjunctive
$\TAGBCF[\approx,\not\approx, \mathbb{N}]$
$\A_1,\ldots,\A_n$ such that $\L(\A)=\L(\A_1)\cup\ldots\cup\L(\A_n)$ and
each of them has measure $\langle 0,0\rangle$.
This kind of automata are, in fact,
$\PCTAGBCF[\approx,\not\approx, |.|_{\mathbb{N}}]$, since measure
$\langle 0,0\rangle$ implies that negative literals
and literals of type $\|.\|_{\mathbb{N}}$ do not occur.
\qed

Now, in order to remove all arithmetic constraints,
it remains to remove the ones of type
$|.|_{\mathbb{N}}$. This is a rather easy task.
For a given $\PCTAGBCF[\approx,\not\approx, |.|_{\mathbb{N}}]$
$\A$ we create
a new  $\PCTAGBCF[\approx,\not\approx]$ $\A_{\not{\mathbb{N}}}$ whose purpose
is to simulate the computations of $\A$. 
To this end,
the states of $\A_{\not{\mathbb{N}}}$ count the number of occurrences of
the states of $\A$ in the simulated computation, up to a certain maximum value. 
This allows $\A_{\not{\mathbb{N}}}$ to check the constraints
of type $|.|_{\mathbb{N}}$ of $\A$ directly through states.
Thus, each state of $\A_{\not{\mathbb{N}}}$
is of the form $q_M$ for a state $q$ of $\A$ and a mapping
$M:Q_\A\to{\mathbb{N}}$, that is, a mapping counting the number
of occurrences of each state.

\begin{defi}
Let $\A = \langle Q, \F, F, \Delta, E, C\rangle$
be a normalized $\PCTAGBCF[\approx,\not\approx, |.|_{\mathbb{N}}]$.

We define ${\tt max}_\A$ as one plus the maximum isolated constant occurring
in the literals of $C$ of type $|.|_{\mathbb{N}}$, i.e.\ one plus the
maximum constant $k$ occurring in a literal of $C$ of the form
$(a_1|q_1|+\ldots+a_n|q_n|\otimes k)$, for $\otimes$ in $\{\geq,\leq,=\}$.

Given two mappings
$M_1:Q\to \{0,\ldots,{\tt max}_\A\}$ and
$M_2:Q\to \{0,\ldots,{\tt max}_\A\}$,
the {\em sum} of $M_1$ and $M_2$ is defined as the mapping
$M_1+M_2:Q\to \{0,\ldots,{\tt max}_\A\}$
satisfying $(M_1+M_2)(q)={\tt min}(M_1(q)+M_2(q),{\tt max}_\A)$.
Given a state $q$ in $Q$ we define $M_q:Q\to \{0,\ldots,{\tt
max}_\A\}$ as the mapping satisfying $M_q(q)=1$ and
$M_{q}(q')=0$ for all $q'\in Q\setminus \{q\}$.

We define $\A_{\not{\mathbb{N}}}$ as the
$\PCTAGBCF[\approx,\not\approx]$
$\langle Q_{\not{\mathbb{N}}}, \F, F_{\not{\mathbb{N}}},
\Delta_{\not{\mathbb{N}}}, E, C_{\not{\mathbb{N}}}\rangle$,
where:
\begin{iteMize}{$\bullet$}
\item $Q_{\not{\mathbb{N}}}$ is
$\{q_M\mid q\in Q\wedge M:Q\to\{0,\ldots,{\tt max}_\A\}\}$.
\item $F_{\not{\mathbb{N}}}$ is $\{q_M\in Q_{\not{\mathbb{N}}}\mid q\in F\wedge
\forall (a_1|q_1|+\ldots+a_n|q_n|\otimes k) \in C,
\otimes\in\{\geq,\leq,=\} : (a_1 M(q_1)+\ldots+a_n M(q_n)\otimes k)\}$.
\item $\Delta_{\not{\mathbb{N}}}$ is
$\{f((q_1)_{M_1},\ldots,(q_m)_{M_m})\upder{D}q_{M_1+\ldots+M_m+M_q}\mid
(f(q_1,\ldots,q_m)\upder{D}q)\in \Delta\}$.
\item $C_{\not{\mathbb{N}}}$ is
$\{\bar{q}_{\bar{M}}\approx\tilde{q}_{\tilde{M}}\mid
(\bar{q}\approx\tilde{q})\in C\}\cup
\{\bar{q}_{\bar{M}}\not\approx\tilde{q}_{\tilde{M}}\mid
(\bar{q}\not\approx\tilde{q})\in C\}$.
\end{iteMize}
\end{defi}

\begin{lem}\label{lemma-mappingspreserveslanguage}
Let $\A = \langle Q, \F, F, \Delta, E, C\rangle$
be a $\PCTAGBCF[\approx,\not\approx, |.|_{\mathbb{N}}]$.

Then, $\L(\A_{\not{\mathbb{N}}})=\L(\A)$.
\end{lem}
\proof
The accepting runs of $\A$ can be converted
into accepting runs of $\A_{\not{\mathbb{N}}}$ and
vice-versa, following the transformations described below.
\begin{iteMize}{$\bullet$}
\item
A run $r_{\not{\mathbb{N}}}$ of $\A_{\not{\mathbb{N}}}$ can be converted
into a run $r$ of $\A$ by replacing each occurrence
of a state $q_M$ by the corresponding state $q$.
\item
A run $r$ of $\A$ can be converted into a
run $r_{\not{\mathbb{N}}}$ of $\A_{\not{\mathbb{N}}}$.
The transformation can be defined recursively as follows.
Let $r$ be a run of the form
$(f(q_1,\ldots,q_m)\upder{D} q)(r_1,\ldots,r_m)$.
Let $(r_1)_{\not{\mathbb{N}}},\ldots,(r_m)_{\not{\mathbb{N}}}$
be the transformations of $r_1,\ldots,r_m$,
and let $(q_1)_{M_1},\ldots,(q_m)_{M_m}$ be the states 
reached by $(r_1)_{\not{\mathbb{N}}},\ldots,(r_m)_{\not{\mathbb{N}}}$,
respectively.
Then, $r_{\not{\mathbb{N}}}$ is
$(f((q_1)_{M_1},\ldots,(q_m)_{M_m})\upder{D}q_{M_1+\ldots+M_m+M_q})
((r_1)_{\not{\mathbb{N}}},\ldots,(r_m)_{\not{\mathbb{N}}})$.
\end{iteMize}
Each one of the two above transformations is the inverse of the other.
Thus, they describe a bijection between runs of $\A$
and runs of $\A_{\not{\mathbb{N}}}$.
Moreover,
for each run $r$ of $\A$, the state
$q_M$ reached by $r_{\not{\mathbb{N}}}$ holds that each $q'\in Q$
satisfies $M(q')={\tt min}(|r^{-1}(q')|,{\tt max}_\A)$
(note that $r^{-1}(q')$ is the set of positions reaching state $q'$).
Hence, by the definition of $F_{\not{\mathbb{N}}}$,
it follows that $q$ is in $F$ and
$r$ satisfies the arithmetic constraints
of $C$ if and only if $q_M$ is in $F_{\mathbb{N}}$.
As a consequence, $r$ is accepting if and only if 
$r_{\not{\mathbb{N}}}$ is accepting.
Thus, $\L(\A_{\not{\mathbb{N}}})=\L(\A)$ holds.
\qed

The following corollary is a consequence
of Corollary~\ref{corollary-removenegativeandclasscounting} combined
with Lemma~\ref{lemma-mappingspreserveslanguage}.

\begin{cor}
Let $\A = \langle Q, \F, F, \Delta, E, C\rangle$ 
be a $\TAGBCF[\approx,\not\approx, \mathbb{N}]$.

Then, one can construct some
$\PCTAGBCF[\approx,\not\approx]$
$\A_1,\ldots,\A_n$ with the same equational theory $E$
such that $\L(\A)=\L(\A_1)\cup\ldots\cup\L(\A_n)$.
\end{cor}

As a final step, we show that $\PCTAGBCF[\approx,\not\approx]$
are closed under union for a fixed $E$.

\begin{lem} \label{lem:union}
Let $\A_1$ and $\A_2$
be $\PCTAGBCF[\approx,\not\approx]$ with the same equational
theory $E$.
Then, a $\PCTAGBCF[\approx,\not\approx]$ $\A$ with
the same equational theory $E$ can
be effectively constructed satisfying $\L(\A)=\L(\A_1)\cup\L(\A_2)$.
\end{lem}
\proof
Let $\A_1$ be $\langle Q_1, \F, F_1, \Delta_1, E, C_1\rangle$
and $\A_2$ be $\langle Q_2, \F, F_2, \Delta_2, E, C_2\rangle$.
Without loss of generality we can assume that the sets of
states $Q_1$ and $Q_2$ are disjoint.

In the case where $C_1$ is just $\false$ the result
follows by defining $\A:=\A_2$. Similarly,
in the case where $C_2$ is just $\false$
the result follows by
defining $\A:=\A_1$.
From now on we assume that these cases do not take place.

We define $\A$ as
$\langle Q_1\uplus Q_2, \F, F_1\uplus F_2, \Delta_1\uplus\Delta_2, E,
C_1\wedge C_2\rangle$.
Note that $\A$ is a $\PCTAGBCF[\approx,\not\approx]$.
It is clear that any accepting run of $\A$ is also an
accepting run of either $\A_1$ or $\A_2$.
Moreover, it can be proved that any accepting run of either $\A_1$
or $\A_2$ is also an accepting run of $\A$. We show this
fact only for $\A_1$, since the case for $\A_2$ is analogous.

Let $r$ be an accepting run of $\A_1$. Then, $r\models C_1$ holds.
In order to see that it
is, in fact, an accepting run of $\A$, it remains to prove
$r\models C_2$. Since
$\A_2$ is a $\PCTAGBCF[\approx,\not\approx]$,
$C_2$ is a conjunction of positive literals of type
$\approx$, $\not\approx$ applied to states of $Q_2$.
Therefore, $r\models C_2$ holds, since $C_2$ is not $\false$ and any
positive literal holds because $r$ uses only states
from $Q_1$.
\qed

\begin{cor}\label{corollary-reductiontopctagc}
Let $\A = \langle Q, \F, F, \Delta, E, C\rangle$ 
be a $\TAGBCF[\approx,\not\approx, \mathbb{N}]$.

Then, one can construct a
$\PCTAGBCF[\approx,\not\approx]$
$\A'$ with the same equational theory $E$
such that $\L(\A')=\L(\A)$.
\end{cor}

\begin{cor} \label{corollary-union}
The class of $\TAGBCF$ languages (modulo the same equational theory) 
is closed under union.
\end{cor}

In order to complete the closure results for $\TAGBCF$ languages
under basic set operations, we show that they are also closed
under intersection, but not under complementation.

\begin{lem} \label{lemma-intersection}
The class of $\TAGBCF$ languages (modulo the same equational theory)
is closed under intersection.
\end{lem}
\proof
We use a classical Cartesian product of sets of states,
with a careful redefinition of constraints on this product.

More precisely,
let $\A_1 = \langle Q_1, \F, F_1, \Delta_1, E, C_1\rangle$ 
and $\A_2 = \langle Q_2, \F, F_2, \Delta_2, E, C_2 \rangle$
be two $\TAGBCF$.
We construct the $\TAGBCF$
$\A = \langle Q_1 \times Q_2, \F, F_1 \times F_2, \Delta, E, C \rangle$
where 
$\Delta = \bigl\{  
f\bigl(\langle q_{1,1},q_{2,1}\rangle,$ $\ldots,\langle q_{1,n},q_{2,n}\rangle\bigr) 
  \to  \langle q_{1},q_{2}\rangle \bigm| 
f(q_{i,1},\ldots,q_{i,n}) \to q_{i} \in \Delta_i \mbox{~for~} i \in\{1,2\} \bigr\}$
and
the constraint $C$ is obtained from $C_1 \wedge C_2$ by replacing 
every atom $q_1 \approx q'_1$ with $q_1, q'_1 \in Q_1$
(respectively $q_2 \approx q'_2$ with $q_2, q'_2 \in Q_2$)
by $\bigwedge_{q_2, q'_2 \in Q_2} \langle q_1, q_2\rangle \approx \langle q'_1, q'_2\rangle$
(respectively $\bigwedge_{q_1, q'_1 \in Q_1} \langle q_1, q_2\rangle \approx \langle q'_1, q'_2\rangle$),
and similarly for the atoms $q_1 \not\approx q'_1$, $q_2 \not\approx q'_2$.
With this construction, $\L(\A) = \L(\A_1) \cap \L(\A_2)$ holds:
the left (respectively right) projection of a successful run of $\A$ on a term $t \in \T(\F)$
is a successful run of $\A_1$ (respectively $\A_2$) on $t$, 
and the product of two successful runs $r_1$ of $\A_1$ and $r_2$ of $\A_2$, 
both on the same term $t \in \T(\F)$, is a a successful run of $\A$ on $t$.
\qed

\begin{lem} \label{lemma-complementation}
The class of $\TAGBCF$ languages is not closed under complementation.
\end{lem}
\proof
To prove the statement it suffices to define a language $L$ such that $L$ is
not recognizable by $\TAGBCF$ but its complement $\overline{L}$ is. In order
to simplify the presentation, we denote terms of the form
$f(g^{n_1}(a),f(g^{n_2}(a),\ldots f(g^{n_{k-1}}(a),g^{n_k}(a))\ldots))$ simply
with $[n_1,n_2,\ldots,n_{k-1},n_k]$.
Let $L$ be the language defined as:
$$\begin{array}{r@{\;}l}
L=\{[n_1,\ldots,n_k] \mid & k,n_1,\ldots,n_k\in\mathbb{N} \wedge{}\\
& \forall i\in\{1,\ldots,k\}\;\exists!j\in\{1,\ldots,k\}\setminus\{i\}:n_i=n_j\}
\end{array}$$
In order to prove that $L$ is not recognizable by $\TAGBCF$, by
Corollary~\ref{corollary-reductiontopctagc}, it suffices to prove it for
$\PCTAGBCF[\approx,\not\approx]$. We proceed by contradiction assuming that
there exists a $\PCTAGBCF[\approx,\not\approx]$ $\A$ such that $\L(\A)=L$. Let
$t\in L$ be the term $[1,\ldots,n,n,\ldots,1]$, where $n>|Q_\A|$, and let $r$
be an accepting run of $\A$ on $t$. By the pigeonhole principle, there exist
$i,j\in\{1,\ldots,n\}$, with $i<j$, such that the positions
$p_i=\overbrace{2.\ldots.2}^{i-1}$ and $p_j=\overbrace{2.\ldots.2}^{j-1}$
satisfy $r(p_i)=r(p_j)$. Let $r'$ be the replacement $r[r|_{p_j}]_{p_i}$. Note
that $r'$ is an accepting run of $\ta(\A)$ on the term
$[1,\ldots,i-1,j,\ldots,n,n,\ldots,1]$, which is not in $L$.
To conclude, it remains to prove that the constraints of $\A$ are satisfied in
$r'$. First, note that this replacement only introduces new subterms at the
positions $\hat{P}=\{\hat{p}\in\Pos(r)\mid\hat{p}<p_i\}$. Moreover, the rules
applied by $r'$ at positions in $\hat{P}$ are the same as in $r$,
and any constraint affecting a position in $\hat{P}$ in $r$ is
necessarily a disequality, since
$\term(r|_{\hat{p}})\not=_{E_\A}\term(r|_{p'})$ holds for $\hat{p}\in \hat{P}$
and $p'\in\Pos(r)\setminus\{\hat{p}\}$. By the definition of $r'$, necessarily
$\term(r'|_{\hat{p}})\not=_{E_\A}\term(r'|_{p'})$ holds also for $\hat{p}\in
\hat{P}$ and $p'\in\Pos(r')\setminus\{\hat{p}\}$. Therefore, $r'$ satisfies
all the constraints, and hence, $r'$ is an accepting run of $\A$, a
contradiction.

It remains to prove that $\overline{L}$ can be recognized by a
$\TAGBCF$. We start by decomposing $\overline{L}$ into simpler languages.
First, let $L_1$ be the language of the malformed terms, i.e.\ the terms over
$\{f:2,g:1,a:0\}$ that are not of the form $[n_1,\ldots,n_k]$. Second, let
$L_2$ be the language of the well-formed terms $[n_1,\ldots,n_k]$ such that for
some $i\in\{1,\ldots,k\}$ there exists no $j\in\{1,\ldots,k\}\setminus\{i\}$
satisfying $n_i=n_j$. Third, let $L_3$ be the language of the well-formed terms
$[n_1,\ldots,n_k]$ such that there exist different
$i_1,i_2,i_3\in\{1,\ldots,k\}$ satisfying $n_{i_1}=n_{i_2}=n_{i_3}$.  It is
easy to see that $\overline{L}=L_1\cup L_2\cup L_3$. Moreover, note that $L_1$
can be recognized by a \TA, $L_2$ can be recognized by a
$\PCTAGBCF[\not\approx, |.|_{\mathbb{N}}]$ and $L_3$ can be recognized by a
$\PCTAGBCF[\approx, |.|_{\mathbb{N}}]$.
By Corollaries~\ref{corollary-reductiontopctagc}
and~\ref{corollary-union}, this concludes the proof.
\qed



\section{Emptiness Decision Algorithm} 
\label{section-emptiness}
In this section we prove the decidability of the \emph{emptiness}
problem for $\PCTAGBCF$. As a consequence of this result and
the results of Section~\ref{section-arithmetic}, it follows the decidability
of emptiness for $\TAGBCF$, and even more, of $\TAGBCF[\approx,\not\approx,{\mathbb{N}}]$.

\medskip

The decidability of emptiness for $\PCTAGBCF$ is proved in three steps.
In Subsection~\ref{subsection-global-pumpings},
we present a new notion of pumping which allows to transform
a run into a smaller run under certain conditions.
In Subsection~\ref{subsection-well-quasiorder},
we define a well quasi-ordering $\leq$ on a certain set $S$.
In Subsection~\ref{subsection-mappingruntoquasiorder},
we connect the two previous subsections by
describing how to compute, for each run $r$ with
height $h= \height(r)$,
a certain sequence $e_h,\ldots,e_1$ of elements of $S$
satisfying the following fact:
there exists a pumping on $r$ if and only if
$e_i \leq e_j$ for some $h\geq i>j \geq 1$. Moreover,
each $e_i$ of the computed sequence is chosen among a finite
number of possibilities.
Finally, all of these constructions are used as follows.
Suppose the existence of an accepting run $r$.
If $r$ is ``too high'', the fact that $\leq$
is a well quasi-ordering and the properties of the sequence
imply the existence of such $i,j$. 
Thus, it follows the existence of a pumping providing
a smaller accepting run $r'$.
We conclude the existence of a computational bound
for the height of a minimum accepting run, and hence,
decidability of emptiness.

\subsection{Global Pumpings}\label{subsection-global-pumpings}

Pumping is a traditional concept in automata theory,
and in particular, it is very useful in order to reason about tree automata.
The basic idea is to convert a given run $r$ into another run
by replacing a subrun at a certain position $p$ in $r$
by a run $r'$, thus obtaining a run $r[r']_p$.
Pumpings are useful for deciding emptiness:
if a ``big'' run can always be reduced by a pumping, then decision
of emptiness is obtained by a search of an accepting ``small'' run.

For plain
tree automata, a necessary and sufficient condition to
ensure that $r[r']_p$ is a run is that the resulting states
of $r|_p$ and $r'$ coincide, since the correct application of a rule
at a certain position depends only on the resulting states
of the subruns of the direct children. In this case,
an accepting run with height bounded by the number of states exists,
whenever the accepted language is not empty.

When the tree automaton
has equality and disequality constraints, the constraints
may be falsified when replacing a subrun by a new run.
For $\PCTAGBCF$, we will define a notion of pumping ensuring
that the constraints are satisfied. This notion of pumping
requires to perform several replacements in parallel.
We first define the sets of positions involved in such
kind of pumping.

\begin{defi} \label{definition-Hi}
Let $\A$ be a $\PCTAGBCF$.
Let $r$ be a run of $\A$.
Let $i$ be an integer between $1$ and $\height(r)$.
We define 
\begin{iteMize}{$\bullet$}
\item[] $H_i$ as $\{p\in\Pos(r) \mid 0<\height(r|_p)=i\}$,
\item[] $\check{H}_i$ as 
$\{p.j\in \Pos(r) \mid 0<\height(r|_{p.j})<i
\wedge \height(r|_p)>i\}$,
\item[] $\mathring{H}_i$ as $\{p.j\in \Pos(r) \mid 0=\height(r|_{p.j})<i
\wedge \height(r|_p)>i\}$.
\end{iteMize}
\end{defi}


\begin{exa} \label{example-Hi}
According to Definition~\ref{definition-Hi},
for our running example (Example~\ref{example-menu}),
we have the $H_i$, $\check{H}_i$ and $\mathring{H}_i$ presented in 
Figure~\ref{figure-Hi}.
\begin{figure}
\begin{center}
\begin{tabular}{|c|c|c|c|}
\hline
$i$ & $H_i$ & $\check{H}_i$ & $\mathring{H}_i$\\
\hline
$5$ 
    & $\{ {\scriptstyle {\scriptstyle \rootp}} \}$ & $\emptyset$ & $\emptyset$\\
$4$ 
    & $\{ {\scriptstyle {\scriptstyle 3}} \}$   
    & $\{ {\scriptstyle {\scriptstyle 2}} \}$
    & $\{ {\scriptstyle {\scriptstyle 1}} \}$\\
$3$ 
    & $\{ {\scriptstyle {\scriptstyle 3.3}} \}$ &
      $\{ {\scriptstyle {\scriptstyle 2},\,
                        {\scriptstyle 3.2}} \}$ &
      $\{ {\scriptstyle {\scriptstyle 1},\,
                        {\scriptstyle 3.1}} \}$\\
$2$ 
    & $\{ {\scriptstyle {\scriptstyle 3.3.3}} \}$ & 
  $\{ {\scriptstyle 
      {\scriptstyle 2},\, 
      {\scriptstyle 3.2},\, {\scriptstyle 3.3.2} 
        }\}$&
  $\{ {\scriptstyle 
      {\scriptstyle 1},\, {\scriptstyle 3.1},\, 
      {\scriptstyle 3.3.1}
        }\}$\\
$1$ 
    & $\{{\scriptstyle
          {\scriptstyle 2},\, {\scriptstyle 3.2},\, 
          {\scriptstyle 3.3.2},\, {\scriptstyle 3.3.3.2} 
          }\}$ 
    & $\emptyset$
    & $\{{\scriptstyle
          {\scriptstyle 1},\, {\scriptstyle 3.1},\, {\scriptstyle 3.3.1},\,
          {\scriptstyle 3.3.3.1} 
          }\}$\\
\hline
\end{tabular}
\end{center}
\caption{$H_i$, $\check{H}_i$ and $\mathring{H}_i$ of Example~\ref{example-Hi}.} 
\label{figure-Hi}
\end{figure}
\end{exa}

The following lemma is rather straightforward from the previous definition.

\begin{lem}\label{lemma-parallel-positions}
Let $\A$ be a $\PCTAGBCF$.
Let $r$ be a run of $\A$.
Let $i$ be an integer between $1$ and $\height(r)$.
Then, any two different positions in $H_i\cup
\check{H}_i\cup\mathring{H}_i$ are parallel,
and for any arbitrary position $p$ in $\Pos(r)$ there is a position
$\bar{p}$ in $H_i\cup \check{H}_i\cup\mathring{H}_i$ such that, either
$p$ is a prefix of $\bar{p}$, or $\bar{p}$ is a prefix of $p$.
\end{lem}
\proof
For the first fact, note that any proper prefix $p$ of
a position $\bar{p}$ in $H_i\cup\check{H}_i\cup\mathring{H}_i$ satisfies
$\height(r|_p)>i$. Thus, such a $p$ is not in
$H_i\cup\check{H}_i\cup\mathring{H}_i$. For the second fact, consider any
$p$ in $ \Pos(r)$. If $\height(r|_p)\leq i$ holds,
then the smallest position $\bar{p}$ satisfying
$\bar{p}\leq p$ and $\height(r|_{\bar{p}})\leq i$ is
in $H_i\cup\check{H}_i\cup\mathring{H}_i$, and we are done. Otherwise,
if $\height(r|_p)>i$ holds, then the smallest position
$\bar{p}$ of the form $p.1.\ldots.1$ and satisfying
$\height(r|_{\bar{p}})\leq i$ is
in $H_i\cup\check{H}_i\cup\mathring{H}_i$, and we are done.
\qed

\begin{defi}
Let $\A$ be a $\PCTAGBCF$. Let $E$ be $E_\A$.
Let $r$ be a run of $\A$.
Let $i,j$ be integers satisfying $1\leq j<i\leq \height(r)$.
A {\em pump-injection} $I:(H_i\cup
\check{H}_i\cup\mathring{H}_i)\to(H_j\cup\check{H}_j\cup\mathring{H}_j)$
is an injective function
such that the following conditions hold:
\begin{enumerate}[($C_1$)]
\item $I(H_i)\subseteq H_j$, $I(\check{H}_i)\subseteq
\check{H}_j$ and $I(\mathring{H}_i)\subseteq \mathring{H}_j$. Moreover,
$I$ restricted to $\mathring{H}_i$ is the identity, 
i.e.\ $I(p)=p$ for each $p$ in $\mathring{H}_i$.
\item[($C_2$)] For each $\bar{p}$ in $H_i\cup\check{H}_i\cup\mathring{H}_i$,
$r(\bar{p})=r(I(\bar{p}))$.
\item[($C_3$)] For each $\bar{p}_1,\bar{p}_2$ in
$H_i\cup\check{H}_i\cup\mathring{H}_i$,
$(\term(r|_{\bar{p}_1})=_E\term(r|_{\bar{p}_2}))\Leftrightarrow 
(\term(r|_{I(\bar{p}_1)})=_E\term(r|_{I(\bar{p}_2)}))$.
\end{enumerate}
Let $\{\bar{p}_1,\ldots,\bar{p}_n\}$ be
$H_i\cup\check{H}_i\cup\mathring{H}_i$ more explicitly written.
The run
$r[r|_{I(\bar{p}_1)}]_{\bar{p}_1}\ldots$ $[r|_{I(\bar{p}_n)}]_{\bar{p}_n}$
is called a {\em global pumping} on $r$ with indexes $i,j$,
and injection $I$.
\end{defi}

By Condition $C_2$,
$r[r|_{I(\bar{p}_1)}]_{\bar{p}_1}\ldots[r|_{I(\bar{p}_n)}]_{\bar{p}_n}$
is a run of $\ta(\A)$, but it
is still necessary to prove that it is a run of $\A$.
By abuse of notation,
when we write
$r[r|_{I(\bar{p}_1)}]_{\bar{p}_1}\ldots[r|_{I(\bar{p}_n)}]_{\bar{p}_n}$,
we sometimes consider that $I$ and $\{\bar{p}_1,\ldots,\bar{p}_n\}$
are still explicit, and say that it is a global pumping with
some indexes $1\leq j<i\leq \height(r)$.

\begin{exa} \label{example-pump}
Following our running example, we define a pump-injection 
$I:(H_4\cup
\check{H}_4\cup\mathring{H}_4)\to(H_3\cup\check{H}_3\cup\mathring{H}_3)$ as follows:
$I(1)=1$, $I(2)=2$, $I(3)=3.3$.
We note that $I$ is a correct pump-injection:
$I(H_4)\subseteq H_3$, $I(\check{H}_4)\subseteq \check{H}_3$
and $I(\mathring{H}_4)\subseteq \mathring{H}_3$ hold,
and $I$ restricted to $\mathring{H}_4$ is, in fact, the identity,
thus ($C_1$) holds.
For ($C_2$), we have $r(1)=r(I(1))=q_{id}$,
$r(2)=r(I(2))=q_{t}$, and 
$r(3)=r(I(3))=q_{L}$.
Regarding ($C_3$), for each different $\bar{p}_1,\bar{p}_2$ in $H_4\cup\check{H}_4\cup\mathring{H}_4$, 
$\term(r|_{\bar{p}_1})\neq\term(r|_{\bar{p}_2})$ and
$\term(r|_{I(\bar{p}_1)})\neq\term(r|_{I(\bar{p}_2)})$ hold.
After applying the pump-injection $I$, we obtain the 
term and run $r'$ of Figure~\ref{figure-pump}.
\begin{figure}
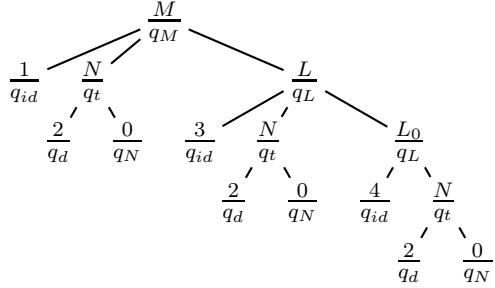

\[
\def\dedge{\ncline[linestyle=dotted]}
\psset{nodesep=2pt,levelsep=8mm,treesep=5mm}
\pstree{\TR{\frac{M}{q_M}}}{%
  \TR{\frac{1}{q_{id}}}
  \pstree{\TR{\frac{N}{q_t}}}{\TR{\frac{2}{q_{d}}} \TR{\frac{0}{q_{N}}}}
  \pstree{\TR{\frac{L}{q_L}}}{%
   \TR{\frac{3}{q_{id}}}
   \pstree{\TR{\frac{N}{q_t}}}{\TR{\frac{2}{q_{d}}} \TR{\frac{0}{q_{N}}}}
   \pstree{\TR{\frac{L_0}{q_L}}}{%
    \TR{\frac{4}{q_{id}}}
    \pstree{\TR{\frac{N}{q_t}}}{\TR{\frac{2}{q_{d}}} \TR{\frac{0}{q_{N}}}}}}}
\]
%
%
\caption{Global pumping of Example~\ref{example-pump}.}
\label{figure-pump}
\end{figure}
\end{exa}

Our goal is to prove that any global pumping
$r[r|_{I(\bar{p}_1)}]_{\bar{p}_1}\ldots[r|_{I(\bar{p}_n)}]_{\bar{p}_n}$
is a run, and in particular, that all equality and disequality
constraints are satisfied. To this end we first state the following
intermediate statement, which determines the height of the terms
pending at some positions after the pumping.
It can be easily proved by induction on the height of the involved term.

\begin{lem}\label{lemma-height}
Let $\A$ be a $\PCTAGBCF$.
Let $r$ be a run of $\A$.
Let $r'$ be the global pumping
$r[r|_{I(\bar{p}_1)}]_{\bar{p}_1}\ldots[r|_{I(\bar{p}_n)}]_{\bar{p}_n}$
on $r$ with indexes $1\leq j<i\leq \height(r)$ and injection $I$.
Let $k\geq 0$ be a natural number
and let $p$ be a position of $r$ such that
$\height(r|_p)$ is $i+k$.

Then, $p$ is also a position of $r'$ and
$\height(r'|_p)$ is $j+k$.
\end{lem}
\proof
Position $p$ is obviously a position of $r'$ since
no position in $H_i\cup \check{H}_i\cup\mathring{H}_i$
is a proper prefix of $p$.
We prove the second part of the statement by induction on $k$.
First, assume $k=0$.
Then, $\height(r|_p)$ is $i$. Thus,
$p$ is in $H_i$, say $p$ is $\bar{p}_1$.
Therefore, $r'|_p$ is $r|_{I(\bar{p}_1)}$.
By Condition~($C_1$) of the definition of pump-injection,
$I(\bar{p}_1)\in H_j$ holds. Hence, $\height(r'|_p)=
\height(r|_{I(\bar{p}_1)})=j$.

Now, assume $k>0$.
Let $m$ be the arity of ${\tt symbol}(r|_p)$.
Thus, $p.1,\ldots,p.m$ are all the child positions
of $p$ in $r$. Since $\height(r|_p)$ is $i+k$,
all $\height(r|_{p.1}),\ldots,
\height(r|_{p.m})$ are smaller than or
equal to $i+k-1$, and at least one of them is
equal to $i+k-1$.

Consider any $\alpha$
in $\{1,\ldots,m\}$.
If $\height(r|_{p.\alpha})$
is $i+k'$ for some $0\leq k'\leq k-1$, then,
by induction hypothesis, $\height(r'|_{p.\alpha})$
is $j+k'$. Otherwise, if $\height(r|_{p.\alpha})$
is strictly smaller than $i$, then
$p.\alpha$ is one of the positions in $\check{H}_i\cup\mathring{H}_i$,
say $\bar{p}_1$. In this case, $r'|_{\bar{p}_1}$ is
$r|_{I(\bar{p}_1)}$, and by Condition~($C_1$) of the definition of $I$,
$I(\bar{p}_1)$ belongs to $\check{H}_j\cup\mathring{H}_j$. Therefore,
$\height(r|_{I(\bar{p}_1)})<j$ holds,
and hence,
$\height(r'|_{p.\alpha})=
 \height(r'|_{\bar{p}_1})=
 \height(r|_{I(\bar{p}_1)})<j\leq j+k-1$ holds.

From the above cases we conclude that,
if $\height(r|_{p.\alpha})$ is $i+k-1$, then
$\height(r'|_{p.\alpha})$ is $j+k-1$,
and if $\height(r|_{p.\alpha})$ is smaller than
$i+k-1$, then $\height(r'|_{p.\alpha})$
is smaller than $j+k-1$.
It follows that all $\height(r'|_{p.1}),\ldots,
\height(r'|_{p.m})$ are smaller than or
equal to $j+k-1$, and at least one of them is
equal to $j+k-1$. As a consequence,
$\height(r'|_p)$ is $j+k$.
\qed

\begin{cor}\label{corollary-height}
Let $\A$ be a $\PCTAGBCF$.
Let $r$ be a run of $\A$.
Let $r'$ be a global pumping on $r$.
Then, $\height(r')<\height(r)$.
\end{cor}

The following lemma states that equality and disequality
relations are preserved, not only for terms pending at
the positions of the domain of $I$, but also for terms
pending at prefixes of positions of such domain.
Again, it is rather easy to prove by induction on
the height of the involved terms.

\begin{lem}\label{lemma-preserves-equal}
Let $\A$ be a $\PCTAGBCF$.
Let $r$ be a run of $\A$.
Let $r'$ be the global pumping
$r[r|_{I(\bar{p}_1)}]_{\bar{p}_1}\ldots[r|_{I(\bar{p}_n)}]_{\bar{p}_n}$
with indexes $1\leq j<i\leq \height(r)$ and injection $I$.
Let $p_1,p_2$ be positions of $r$ satisfying
that each of them is a prefix of a position in
$H_i\cup\check{H}_i\cup\mathring{H}_i$.

Then, $p_1,p_2$ are positions of $r'$ and
$(\term(r|_{p_1})=_E\term(r|_{p_2}))\Leftrightarrow
(\term(r'|_{p_1})=_E\term(r'|_{p_2}))$ holds.
\end{lem}
\proof
The first statement follows by Lemma~\ref{lemma-height}.
We prove the second part of the statement by induction on
$\height(r|_{p_1})+\height(r|_{p_2})$.
We distinguish the following cases:

\begin{iteMize}{$\bullet$}
\item 
Assume that both $p_1$ and $p_2$ are positions in
$H_i\cup\check{H}_i\cup\mathring{H}_i$, say
$\bar{p}_1$ and $\bar{p}_2$, respectively.
Therefore, $r'|_{p_1}$ is $r|_{I(\bar{p}_1)}$
and $r'|_{p_2}$ is $r|_{I(\bar{p}_2)}$.
By Condition~($C_3$) of the definition of pump-injection,
$(\term(r|_{\bar{p}_1})=_E\term(r|_{\bar{p}_2}))\Leftrightarrow 
(\term(r|_{I(\bar{p}_1)})=_E\term(r|_{I(\bar{p}_2)}))$ holds.
Thus,
$(\term(r|_{p_1})=_E\term(r|_{p_2}))\Leftrightarrow
(\term(r'|_{p_1})=_E\term(r'|_{p_2}))$ holds, and we are done.

\item
Assume that one of $p_1$ or $p_2$, say $p_1$, is a proper prefix of
a position in $H_i\cup\check{H}_i\cup\mathring{H}_i$,
and $p_2$ is a position in 
$H_i\cup\check{H}_i\cup\mathring{H}_i$.
Then, $\height(r|_{p_1})=i+k$ for some $k>0$, and
$\height(r|_{p_2})\leq i$ holds.
Thus, $(\term(r|_{p_1})\not=_E\term(r|_{p_2}))$ holds.
By Lemma~\ref{lemma-height},
$\height(r'|_{p_1})=j+k$. By the definition of pump-injection,
$\height(r'|_{p_2})\leq j$.
Thus, also $(\term(r'|_{p_1})\not=_E\term(r'|_{p_2}))$ holds,
and we are done.

\item
Assume that both $p_1$ and $p_2$ are proper prefixes of positions in
$H_i\cup\check{H}_i\cup\mathring{H}_i$.
Note that, in this case, ${\tt symbol}(r'|_{p_1})={\tt symbol}(r|_{p_1})$
and ${\tt symbol}(r'|_{p_2})={\tt symbol}(r|_{p_2})$ hold.
Let ${\tt symbol}(r|_{p_1})$ and ${\tt symbol}(r|_{p_2})$
be $f$ and $g$, with arities $n$ and $m$, respectively.
Recall that $I$ is the identity for the positions
in $\mathring{H}_i$, and hence, a position $\alpha$ in $\{1,\ldots,n\}$
satisfies ${\tt symbol}(r|_{p_1.\alpha})\in\F_0\Leftrightarrow
{\tt symbol}(r'|_{p_1.\alpha})\in\F_0$,
and ${\tt symbol}(r|_{p_1.\alpha}),{\tt symbol}(r'|_{p_1.\alpha})\in\F_0
\Rightarrow {\tt symbol}(r|_{p_1.\alpha})={\tt symbol}(r'|_{p_1.\alpha})$.
Similarly,
a position $\beta$ in $\{1,\ldots,m\}$
satisfies ${\tt symbol}(r|_{p_2.\beta})\in\F_0\Leftrightarrow
{\tt symbol}(r'|_{p_2.\beta})\in\F_0$,
and ${\tt symbol}(r|_{p_2.\beta}),{\tt symbol}(r'|_{p_2.\beta})\in\F_0
\Rightarrow {\tt symbol}(r|_{p_2.\beta})={\tt symbol}(r'|_{p_2.\beta})$.
Moreover, since such positions
$p_1.\alpha$ and $p_2.\beta$ are prefixes of positions in
$H_i\cup\check{H}_i\cup\mathring{H}_i$, by induction hypothesis,
$(\term(r|_{p_1.\alpha})=_E\term(r|_{p_2.\beta}))\Leftrightarrow
(\term(r'|_{p_1.\alpha})=_E\term(r'|_{p_2.\beta}))$
for all such $\alpha$ in $\{1,\ldots,n\}$ and $\beta$ in $\{1,\ldots,m\}$.
By Lemma~\ref{lemma-nightmare},
$(\term(r|_{p_1})=_E\term(r|_{p_2}))\Leftrightarrow
(\term(r'|_{p_1})=_E\term(r'|_{p_2}))$ follows, and we are done.
\qed\end{iteMize}

\medskip

\noindent Now we prove that the result of a global pumping preserves
the satisfaction of the global constraints.

\begin{lem}
Let $\A$ be a $\PCTAGBCF$.
Let $r$ be a run of $\A$.
Let $r'$ be the global pumping
$r[r|_{I(\bar{p}_1)}]_{\bar{p}_1}\ldots
[r|_{I(\bar{p}_n)}]_{\bar{p}_n}$
with indexes $1\leq j<i\leq \height(r)$ and injection $I$.

Then, $r'$ satisfies all global constraints of $\A$.
\end{lem}

\proof
Let us consider two different positions
$p_1,p_2$ of $\Pos(r')$ involved in the constraint $C_\A$,
i.e.\ either $r'(p_1)\approx r'(p_2)$
or $r'(p_1)\not\approx r'(p_2)$ occurs in
$C_\A$.
According to Lemma~\ref{lemma-parallel-positions},
we can distinguish the following cases:

\begin{iteMize}{$\bullet$}
\item 
Suppose that a position in $H_i\cup\check{H}_i\cup\mathring{H}_i$,
say $\bar{p}_1$, is a prefix of both $p_1,p_2$.
Then, $r'|_{p_1}=r|_{I(\bar{p}_1).(p_1-\bar{p}_1)}$ and
$r'|_{p_2}=r|_{I(\bar{p}_1).(p_2-\bar{p}_1)}$ hold.
Hence, $r'|_{p_1}$ and $r'|_{p_2}$
are also subruns of $r$ occurring at different positions.
Thus, since $r$ is a run, they satisfy the atom involving
$r'(p_1)$ and $r'(p_2)$.

\item
Suppose that two different positions in $H_i\cup\check{H}_i\cup\mathring{H}_i$,
say $\bar{p}_1$ and $\bar{p}_2$,
are prefixes of $p_1$ and $p_2$, respectively.
Then, $r'|_{p_1}=r|_{I(\bar{p}_1).(p_1-\bar{p}_1)}$ and
$r'|_{p_2}=r|_{I(\bar{p}_2).(p_2-\bar{p}_2)}$ hold.
By the injectivity of $I$,
$I(\bar{p}_1)\not=I(\bar{p}_2)$ holds.
Moreover,
by Lemma~\ref{lemma-parallel-positions}, $I(\bar{p}_1)\parallel
I(\bar{p}_2)$ holds.
Hence, as before, $r'|_{p_1}$ and $r'|_{p_2}$
are subruns of $r$ occurring at different (in fact, parallel) positions.
Thus, they satisfy the atom involving
$r'(p_1)$ and $r'(p_2)$.

\item
Suppose that one of $p_1,p_2$,
say $p_1$, is a proper prefix of a position in
$H_i\cup\check{H}_i\cup\mathring{H}_i$,
and that $p_2$ satisfies
that some position in $H_i\cup\check{H}_i\cup\mathring{H}_i$ is a prefix of $p_2$.
It follows that $\height(r'|_{p_2})$ is smaller than or equal to
$j$, and $r'|_{p_2}$ is also a subrun of $r$.
Moreover, $p_1$ is also a position of $r$,
$r'(p_1)=r(p_1)$ holds,
and $\height(r|_{p_1})=i+k$ holds for some $k>0$.
Hence, $\term(r|_{p_1})\not=_E \term(r'|_{p_2})$ holds.
Since $r$ is a run and $r'|_{p_2}$ is a subrun of $r$, the atom involving
$r(p_1)$ and $r'(p_2)$
is necessarily of the form
$r(p_1)\not\approx r'(p_2)$.
Thus, the atom involving
$r'(p_1)$ and $r'(p_2)$
is necessarily of the form
$r'(p_1)\not\approx r'(p_2)$.
By Lemma~\ref{lemma-height}, $\height(r'|_{p_1})$
is $j+k$. Therefore, also
$\term(r'|_{p_1})\not=_E \term(r'|_{p_2})$ holds,
and hence, such an atom is satisfied for such positions in $r'$.

\item
Suppose that both $p_1,p_2$ are proper prefixes of positions
in $H_i\cup\check{H}_i\cup\mathring{H}_i$.
Then, $p_1,p_2$ are positions of $r$ satisfying
$\height(r|_{p_1}),\height(r|_{p_2})\geq i$.
Moreover, $r(p_1)=r'(p_1)$
and $r(p_2)=r'(p_2)$ hold.
Since $r$ is a run, the atom involving
$r(p_1)$ and $r(p_2)$
is satisfied in the run $r$ for positions $p_1$ and $p_2$.
By Lemma~\ref{lemma-preserves-equal},
$(\term(r|_{p_1})=_E \term(r|_{p_2}))\Leftrightarrow
(\term(r'|_{p_1})=_E \term(r'|_{p_2}))$ holds.
Thus, the atom involving
$r'(p_1)$ and $r'(p_2)$
is satisfied in the run $r'$ for positions $p_1$ and $p_2$.
\qed\end{iteMize}

\noindent Finally, we prove that the result of a global pumping preserves
the satisfaction of the constraints between brothers.

\begin{lem}
Let $\A$ be a $\PCTAGBCF$.
Let $r$ be a run of $\A$.
Let $r'$ be the global pumping
$r[r|_{I(\bar{p}_1)}]_{\bar{p}_1}\ldots
[r|_{I(\bar{p}_n)}]_{\bar{p}_n}$
with indexes $1\leq j<i\leq \height(r)$ and injection $I$.

Then, $r'$ satisfies all constraints between brothers of $\A$.
\end{lem}

\proof
Let us consider a position
$p$ of $\Pos(r')$ and two positions $i_1,i_2$ involved in
a constraint of the rule used at position $p$ in $r'$,
i.e.\ either $\gamma=(i_1\approx i_2)$
or $\gamma=(i_1\not\approx i_2)$ occur in this constraint.
According to Lemma~\ref{lemma-parallel-positions},
we can distinguish the following cases:

\begin{iteMize}{$\bullet$}
\item 
Suppose that a position in $H_i\cup\check{H}_i\cup\mathring{H}_i$,
is a prefix of $p$.
Then, $r'|_p$ is also a subrun of $r$.
Thus, since $r$ is a run, the constraint is satisfied.

\item
Suppose that $p$ is a proper prefix of a position in
$H_i\cup\check{H}_i\cup\mathring{H}_i$.
Then, $p.i_1$ and $p.i_2$
are prefixes of positions in $H_i\cup\check{H}_i\cup\mathring{H}_i$.
By Lemma~\ref{lemma-preserves-equal},
$(\term(r|_{p.i_1})=_E \term(r|_{p.i_2}))\Leftrightarrow
(\term(r'|_{p.i_1})=_E \term(r'|_{p.i_2}))$ holds.
Since $r$ is a run, it follows that $(\term(r|_{p.i_1})=_E \term(r|_{p.i_2}))
\Leftrightarrow \gamma=(i_1\approx i_2)$.
Thus, $(\term(r'|_{p.i_1})=_E \term(r'|_{p.i_2}))
\Leftrightarrow \gamma=(i_1\approx i_2)$ holds.
Thus, the atom involving
$i_1$ and $i_2$
is satisfied in the run $r'$ for position $p$.\vspace{3 pt}
\qed\end{iteMize}

\noindent As a consequence of the previous lemmas, we have that
the result of a global pumping satisfies all constraints.

\begin{cor}\label{corollary-pumpingisrun}
Let $\A$ be a $\PCTAGBCF$.
Let $r$ be a run of $\A$.
Let $r'$ be the global pumping
$r[r|_{I(\bar{p}_1)}]_{\bar{p}_1}\ldots
[r|_{I(\bar{p}_n)}]_{\bar{p}_n}$
with indexes $1\leq j<i\leq \height(r)$ and injection $I$.

Then, $r'$ is a run of $\A$.
\end{cor}

\subsection{A well quasi-ordering}\label{subsection-well-quasiorder}

In this subsection we define a well
quasi-ordering. It assures the existence of a computational bound for
certain sequences of elements of the corresponding well quasi-ordered set.
It will be connected with global pumpings in
the next subsection.

\begin{defi}
Let $\leq$ denote the usual
quasi-ordering on natural numbers.
Let $n$ be a natural number.

We define the extension of $\leq$ to $n$-tuples
of natural numbers
as $\langle x_1,\ldots,x_n\rangle\leq\langle y_1,\ldots,y_n\rangle$
if $x_i\leq y_i$ for each $i$ in $\{1,\ldots,n\}$.
We define ${\tt sum}(\langle x_1,\ldots,x_n\rangle):=x_1+\cdots+x_n$.

We define the extension of $\leq$ to multisets of $n$-tuples
of natural numbers as
$[e_1,\ldots,e_\alpha]\leq [e_1',\ldots,e_\beta']$ if
there is an injection $I:\{1,\ldots,\alpha\}\to\{1,\ldots,\beta\}$
satisfying $e_i\leq e_{I(i)}'$ for each $i$ in $\{1,\ldots,\alpha\}$.
We define ${\tt sum}([e_1,\ldots,e_\alpha]):={\tt sum}(e_1)+
\cdots+{\tt sum}(e_\alpha)$.

We define the extension of $\leq$ to pairs of multisets of
$n$-tuples of natural numbers as
$\langle P_{1},\check{P}_{1}\rangle\leq \langle P_{2},\check{P}_{2}\rangle$
if $P_{1}\leq {P}_{2}$ and $\check{P}_{1}\leq \check{P}_{2}$.
\end{defi}

As a direct consequence of Higman's Lemma~\cite{Gallier91}
we have the following:

\begin{lem}
Given $n$,
$\leq$ is a well quasi-ordering for pairs of multisets
of $n$-tuples of natural numbers.
\end{lem}

In any infinite sequence $e_1,e_2,\ldots$
of elements from a well quasi-ordered
set there always exist two indexes $i<j$ satisfying
$e_i\leq e_j$. In general, this fact does not imply
the existence of a bound for the length of sequences
without such indexes.
For example, the relation $\leq$ between natural numbers
is a well quasi-ordering, but there may exist arbitrarily
long sequences $x_1,\ldots,x_k$ of natural numbers such that 
$x_i > x_j$ for all $1 \leq i < j \leq k$. 
In order to bound the length of such sequences,
it is sufficient to force that the first element and each
next element of the sequence are chosen among a finite
number of possibilities. 
Indeed in this this case, by K{\"o}nig's lemma,
the prefix trees describing all such (finite) sequences is finite.
As a particular case of this fact we have the following
result (the proof is standard, but we include it for completeness).

\begin{lem} \label{lemma-computational-bound}
There exists a computable function
$B:\mathbb{N}\times\mathbb{N}\to\mathbb{N}$
such that, given two natural numbers $a,n$,
$B(a,n)$ is a bound for the length $\ell$ of
any sequence
$\langle T_{1}, \check{T}_{1}\rangle, \ldots, \langle T_{\ell}, \check{T}_{\ell}\rangle$
of pairs of multisets of $n$-tuples
of natural numbers such that the following conditions hold:
\begin{enumerate}[\em(1)] 
\item The tuple $\langle 0,\ldots,0\rangle$ does not
occur in any $T_{i}$, $\check{T}_{i}$ 
for $i$ in $\{1,\ldots,\ell\}$.
\item ${\tt sum}(T_{1}) = 1$ and ${\tt sum}(\check{T}_{1}) = 0$.
\item For each $i$ in $\{1,\ldots,\ell-1\}$,
 $a \cdot {\tt sum}(T_{i}) + {\tt sum}(\check{T}_{i})\geq
  {\tt sum}(T_{i+1}) + {\tt sum}(\check{T}_{i+1})$.
\item There are no $i,j$ satisfying $1\leq i < j \leq \ell$ and 
 $\langle T_{i},\check{T}_{i}\rangle\leq \langle T_{j},\check{T}_{j}\rangle$.
\end{enumerate}
\end{lem}
\proof
For proving the statement, we first construct 
a rooted tree $S = (V, E)$ labelled by sequences of
pairs of multisets of $n$-tuples, where the depth of each node is
equal to the length of the sequence labeling it and such that the set
of internal nodes of $S$ corresponds exactly to the set of sequences
satisfying conditions (1) to (4). Second, we show that $S$
is finite. This concludes the proof, since finiteness of $S$
and its constructive definition
imply that $S$ is computable, and $B(a,n)$ can be
defined as the maximal depth of $S$.

We define $V$ as the set of all the
sequences $\langle T_{1}, \check{T}_{1} \rangle, \dots,
           \langle T_{\ell}, \check{T}_{\ell} \rangle$
of pairs of multisets of $n$-tuples satisfying the conditions
(1) to (3) and such that there are no $i,j$ satisfying
$1\leq i < j < \ell$ and 
$\langle T_{i}, \check{T}_{i} \rangle \leq \langle T_{j}, \check{T}_{j} \rangle$.
This last condition, that we will refer to as (5), is weaker than (4)
since in (5) we have $j<\ell$ instead of $j\leq\ell$.
Thus, all sequences satisfying conditions (1) to (4) belong to $V$.
Note that $V$ contains the empty sequence, which we denote as $\varepsilon$.
We define $E \subseteq V^2$ as the set of edges containing
$\langle T_{1}, \check{T}_{1} \rangle, \dots, \langle T_{i}, \check{T}_{i}\rangle
      \longrightarrow
      \langle T_{1}, \check{T}_{1} \rangle, \dots, 
       \langle T_{i}, \check{T}_{i}\rangle,
       \langle T_{i+1}, \check{T}_{i+1}\rangle$
     for every such couple of sequences in $V$.

It is quite obvious that $S = (V, E)$ is a tree rooted at $\varepsilon$,
since $\varepsilon$ does not have an input edge, each sequence of length $1$ has
a unique input edge coming from $\varepsilon$, and each sequence of
length $i > 1$ has a unique input edge coming from its unique prefix sequence
of length $i-1$. Also, the set
of internal nodes of $S$ is exactly the set of sequences satisfying
conditions (1) to (4), and the set of leaves of $S$ is exactly the
set of sequences satisfying conditions (1) to (3), and (5), but not (4).

It remains to show that $S$ is finite. To this end, it suffices
to see that $S$ is finitely branching and that there is
no path with infinite length.

First, we prove that each node $v\in V$ has a finite branching: 
$\varepsilon$ links to all the sequences of length $1$, 
the number of which is bounded  by conditions (1) and (2);
and each sequence
$\langle T_{1}, \check{T}_{1} \rangle, \dots, \langle T_{i}, \check{T}_{i}\rangle$ 
can only link to sequences of the form
$\langle T_{1}, \check{T}_{1} \rangle, \dots, 
\langle T_{i}, \check{T}_{i}\rangle, \langle T_{i+1}, \check{T}_{i+1}\rangle$,
the number of which is bounded by conditions (1) and (3).

Second, we prove that there is no path with infinite length in $S$
in a standard way. 
We proceed by contradiction
by assuming that we have an infinite path $v_0, v_1, v_2, v_3, \dots$
By construction, we have $v_0 = \varepsilon$, 
and for all $i\geq 1$ and all $j\geq i$, the
prefix of length $i$ of the sequence $v_j$ is equal to $v_i$.
Consider the infinite sequence
$\langle T_{1}, \check{T}_{1}\rangle, \langle T_{2}, \check{T}_{2}\rangle, \dots$
where for all $i \geq 1$, $\langle T_{i}, \check{T}_{i}\rangle$ 
is the last element of the sequence $v_i$. 
Since $\leq$ on pairs of multisets of $n$-tuples is a well quasi-ordering, 
there exist two indexes $i, j$ satisfying $i < j$ and 
$\langle T_{i}, \check{T}_{i}\rangle \leq \langle T_{j}, \check{T}_{j}\rangle$.
Hence, all sequences $v_{k}$ for $k > j$ do not satisfy
condition (5), and hence they do not belong to $V$, contradicting
the infiniteness of the path.
\qed

In order to bound the height of a term accepted by a given $\PCTAGBCF$ $\A$
(and of minimum height),
Lemma~\ref{lemma-computational-bound} will be used
by making $a$ to be the maximum arity of the signature
of $\A$, and making $n$ to be the number of states of $\A$.

\subsection{Mapping a run to a sequence of the well quasi-ordered set}
\label{subsection-mappingruntoquasiorder}

We will associate, to each number $i$ in $\{1,\ldots,\height(r)\}$,
a pair of multisets of $n$-tuples of natural numbers, which can be compared
with other pairs according to the definition
of $\leq$ in the previous subsection.
To this end, we first associate 
$n$-tuples to terms and 
multisets of $n$-tuples to sets of positions.

\begin{defi}
Let $\A$ be a $\PCTAGBCF$. Let $E$ be $E_\A$.
Let $q_1,\ldots,q_n$ be the states of $\A$.
Let $r$ be a run of $\A$.
Let $P$ be a set of positions of $r$.
Let $t$ be a term. We define $r_{t,P}$ as the following tuple
of natural numbers:
$
\bigl\langle \big|\{p\in P \mid \term(r|_p)=_E t\wedge r(p)=q_1\}\big|,
\ldots,
\big|\{p\in P \mid \term(r|_p)=_E t\wedge r(p)=q_n\}\big|\bigr\rangle
$
\end{defi}

\begin{defi}
Let $\A$ be a $\PCTAGBCF$. Let $E$ be $E_\A$.
Let $r$ be a run of $\A$.
Let $P$ be a set of positions of $r$.
Let $\{[t_1],\ldots,[t_k]\}$ be the set of equivalence
classes modulo $E$ of the set of terms
$\{\term(r|_p) \mid p\in P\}$
with representatives $t_1,\ldots,t_k$.
We define $r_P$ as the multiset
$[r_{t_1,P},\ldots,r_{t_k,P}]$.
\end{defi}

\begin{exa} \label{example-rHi}
Following our running example, for the representation
of the $n$-tuples of natural numbers we order the states
as $\langle q_d, q_N, q_{id}, q_t, q_L, q_M \rangle$.
The multisets $r_{H_i}$, $r_{\check{H}_i}$ and $r_{\mathring{H}_i}$
are presented in Figure~\ref{figure-rHi}.
\begin{figure}
\begin{center}
\begin{tabular}{|c|c|c|c|}
\hline
$i$ & $r_{H_i}$ & $r_{\check{H}_i}$ & $r_{\mathring{H}_i}$\\
\hline
$5$ & $[{\scriptstyle\langle 0,0,0,0,0,1\rangle}]$ & $[\ ]$ & $[\ ]$\\
$4$ & $[{\scriptstyle\langle 0,0,0,0,1,0\rangle}]$ &
$[{\scriptstyle\langle 0,0,0,1,0,0\rangle}]$ &
$[{\scriptstyle\langle 0,0,1,0,0,0\rangle}]$ \\
$3$ & $[{\scriptstyle\langle 0,0,0,0,1,0\rangle}]$ &
$[{\scriptstyle\langle 0,0,0,2,0,0\rangle}]$&
$[{\scriptstyle{\scriptstyle\langle 0,0,1,0,0,0\rangle},
{\scriptstyle\langle 0,0,1,0,0,0\rangle}}]$\\
$2$ & $[{\scriptstyle\langle 0,0,0,0,1,0\rangle}]$ &
$[{\scriptstyle\langle 0,0,0,3,0,0\rangle}]$&
$[{\scriptstyle{\scriptstyle\langle 0,0,1,0,0,0\rangle},{\scriptstyle\langle 0,0,1,0,0,0\rangle},}$\\
&&&${\scriptstyle\langle 0,0,1,0,0,0\rangle}]$\\
$1$ & $[{\scriptstyle\langle 0,0,0,4,0,0\rangle}]$ & $[\ ]$ &
$[{\scriptstyle{\scriptstyle\langle 0,0,1,0,0,0\rangle},{\scriptstyle\langle 0,0,1,0,0,0\rangle},}$\\
&&&${\scriptstyle{\scriptstyle\langle 0,0,1,0,0,0\rangle},{\scriptstyle\langle 0,0,1,0,0,0\rangle}}]$\\
\hline
\end{tabular}
\end{center}
\caption{Multisets $r_{H_i}$, $r_{\check{H}_i}$ and
$r_{\mathring{H}_i}$ of Example \ref{example-rHi}.}
\label{figure-rHi}
\end{figure}
\end{exa}



The following lemma connects the existence of a pump-injection
with the quasi-ordering relation.

\begin{lem}\label{lemma-leqiffpumpinjection}
Let $\A$ be a $\PCTAGBCF$.
Let $r$ be a run of $\A$.
Let $i,j$ be integers satisfying $1\leq j<i\leq \height(r)$.

Then, there exists a pump-injection
$I:(H_i\cup\check{H}_i\cup\mathring{H}_i)\to
(H_j\cup\check{H}_j\cup\mathring{H}_j)$
if and only if $\langle r_{H_i},r_{\check{H}_i}\rangle
\leq\langle r_{H_j},r_{\check{H}_j}\rangle$.
\end{lem}

\proof
Although we prove both directions of the double implication,
the left-to-right one
is technical but not conceptually difficult,
and it is not necessary for the rest of the paper.
In the following, we write $E$ for $E_\A$.

\noindent
{$\Rightarrow)$}
Assume that there exists a pump-injection
$I:(H_i\cup\check{H}_i\cup\mathring{H}_i)
\to(H_j\cup\check{H}_j\cup\mathring{H}_j)$. We just
prove $r_{H_i}\leq r_{H_j}$, since $r_{\check{H}_i}\leq r_{\check{H}_j}$ can
be proved analogously.
By Condition~($C_1$) of the definition of pump-injection,
$I(H_i)\subseteq H_j$ holds.
We write the equivalence classes of
$\{\term(r|_p)\;|\;p\in H_i\}$ and
$\{\term(r|_p)\;|\;p\in H_j\}$ modulo $E$ more explicitly as
$\{[t_{i,1}],\ldots,[t_{i,\alpha}]\}$ and
$\{[t_{j,1}],\ldots,[t_{j,\beta}]\}$,
respectively.
Hence, it remains to prove that
$[r_{t_{i,1},H_i},\ldots,r_{t_{i,\alpha},H_i}]
\leq [r_{t_{j,1},H_j},\ldots,r_{t_{j,\beta},H_j}]$.
To this end we define the function
$I':\{1,\ldots,\alpha\}\to\{1,\ldots,\beta\}$ as follows.
For each $\gamma$ in $\{1,\ldots,\alpha\}$, we choose a position
$p$ in $H_i$
satisfying $\term(r|_p)=_E t_{i,\gamma}$, determine the index $\delta$
of the term $t_{j,\delta}$ satisfying $t_{j,\delta}=_E \term(r|_{I(p)})$,
and define $I'(\gamma):=\delta$. This function $I'$ is injective
due to Condition~($C_3$) of the definition of pump-injection. 
In order to conclude, it suffices to prove
$r_{t_{i,\gamma},H_i}\leq r_{t_{j,I'(\gamma)},H_j}$ for each
$\gamma$ in $\{1,\ldots,\alpha\}$. We just prove it for
$\gamma=1$. For proving $r_{t_{i,1},H_i}\leq r_{t_{j,I'(1)},H_j}$ it suffices
to prove the following statement for each state $q$ of $\A$:
$
\big|\{ p\in H_i \mid \term(r|_p)=_E t_{i,1}\wedge r(p)=q \}\big| 
\leq
\big|\{p\in H_j  \mid \term(r|_p)=_E t_{j,I'(1)}\wedge r(p)=q \}\big|
$.

To this end, since $I$ is injective, it suffices
to prove that
$I(\{p\in H_i \mid \term(r|_p)=_E t_{i,1}\wedge r(p)=q\})$
is included in
$\{p\in H_j \mid \term(r|_p)=_E t_{j,I'(1)}\wedge r(p)=q\}$
for each state $q$ of $\A$.
Thus, consider any $\bar{p}$ of
$\{p\in H_i \mid \term(r|_p)=_E t_{i,1}\wedge r(p)=q\}$.
Let $p'$ be the chosen position for defining $I'(1)$.
In particular, $\term(r|_{p'})=_E t_{i,1}$ and
$\term(r|_{I(p')})=_E t_{j,I'(1)}$ hold.
Note that $\term(r|_{\bar{p}})=_E \term(r|_{p'})=_E t_{i,1}$ holds.
Thus, by Condition~($C_3$) of the definition of pump-injection,
$\term(r|_{I(\bar{p})})=_E \term(r|_{I(p')})$ holds.
Therefore, $\term(r|_{I(\bar{p})})=_E t_{j,I'(1)}$ holds.
In order to show the inclusion $I(\bar{p})\in
\{p\in H_j\mid \term(r|_p)=_E t_{j,I'(1)}\wedge r(p)=q\}$
it remains to see $r({I(\bar{p})})=q$.
Note that, since $\bar{p}$
belongs to
$\{ p \in H_i\mid \term(r|_p)=_E t_{i,1}\wedge r(p) = q \}$, 
$r({\bar{p}}) = q$ holds.
By Condition~($C_2$) of the definition of pump-injection,
$r({I(\bar{p})}) = r({\bar{p}}) = q$ holds,
and we are done.

\medskip\noindent
{$\Leftarrow)$}
Assume that
 $\langle r_{H_i},r_{\check{H}_i}\rangle
\leq\langle r_{H_j},r_{\check{H}_j}\rangle$ holds.
We have to construct a pump-injection
$I:(H_i\cup\check{H}_i\cup\mathring{H}_i)\to
(H_j\cup\check{H}_j\cup\mathring{H}_j)$.
By the definition of pump-injection,
the restriction $I:\mathring{H}_i\to\mathring{H}_j$ must
be defined as the identity, which is not a problem since
$\mathring{H}_i$ is always included in $\mathring{H}_j$.
Conditions ($C_2$) and ($C_3$) are satisfied
for free for these positions. 
Moreover, for positions
$\bar{p}_1'\in H_i\cup\check{H}_i$ and $\bar{p}_2'\in \mathring{H}_i$,
Condition~($C_3$) holds whenever Condition~($C_1$) holds since
in this case $\term(r|_{\bar{p}_1'})\not=_E\term(r|_{\bar{p}_2'})$
and $\term(r|_{I(\bar{p}_1')})\not=_E\term(r|_{I(\bar{p}_2')})$
hold.

Hence, it remains to define $I:(H_i\cup\check{H}_i)\to
(H_j\cup\check{H}_j)$.
We just define $I:H_i\to H_j$ and prove Conditions
($C_2$) and ($C_3$) for $\bar{p}, \bar{p}_1, \bar{p}_2$ in $H_i$.
This is because $I:\check{H}_i\to \check{H}_j$ can be defined analogously,
and Conditions ($C_2$) and ($C_3$) for the corresponding positions
can be checked analogously. Moreover, for positions
$\bar{p}_1'\in H_i$ and $\bar{p}_2'\in \check{H}_i$,
Condition~($C_3$) holds whenever Condition~($C_1$) holds since
in this case $\term(r|_{\bar{p}_1'})\not=_E\term(r|_{\bar{p}_2'})$
and $\term(r|_{I(\bar{p}_1')})\not=_E\term(r|_{I(\bar{p}_2')})$
hold. Hence, this simple case is enough to prove the whole statement.

We write the set of equivalence classes of
$\{\term(r|_p) \mid p\in H_i\}$ and
$\{\term(r|_p) \mid p\in H_j\}$ modulo $E$ more explicitly as
$\{[t_{i,1}],\ldots,[t_{i,\alpha}]\}$ and $\{[t_{j,1}],\ldots,[t_{j,\beta}]\}$,
respectively.
Since
$\langle r_{H_i},r_{\check{H}_i}\rangle
\leq\langle r_{H_j},r_{\check{H}_j}\rangle$ holds,
$r_{H_i}\leq r_{H_j}$ also holds.
Thus, there exists an injective function
$I':\{1,\ldots,\alpha\}\to\{1,\ldots,\beta\}$
satisfying the following statement for each
$\delta$ in $\{1,\ldots,\alpha\}$
and each state $q$ of $\A$:
$
\big|\{ p\in H_i \mid \term(r|_p)=_E t_{i,\delta}
\wedge r(p)=q\}\big|
\leq
\big|\{p \in H_j \mid \term(r|_p)=_E t_{j,I'(\delta)}\wedge r(p)=q\}\big|
\quad (\dagger)
$.

In order to define
$I:H_i\to H_j$, we define $I$ for each of such
sets $\{p \in H_i \mid \term(r|_p)=_E t_{i,\delta}\wedge r(p)=q\}$
as any injective function
$
I:\{p \in H_i \mid \term(r|_p) =_E t_{i,\delta}\wedge r(p) = q \}
\to
\{p \in H_j \mid \term(r|_p) =_E t_{j,I'(\delta)}\wedge r(p) = q \} $, 
which is possible by the above inequality ($\dagger$).
The global $I$ is then injective thanks to the injectivity of $I'$.
Conditions ($C_2$) and ($C_3$) trivially follow from this definition.
\qed

\begin{exa}
Following our running example,
we first prove $\langle r_{H_4},r_{\check{H}_4}\rangle
\leq \langle r_{H_3},r_{\check{H}_3}\rangle$.
To this end just note that
$[\langle 0,0,0,0,1,0\rangle]\leq [\langle 0,0,0,0,1,0\rangle]$
and that $[\langle 0,0,0,1,0,0\rangle]\leq [\langle 0,0,0,2,0,0\rangle]$ hold.
We can define $I:(H_4\cup \check{H}_4\cup\mathring{H}_4)\to(H_3\cup\check{H}_3\cup\mathring{H}_3)$
from this relation according to Lemma~\ref{lemma-leqiffpumpinjection}.
Doing the  adequate guess we obtain the following definition:
$I(1)=1$,
$I(2)=2$,
$I(3)=3.3$
which is the pump-injection considered in Example~\ref{example-pump}
for our running example.
\end{exa}

The following lemma follows directly from the definition
of the sets $H_i$ and $\check{H}_i$, and allows to connect
such definitions with Lemma~\ref{lemma-computational-bound}.

\begin{lem}\label{lemma-boundconditions}
Let $\A$ be a $\PCTAGBCF$.
Let $a$ be the maximum arity of the symbols in the signature of $\A$.
Let $r$ be a run of $\A$. Then, the following conditions hold:
\begin{enumerate}[\em(1)] \topsep-1pt \itemsep-2pt
\item $|H_{\height(r)}|=1$ and $|\check{H}_{\height(r)}|=0$.
\item For each $i$ in $\{2,\ldots,\height(r)\}$,
$a\cdot |H_i|+|\check{H}_i|\geq |H_{i-1}|+|\check{H}_{i-1}|$.
\item For each $i$ in $\{1,\ldots,\height(r)\}$,
$|H_{i}|={\tt sum}(r_{H_i})$ and $|\check{H}_{i}|={\tt sum}(r_{\check{H}_i})$.
\end{enumerate}
\end{lem}
\proof
Item (1) is trivial by definition of $H_i$ and $\check{H}_i$
for $i=\height(r)$. 
For Item (2), it suffices to observe that the positions
in $H_{i-1} \cup \check{H}_{i-1}$ are all the positions in
$\check{H}_i$ plus a subset of all
child positions of positions in $H_i$,
and that each position has at most $a$ children. 
For Item (3) we just prove
$|H_{i}| = {\tt sum}(r_{H_i})$, since
$|\check{H}_{i}| = {\tt sum}(r_{\check{H}_i})$ can be proved analogously. 
We write the equivalence classes of the set
$\{ \term(r|_p) \mid p\in H_i \}$ modulo $E=E_\A$
more explicitly as $\{[t_1],\ldots,[t_\alpha]\}$.

Note that
$H_i$ is the disjoint union
$\{p\in H_i \mid \term(r|_p)=_E t_1\}\cup\ldots\cup
\{p\in H_i \mid \term(r|_p)=_E t_\alpha\}$.
Thus,
$|H_i|$ equals
$|\{ p\in H_i \mid \term(r|_p)=_E t_1\}|+\ldots+
|\{p\in H_i \mid \term(r|_p)=_E t_\alpha\}|$.
We conclude by observing that
$|\{p\in H_i \mid \term(r|_p)=_E t_1\}| = {\tt sum}(r_{t_1,H_i})$,
\ldots,
$|\{p\in H_i \mid \term(r|_p)=_E t_\alpha\}| =
{\tt sum}(r_{t_\alpha,H_i})$ hold.
\qed

\begin{lem}\label{lemma-thereisglobalpumping}
Let $B:\mathbb{N}\times\mathbb{N}\to\mathbb{N}$ be the
computable function of Lemma~\ref{lemma-computational-bound}.
Let $\A$ be a $\PCTAGBCF$.
Let $a$ be the maximum arity of the symbols in the signature of $\A$.
Let $n$ be the number of states of $\A$.
Let $r$ be a run of $\A$ satisfying $\height(r)> B(a,n)$.
\quad
Then, there is a global pumping on $r$.
\end{lem}

\proof
Consider the sequence
$\langle r_{H_{\height(r)}}, r_{\check{H}_{\height(r)}}\rangle$,\ldots,
$\langle r_{H_1},r_{\check{H}_1}\rangle$.
Note that the $n$-tuple $\langle 0,\ldots,0\rangle$
does not appear in the multisets of the pairs
of this sequence.
By Lemma~\ref{lemma-boundconditions},
$|H_{\height(r)}|=1$ and $|\check{H}_{\height(r)}|=0$ hold,
and for each $i$ in $\{2,\ldots,\height(r)\}$,
$a\cdot |H_i|+|\check{H}_i|\geq |H_{i-1}|+|\check{H}_{i-1}|$ holds.
Moreover,
for each $i$ in $\{1,\ldots,\height(r)\}$,
$|H_{i}|={\tt sum}(r_{H_i})$ and
$|\check{H}_{i}|={\tt sum}(r_{\check{H}_i})$ hold.
Thus,
${\tt sum}(r_{H_{\height(r)}})=1$,
${\tt sum}(r_{\check{H}_{\height(r)}})=0$,
and  for each $i$ in $\{2,\ldots,\height(r)\}$,
$a\cdot{\tt sum}(r_{H_i})+{\tt sum}(r_{\check{H}_i})\geq
{\tt sum}(r_{H_{i-1}})+{\tt sum}(r_{\check{H}_{i-1}})$ hold.
Hence, since
$\height(r)\geq B(a,n)$ holds, by Lemma~\ref{lemma-computational-bound}
there exist $i,j$ satisfying
$\height(r)\geq i>j\geq 1$ and
$\langle r_{H_i},r_{\check{H}_i}\rangle\leq
\langle r_{H_j},r_{\check{H}_j}\rangle$.
By Lemma~\ref{lemma-leqiffpumpinjection},
there exists a pump-injection
$I:(H_i\cup\check{H}_i\cup\mathring{H}_i)\to
(H_j\cup\check{H}_j\cup\mathring{H}_j)$.
Therefore, there exists a global pumping on $r$.
\qed

\begin{thm} \label{th-emptiness} \label{theorem-emptiness}
Emptiness is decidable for $\PCTAGBCF$.
\end{thm}
\proof
Let $a$ be the maximum arity of the symbols in the signature of $\A$.
Let $n$ be the number of states of $\A$.
Let $r$ be an accepting run of $\A$ with minimum height.

Suppose that $\height(r)\geq B(a,n)$ holds.
Then, by Lemma~\ref{lemma-thereisglobalpumping},
there exists
a global pumping $r'$ on $r$. By
Corollary~\ref{corollary-height}, $\height(r')<\height(r)$
holds. Moreover, by the definition of global pumping,
$r'(\rootp)=r(\rootp)$ holds.
Finally, by Corollary~\ref{corollary-pumpingisrun},
$r'$ is a run of $\A$. Thus, $r'$ contradicts the minimality of $r$.
We conclude that $\height(r) < B(a,n)$ holds.

The decidability of emptiness of $\A$ follows, 
since the existence of successful runs implies that one of them can be found among
a computable and finite set of possibilities.
\qed

Using Corollary~\ref{corollary-reductiontopctagc} and
Theorem~\ref{theorem-emptiness}, we can conclude the decidability of
emptiness for $\TAGBCF$, and more generally for
$\TAGBCF[\approx,\not\approx,{\mathbb{N}}]$.

\begin{cor} \label{corollary-emptiness}
Emptiness is decidable for $\TAGBCF$.
\end{cor}

\begin{cor} \label{corollary-arithmetic}
Emptiness is decidable for 
$\TAGBCF[\approx,\not\approx, \mathbb{N}]$. 
\end{cor}

%


\section{Unranked Ordered Trees} \label{section-unranked}
Our tree automata models and results can be generalized from
ranked to unranked ordered terms.
In this setting, $\F$ is called an \emph{unranked signature}, 
meaning that there is no arity fixed for its symbols,
i.e.\ that in a term $a(t_1,\ldots,t_n)$, 
the number $n$ of children is arbitrary
and does not depend on $a$.
Let us denote by $\U(\F)$ the set of unranked ordered terms over $\F$.
The notions of positions, subterms, etc., are defined for
unranked terms of $\U(\F)$ as for ranked terms of $\T(\F)$.

We extend the definition of automata for unranked ordered terms,
called hedge automata~\cite{Murata99}, with global constraints.
We do not consider constraints between brothers nor flat theories
in this setting.

\begin{defi} \label{definition-HAG}
A \emph{hedge automaton with global constraints} ($\HAGC$) 
over an unranked signature $\F$
is a tuple $\A = \langle Q, \F, F, \Delta, C\rangle$
where 
$Q$ is a finite set of {states}, 
$F \subseteq Q$ is the subset of final states,
$C$ is a Boolean combination of atomic constraints
of the form $q \approx q'$ or $q \not\approx q'$, with $q, q' \in Q$,
and $\Delta$ is a set of transition rules of the form
$a(L) \to q$ where $a \in \F$, $q \in Q$ and $L$ is a regular
(word) language over $Q^*$, assumed given by a 
finite state automaton with input alphabet $Q$.
\end{defi}

\noindent
We still use the notation $\HAGC[\tau_1,\ldots,\tau_n]$
where the types $\tau_i$ can be $\approx$, $\not\approx$,
${|.|}_\mathbb{N}$, $\|.\|_\mathbb{N}$, ${\mathbb N}$.

The notion of run of $\TAGC$ is extended to $\HAGC$ in the natural way.
A \emph{run} of a $\HAGC$ $\A$ is a pair $r = \langle t,M\rangle$ where
$t \in \U(\F)$ is an unranked ordered term and
$M$ is a mapping from $\Pos(t)$ into $\Delta_\A$ such that 
for each position $p \in \Pos(t)$ with $n$ children, 
if $M(p.1),\ldots,M(p.n)$ are rules with right-hand side states
$q_1,\ldots,q_n\in Q_\A$, respectively, 
then $M(p)$ is a transition rule of the form $t(p)(L) \to q$ in $\Delta$,
and the word $q_1\cdots q_n$ belongs to $L$.
Moreover, $r \models C_\A$, 
where satisfiability of $C_\A$ by $r$ is defined like in Section~\ref{section-definition}.
A run $r$ is called \emph{successful} (or \emph{accepting})  
if $r(\rootp) \in F_\A$.

\medskip
The emptiness decision results of Corollary~\ref{corollary-arithmetic}
can be transposed  
from $\TAGC$ into $\HAGC$
using a standard transformation from unranked to ranked binary terms,
like the \emph{extension} encoding described in~\cite{tata}, Chapter~8. 

\begin{figure}
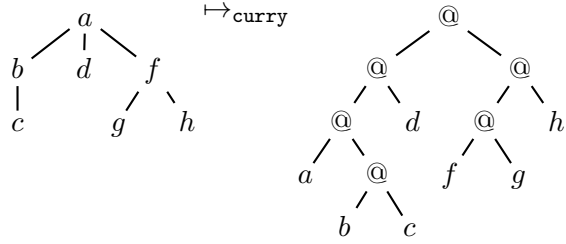

\[
\psset{nodesep=3pt, treesep=7mm, levelsep=7mm} 
\pstree{\TR{a}}{%
  \pstree{\TR{b}}{ \TR{c} }  
  \TR{d}
  \pstree{\TR{f}}{ \TR{g} \TR{h} }
}
\mapsto_\ext
\pstree{\TR{@}}{%
  \pstree{\TR{@}}{%
    \pstree{\TR{@}}{ \TR{a} \pstree{\TR{@}}{ \TR{b} \TR{c} } } 
    \TR{d} }
  \pstree{\TR{@}}{ \pstree{\TR{@}}{ \TR{f} \TR{g} } \TR{h} }
}
\]
\caption{Currying an unranked term.}
\label{figure-curry}
\end{figure}

Let us associate to the unranked signature $\Sigma$
the (ranked) signature 
$\F_@ := \{ a \ofarity 0 \mid a \in \F \} \cup \{ @ \ofarity 2 \}$
where $@$ is a new symbol not in $\F$.
The operator $\ext$ 
is a bijection from $\U(\F)$ into $\T(\F_@)$ recursively defined
as follows:
\[
\begin{array}{rcl}
\ext(a) & = & a \quad \mbox{for all}\; a \in \F\\
\ext\bigl(a(t_1,\ldots, t_n)\bigr) & = &
@\bigl(\ext\bigl(a(t_1,\ldots, t_{n-1})\bigr), \ext(t_n)\bigr)
\end{array}
\]
An example of application of this operator is presented in
Figure~\ref{figure-curry}. We extend the application
of the operator $\ext$ to sets of unranked ordered terms by 
$\ext(L) = \{ \ext(t) \mid t \in L \}$.

%
\begin{prop} \label{proposition-curry}
For all $\HAGC[\approx, \not\approx, \mathbb{N}]$ 
$\A$ over $\F$, 
one can construct effectively in PTIME a 
$\TAGC[\approx, \not\approx, \mathbb{N}]$ 
$\A'$ over $\F_@$
such that $\L(\A') = \ext\bigl(\L(\A)\bigr)$.
\end{prop}
\proof
Let $\A$ be $\langle Q, \F, F, \Delta, C \rangle$ more explicitly written.
Without loss of generality,
we assume that for each $a\in\F,q\in Q$,
the set of rules $\Delta$ contains exactly one transition
of the form $a(L) \to q$, and
we denote by $\bar{A}_{a, q}$ the NFA recognizing the corresponding language
$L$.
Recall that such automata have $Q$ as input alphabet.
Without loss of generality, we assume that the sets of states
of $\A$ and all $\bar{A}_{a,q}$ are pairwise disjoint.
Let $\bar{Q}$ be the union of all states of all the automata $\bar{A}_{a,q}$.
Intuitively, the transitions of the automaton $\A'$
will simulate both the transitions of $\A$
and the transitions of the NFAs $\bar{A}_{a, q}$,
when running on $\ext(t)$ for some $t \in \U(\F)$.

Let $\A' = \langle Q \cup \bar{Q}, \F, F, \Delta', C \rangle$
where $\Delta'$ contains the following transitions for each $a\in\F,q\in Q$:
\begin{iteMize}{$\bullet$}
\item[$\bullet$] $a \to  q$  if $\bar{A}_{a,q}$ recognizes the empty word,
\item[$\bullet$] $a \to \bar{q}$
where $\bar{q}$ is the initial state of $\bar{A}_{a,q}$, 
\item[$\bullet$] $@(\bar{q}, q') \to \bar{q}'$ if there is a transition
$\bar{q} \lrstep{q'}{} \bar{q}'$ in $\bar{A}_{a,q}$,
and 
\item[$\bullet$] $@(\bar{q},q') \to q$  if there is a transition
$\bar{q} \lrstep{q'}{} \bar{q}'$ 
in $\bar{A}_{a,q}$ and $\bar{q}'$ is a final state of $\bar{A}_{a,q}$.
\end{iteMize}

It is not difficult to see that there exists an accepting run
of $\A$ if and only if there exists an accepting run of $\A'$.
\qed

\begin{figure}
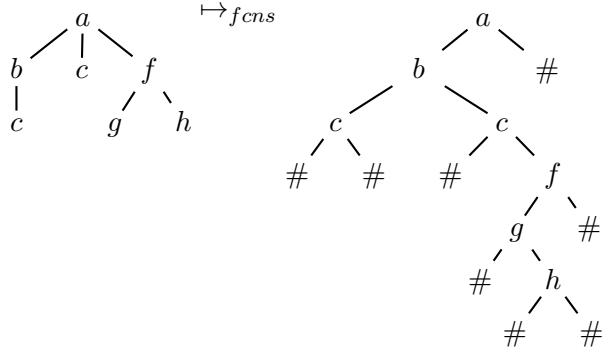

\[
\psset{nodesep=3pt, treesep=7mm, levelsep=7mm} 
\pstree{\TR{a}}{%
  \pstree{\TR{b}}{ \TR{c} }  
  \TR{c}
  \pstree{\TR{f}}{ \TR{g} \TR{h} }
}
\mapsto_{fcns}
\pstree{\TR{a}}{%
  \pstree{\TR{\mbox{~~~}b\mbox{~~~}}}{%
   \pstree{\TR{c}}{%
    \TR{\#}
    \TR{\#}
   }
   \pstree{\TR{c}}{%
    \TR{\#}
    \pstree{\TR{\mbox{~~~}f\mbox{~~~}}}{%
     \pstree{\TR{g}}{%
      \TR{\#}
      \pstree{\TR{h}}{%
       \TR{\#}
       \TR{\#}
      }
     }
     \TR{\#}
    }
   }
  }
  \TR{\mbox{~~~}\#\mbox{~~~}}
}
\]
\caption{First-child next-sibling encoding of an unranked term.}
\label{figure-fcns}
\end{figure}

There exist alternative encodings from unranked to ranked trees in the literature,
e.g., the first-child next-sibling encoding: see Figure~\ref{figure-fcns} for
an example of this transformation. This alternative encoding
makes the representation of equality and disequality between
subterms of the original unranked term difficult, since the transformed subterms
may have original siblings occurring now as their subterms.
For example, in Figure~\ref{figure-fcns}, the two occurrences of the subterm $c$
correspond to different terms in the result of the transformation.

The following emptiness decision result is a direct consequence of 
Proposition~\ref{proposition-curry} and Corollary~\ref{corollary-arithmetic}.
\begin{cor}
Emptiness is decidable for $\HAGC[\approx, \not\approx, \mathbb{N}]$. 
\end{cor}

\section{Logics on Trees}
\label{section-EMSO}
\label{section-logic}
In this section, we discuss the application of our results
to second order logics interpreted over domains defined by terms.
We propose a strict extension of the second order monadic logic of the tree
with equality, disequality and arithmetic constraints,
and show that satisfiability is decidable for this extension
thanks to a correspondence with $\TAGC[\approx,\not\approx, \mathbb{N}]$.
%

\subsection{MSO on Ranked Terms} \label{section-MSO-ranked}
A ranked term $t \in \T(\Sigma)$ over $\Sigma$ can be seen as 
a model for logical formulae, with an interpretation domain 
which is the set of positions $\Pos(t)$.
We consider monadic second order formulae interpreted on such models,
built with the usual Boolean connectors, 
with quantifications over first order variables (interpreted as positions),
denoted $x, y\ldots$
and over unary predicates 
(i.e.\ second order variables interpreted as sets of positions),
denoted $X, Y\ldots$,
and with the following predicates,

\begin{iteMize}{$\bullet$}
\item 
equality: $x = y$,
\item 
membership: $X(x)$, 
\item 
labeling: $a(x)$, for $a \in \F$
%
\item 
navigation: $S_i(x,y)$, for all $i$ smaller than or equal to the maximal arity of symbols of $\F$
(we call $+1$ the type of such predicates),
\item 
term equality: $X \approx Y$, 
term disequality: $X \not\approx Y$
(predicate types $\approx$ and $\not\approx$),

\item 
linear inequalities: 
$\sum a_i \cdot | X_i | \geq a$
or 
$\sum a_i \cdot \| X_i \| \geq a$,
where every $a_i$ and $a$ belong to $\mathbb{Z}$
(predicate types $|.|_{\mathbb{Z}}$ and
$\|.\|_{\mathbb{Z}}$).\vspace{3 pt}
\end{iteMize}

\noindent We write $\MSO[\tau_1,\ldots,\tau_k]$ for the set of monadic second order logic 
formulae with equality, membership, labeling predicates and 
other predicates of types
$\tau_1,\ldots,\tau_k$, amongst the above types $+1$, 
$\approx$, $\not\approx$,
and $|.|_{\mathbb{Z}}$, $\|.\|_{\mathbb{Z}}$.
We also use the notations 
$|.|_{\mathbb{N}}$ and $\|.\|_{\mathbb{N}}$ 
for natural linear inequalities
(linear inequalities whose coefficient all have the same sign)
and the abbreviations 
$\mathbb{Z}$ 
and 
$\mathbb{N}$ 
of Section~\ref{section-arithmetic}.

Let 
$\EMSO[\tau_1,\ldots,\tau_k]$ 
be the fragment of 
$\MSO[\tau_1,\ldots,\tau_k]$ 
containing the formulae of the form
$\exists X_1 \ldots \exists X_n\, \phi$
such that all the atoms of type $\approx$, $\not\approx$,
$\mathbb{Z}$ or $\mathbb{N}$ in $\phi$ 
involve only second order variables amongst $X_1,\ldots,X_n$.

A variable assignment into a term $t \in \T(\Sigma)$ 
is a function $\sigma$ mapping first order variables into positions of $\Pos(t)$ 
and second order variables into subsets of $\Pos(t)$.
The satisfiability of a formula $\phi$ by a term $t \in \T(\Sigma)$ 
and a variable assignment $\sigma$, 
denoted $t, \sigma \models \phi$ is defined in the usual Tarskian manner, with:

\begin{tabular}{l@{~}l@{~}lcl}
$t, \sigma$ & $\models$ & $x = y$ & iff & $\sigma(x) = \sigma(y)$\\
$t, \sigma$ & $\models$ & $X(x)$ & iff & $\sigma(x) \in \sigma(X)$\\
$t, \sigma$ & $\models$ & $a(x)$ & iff & $t(\sigma(x)) = a$\\
$t, \sigma$ & $\models$ & $S_i(x,y)$ & iff & $\sigma(x).i=\sigma(y)$\\
$t, \sigma$ & $\models$ & $X \approx Y$ & iff & 
$\forall p\in\sigma(X),p'\in\sigma(Y),p\not=p':\;t|_p = t|_{p'}$\\
$t, \sigma$ & $\models$ & $X \not\approx Y$ & iff & 
$\forall p\in\sigma(X),p'\in\sigma(Y),p\not=p':\;t|_p \not= t|_{p'}$\\
$t, \sigma$ & $\models$ & $\sum a_i \cdot | X_i | \geq a$ & iff &
$\sum_i a_i \cdot | \sigma(X_i) | \geq a$\\
$t, \sigma$ & $\models$ & $\sum a_i \cdot \| X_i \| \geq a$ & iff &
$\sum_i a_i \cdot \bigl| \{ t|_p \mid p \in \sigma(X_i) \} \bigr| \geq a$\\
\end{tabular}

\begin{exa}
The following formula of $\EMSO[\approx, \not\approx]$
expresses that all the subterms headed by $a$ in a term $t$ are
pairwise different:
$\exists X_a\, ((\forall x\, X_a(x) \leftrightarrow a(x)) \wedge X_a \not\approx X_a)$.
In other words, $a$ is used to mark monadic keys in $t$ (see Example~\ref{example-menu}).
\end{exa}

A seminal result of~\cite{ThatcherWright68} shows that
$\MSO[+1]$ has exactly the same expressiveness as \TA,
and therefore it is decidable.
The extension $\MSO[+1,\approx]$ is undecidable, 
see e.g.~\cite{FiliotTalbotTison07}.
The extension $\MSO[+1,|.|_\mathbb{Z}]$ is undecidable as well~\cite{Klaedtke02parikhautomata}.

On the other side, the fragment $\EMSO[+1,|.|_\mathbb{Z}]$ 
is decidable~\cite{Klaedtke02parikhautomata},
and a fragment of $\EMSO[+1,\approx,\not\approx]$
is shown decidable in \cite{FiliotTalbotTison08}
for a restricted variant of $\not\approx$,
using a two way correspondence 
between these formulae and a decidable subclass of \TAGED.

This latter construction can be straightforwardly adapted to establish 
a two way correspondence between $\EMSO[+1,\approx,\not\approx,\mathbb{N}]$
and $\TAGC[\approx,\not\approx,\mathbb{N}]$.

\begin{thm} \label{theorem-EMSO}
$\EMSO[+1,\approx,\not\approx,\mathbb{N}]$
is decidable on ranked terms.
\end{thm}
\proof
Following the same proof scheme as~\cite{FiliotTalbotTison08},
we show that for every closed formula $\phi$ in 
$\EMSO[+1,\approx,\not\approx, \mathbb{N}]$,
we can construct a 
$\TAGC[\approx,\not\approx, \mathbb{N}]$
recognizing exactly the set of models of $\phi$.
Then, the decidability of the logic follows from 
Theorem~\ref{corollary-arithmetic}.

Without loss of generality, we may assume
that $\phi$ is of the form
\[\exists X_1\ldots \exists X_n \;
  (\phi_0(\overline{X}) \wedge \phi_\approx(\overline{X}) \wedge \phi_\mathbb{N}(\overline{X}))
\]
where $\phi_0(\overline{X})$ is a $\MSO[+1]$ formula with free variables
$\overline{X} = X_1,\ldots, X_n$,
and
$\phi_\approx(\overline{X})$ 
and $\phi_\mathbb{N}(\overline{X})$
are Boolean combinations of atoms of the respective form 
$X_i \approx X_j$, $X_i \not\approx X_j$
and
$\sum a_i \cdot | X_i | \geq a$, $\sum a_i \cdot \| X_i \| \geq a$.
Moreover, we shall also assume that $\phi_\approx(\overline{X})$ 
and $\phi_\mathbb{N}(\overline{X})$ are conjunctions of
atoms or negations of atoms of the above form.
Otherwise, we put them into disjunctive normal form
and then split $\phi$ into an equivalent formula $\phi_1 \vee \ldots \vee \phi_k$,
where each $\phi_i$, $i \leq k$, is of the form requested:
$\phi_i = \exists X_1\ldots \exists X_n \;
  (\phi^i_0(\overline{X}) \wedge \phi^i_\approx(\overline{X}) \wedge \phi^i_\mathbb{N}(\overline{X}))$,
where $\phi^i_0(\overline{X}) \in \MSO[+1]$ and 
$\phi^i_\approx(\overline{X})$ and $\phi^i_\mathbb{N}(\overline{X})$
are conjunctions of atoms or negations of atoms as above,
and we solve satisfiability separately for each $\phi_i$.

First, we recall the definitions of~\cite{ThatcherWright68} of
the signature $\F \times \{ 0, 1 \}^n$,
where the arity of a symbol $\langle f, b_1,\ldots, b_n \rangle$
is the arity of $f$, and of the term $t\otimes\sigma$ over this
signature obtained, from a term $t$ over $\F$ and
a mapping $\sigma:\{X_1,\ldots,X_n\}
\to 2^{\Pos(t)}$, by
relabeling every position $p \in \Pos(t)$
by $\langle t(p), b_1,\ldots,b_n \rangle$, where for each $i \leq n$, 
$b_i = 1$ if $p \in \sigma(X_i)$ and $b_i = 0$ otherwise.
Also, from~\cite{ThatcherWright68} we get the construction of
a \TA $\A_0 = \langle Q, \F \times \{ 0, 1 \}^n, F, \Delta_0 \rangle$
which recognizes the set of terms 
$\{t \otimes \sigma \in \T\bigl(\F \times \{ 0, 1 \}^n\bigr) \mid 
 t, \sigma \models \phi_0(\overline{X})\}$.

Second, following a construction in~\cite{NiehrenPlanqueTalbotTison05},
we shift in $\A_0$ the bit-vectors from the signature into the state symbols,
obtaining a \TA 
$\A'_0 = \langle Q\times \{ 0, 1 \}^n, \Sigma, F\times \{ 0, 1 \}^n, \Delta \rangle$
where $\Delta$
contains all the transition rules 
\[
f\bigl( 
    \langle q_1, b_{1,1},\ldots, b_{1,n} \rangle, \ldots, 
    \langle q_m, b_{m,1},\ldots, b_{m,n} \rangle\bigr) 
   \to \langle q, b_1,\ldots,b_n \rangle
\]   
such that $f \in \F$,    
$\langle f, b_1,\ldots,b_n \rangle \bigl(q_1,\ldots,q_m\bigr) \to q \in \Delta_0$ 
and $b_{1,1},\ldots, b_{1,n}$, \ldots, $b_{m,1},\ldots, b_{m,n} \in \{ 0,1\}$.
This automaton $\A'_0$ recognizes the projection (on the first components)
of the terms recognized by $\A_0$,
i.e.\ it recognizes the set of terms
$t \in \T(\F)$ such that there exists 
$\sigma: \{ X_1,\ldots, X_n\} \to 2^{\Pos(t)}$
satisfying $t, \sigma \models \phi_0(\overline{X})$.

Third, we obtain a constraint $C$ 
by rewriting all the atoms of
$\phi_\approx(\overline{X})\wedge \phi_\mathbb{N}(\overline{X})$
with the following rules:
\[
\begin{array}{rcl}
X_i \approx X_j & \mapsto 
& \displaystyle\bigwedge_{b_i = b'_j = 1} 
  \langle q, b_1,\ldots,b_n \rangle \approx \langle q', b'_1,\ldots,b'_n \rangle\\
X_i \not\approx X_j & \mapsto 
& \displaystyle\bigwedge_{b_i = b'_j = 1} 
  \langle q, b_1,\ldots,b_n \rangle \not\approx \langle q', b'_1,\ldots,b'_n \rangle\\
\displaystyle\sum_{i} 
  a_i \cdot | X_i | \geq a & \mapsto 
& \displaystyle\sum_{i} \sum_{b_i = 1}
  a_i \cdot | \langle q, b_{1},\ldots, b_{n} \rangle | \geq a\\
\displaystyle\sum_{i} 
  a_i \cdot \| X_i \| \geq a & \mapsto 
& \displaystyle\sum_{i} \sum_{b_i = 1}
  a_i \cdot \| \langle q, b_{1},\ldots, b_{n} \rangle \| \geq a 
\end{array}
\]
The $\TAGC[\approx,\not\approx, \mathbb{N}]$
$\A = \langle Q\times \{ 0, 1 \}^n, \Sigma, F\times \{ 0, 1 \}^n, \Delta, C \rangle$
recognizes $\{ t \in \L(\A) \mid t \models \phi \}$.
\qed

The above transformation also works in the other direction
(this result is not necessary for the proof of Theorem~\ref{theorem-EMSO} though):
for every 
$\TAGC[\approx,\not\approx, \mathbb{N}]$,
we can construct a formula $\phi$ in 
$\EMSO[+1,\approx,\not\approx, \mathbb{N}]$,
whose set of models is $\L(\A)$.

Note that $\EMSO[+1,\approx]$ is strictly more expressive than 
$\MSO$, since the equality between subterms is not expressible in $\MSO$
(see e.g.~\cite{tata}).
The \TA construction of~\cite{ThatcherWright68}
for the decidability of $\MSO[+1]$
involves the closure under projection on components for
\TA languages over signatures made of tuples of symbols
(for the elimination of $\exists$ quantifiers).
$\TAGC{}$ languages are not closed under projection on 
some components of tuples, 
as it is already the case for simpler form tree automata 
with equality~\cite{treinen00fossacs}. Thus, the same approach
cannot be used to prove decidability of emptiness of $\TAGC{}$.

\subsection{MSO on Unranked Ordered Terms}  \label{section-MSO-unranked}
In unranked ordered terms of $\U(\Sigma)$, the number of children of a position is unbounded.
Therefore, for navigating in such terms with logical formulae, 
the successor predicates $S_i(x,y)$ of Section~\ref{section-MSO-ranked}
are not sufficient.
In order to describe unranked ordered terms as models,
we replace these above predicates $S_i$ by:
\begin{iteMize}{$\bullet$}
\item
$S_{\downarrow}(x,y)$ 
($y$ is a child of $x$),
\item 
$S_{\rightarrow}(x,y)$ 
($y$ is the successor sibling of $x$).
\end{iteMize}
The type of these predicates is still called $+1$.
Note that the above predicates $S_1,S_2,\ldots$ can be expressed
using these two predicates only.

The satisfiability of the above atoms by a term $t \in \U(\Sigma)$ 
and a variable assignment $\sigma$ is defined as follows:

\medskip
\begin{tabular}{l@{~}l@{~}lcl}
$t, \sigma$ & $\models$ & $S_{\downarrow}(x,y)$ & iff & 
there exists $i$ such that $\sigma(x).i=\sigma(y)$,\\
$t, \sigma$ & $\models$ & $S_{\rightarrow}(x,y)$ & iff &
there exists $p \in \Pos(t)$ and $i$ such that $\sigma(x) = p.i$\\ 
 & & & & and $\sigma(y) = p.(i+1)$.\\
\end{tabular}

\medskip
\noindent It is shown in~\cite{MuschollSS-PODS03} that the extension
$\MSO[+1,|.|_\mathbb{Z}]$ is undecidable for unranked ordered terms
when counting constraints are applied to sibling positions.

Using the results of Section~\ref{section-unranked}, 
and an easy adaptation of the automata construction in the proof of Theorem~\ref{theorem-EMSO},
we can generalize Theorem~\ref{theorem-EMSO} to $\EMSO$ over 
unranked ordered terms.
\begin{thm}
$\EMSO[+1,\approx,\not\approx,\mathbb{N}]$
is decidable on unranked ordered terms.
\end{thm}

\section{Conclusion}
We have answered (positively) the open problem of 
decidability of the emptiness problem for the 
$\TAGED$~\cite{FiliotTalbotTison08}, 
by proposing a decision algorithm for 
a class $\TAGBCF$ of tree automata with global constraints 
strictly extending the global constraints of $\TAGED$ in several directions.
Moreover, the $\TAGBCF$ combine the global constraints 
with local tests between brother subterms a la~\cite{BogaertTison92}
and equality interpreted modulo flat theories.
Our method for emptiness decision, presented in Section~\ref{section-emptiness}
appeared to be robust enough to deal with several extensions 
like global counting constraints, 
and generalization to unranked terms.

A challenging question would be to investigate 
the precise complexity of the emptiness problem,
avoiding the use of Higman's Lemma in the algorithm.
For instance, in~\cite{FiliotTalbotTison08}, it is shown,
using a direct reduction into solving positive and negative set 
constraints~\cite{CharatonikPacholski94b,GilleronTisonTommasi94,Stefansson94},
that emptiness is decidable in NEXPTIME for $\TAGED$
(i.e.\ for $\PCTAGC[\not\approx]$ modulo an empty theory and 
such that in every atomic constraint $q \not\approx q'$,
$q$ and $q'$ are distinct states).
On the other hand, the best known lower bound 
for emptiness decision for $\TAGBCF$ is EXPTIME-hardness
(this holds already for $\PCTAGC[\approx]$ as shown in~\cite{FiliotTalbotTison08}).

Another interesting problem mentioned in the introduction is
the combination of the $\HAGC$ of Section~\ref{section-unranked}
with the unranked tree automata with tests between siblings,
$\UTASC$~\cite{KL07,LW09}.
Perhaps, the techniques of Section~\ref{section-emptiness}
could help for the emptiness decision for a
formalism using for instance $\MSO$ binary querying 
(following e.g.~\cite{NiehrenPlanqueTalbotTison05})
for selecting the test position of global constraints.


Finally, another branch of research related to $\TAGBCF$ concerns 
automata and logics for \emph{data trees},
i.e.\ trees labeled over an infinite (countable) alphabet
(see~\cite{Segoufin06csl} for a survey).
Indeed, data trees 
can be represented by terms over a finite alphabet, 
with an encoding of the data values into terms.
This can be done in several ways, 
and with such encodings, the data equality relation
becomes the equality between subterms.
Therefore, this could be worth studying
in order to relate our results on $\TAGC$
to decidability results  on automata 
or logics on data trees like those in~\cite{JurdzinskiL07,BojanczykMSS09}.


\section*{Acknowledgement}

We thank Luc Segoufin for many valuable discussions and the anonymous referees 
at LICS 2010 for their useful comments and suggestions.


\bibliographystyle{alpha}
\bibliography{globalconstraints}

\end{document}